\documentclass[useAMS,usenatbib]{mn2e}
\usepackage{color}
\usepackage{aas_macros}
\usepackage{dcolumn}
\usepackage{amsmath}
\usepackage{amssymb}
\usepackage{longtable}
\usepackage{rotating}
\usepackage{graphicx}
\usepackage{subfigure}

\newcommand{\bagatelle}{\textsc{bagatelle}}
\newcommand{\galfit}{\textsc{galfit}}
\newcommand{\sex}{\textsc{SExtractor}}

\def\Reff{\ifmmode{R_\mathrm{eff}}\else{$R_\mathrm{eff}$}\fi}
\def\mueff{\ifmmode{\mu_\mathrm{eff}}\else{$\mu_\mathrm{eff}$}\fi}
\def\kms{\ifmmode{\mbox{km s}^{-1}}\else{$\mbox{km s}^{-1}$}\fi}
\def\mglob{\ifmmode{M_\mathrm{GC}}\else{$M_\mathrm{GC}$}\fi}
\def\mgal{\ifmmode{M_\mathrm{gal}}\else{$M_\mathrm{gal}$}\fi}
\def\menc{\ifmmode{M(r)}\else{$M(r)$}\fi}
\def\msun{\ifmmode{M_\odot}\else{$M_\odot$}\fi}
\newcommand\changed[1]{#1}

\begin{document}
\title[Nuclear star clusters in Coma dEs I.]{The HST/ACS Coma Cluster Survey. X. Nuclear star clusters in low-mass early-type galaxies: scaling relations}
\author[M. den Brok et al.]{Mark~den~Brok,$^{1,2}$\thanks{denbrok@physics.utah.edu} Reynier F.~Peletier,$^1$ Anil~Seth,$^2$ Marc~Balcells,$^{3,4,5}$
\newauthor Lilian~Dominguez,$^{3}$ Alister W.~Graham,$^{6}$ David~Carter,$^{7}$ Peter~Erwin,$^{8}$ 
\newauthor Henry~C.~Ferguson,$^{9}$ Paul~Goudfrooij,$^{9}$ Rafael~Guzm\'an,$^{10}$  Carlos~Hoyos,$^{11}$ 
\newauthor Shardha~Jogee,$^{12}$  John~Lucey,$^{13}$ Steven~Phillipps,$^{14}$ Thomas~Puzia,$^{15}$ 
\newauthor Edwin~Valentijn,$^1$  Gijs~Verdoes~Kleijn$^1$ and Tim~Weinzirl$^{12}$\\ 
$^1$Kapteyn Astronomical Institute, University of Groningen, P.O. Box 800, 9700AV Groningen, The Netherlands.\\
$^2$Department of Physics and Astronomy, University of Utah, Salt Lake City, UT 84112, USA.\\
$^{3}$Isaac Newton Group of Telescopes, Apartado 321, 38700 Santa Cruz de La Palma, Canary Islands, Spain.\\
$^{4}$Instituto de Astrof\'{\i}sica de Canarias, C. V\'{\i}a L\'{a}ctea, S/N, 38200 La Laguna, Tenerife, Spain.\\
$^{5}$Departamento de Astrof\'{\i}sica, Universidad de La Laguna, 38200 La Laguna, Tenerife, Spain.\\
$^{6}$Centre for Astrophysics and Supercomputing, Swinburne University of Technology, Hawthorn, Victoria 3122, Australia.\\
$^{7}$Astrophysics Research Institute, Liverpool John Moores University, IC2 Liverpool Science Park, 146 Brownlow Hill, L3, 5RF, UK.\\
$^{8}$Max-Planck-Insitut f\"ur extraterrestrische Physik, Giessenbachstrasse, D-85748 Garching, Germany\\
$^{9}$Space Telescope Science Institute, 3700 San Martin Drive, Baltimore, MD 21218, USA.\\
$^{10}$Department of Astronomy, University of Florida, P.O. Box 112055, Gainesville, FL 32611, USA.\\
$^{11}$Departamento de F\'isica Te\'orica, Facultad de Ciencias, Universidad Aut\'onoma de Madrid, Cantoblanco, 28049 Madrid, Spain\\
$^{12}$Department of Astronomy, University of Texas at Austin, Austin, TX, USA\\
$^{13}$Department of Physics, University of Durham, South Road, Durham DH1 3LE, UK\\
$^{14}$Astrophysics Group, H. H. Wills Physics Laboratory, Bristol University, Tyndall Avenue, Bristol BS8 1TL, UK\\
$^{15}$Institute of Astrophysics, Pontificia Universidad Cat\'olica de Chile, Avenida Vicu\~na Mackenna 4860, Macul, 7820436, Santiago, Chile\\
}
\maketitle
\begin{abstract}
We present scaling relations between structural properties of nuclear star clusters and their host galaxies for a sample of early-type dwarf galaxies observed as part of the Hubble Space Telescope (HST) Advanced Camera for Surveys (ACS) Coma Cluster Survey. We have analysed the light profiles of 200 early-type dwarf galaxies in the magnitude range $16.0 < m_{F814W} < 22.6 $ mag, corresponding to  $-19.0 < M_{F814W} < -12.4 $ mag.

Nuclear star clusters are detected in 80\% of the galaxies, thus doubling the sample of HST-observed early-type dwarf galaxies with nuclear star clusters. \changed{We confirm that the} nuclear star cluster detection fraction decreases strongly toward faint magnitudes. 
The luminosities of nuclear star clusters do not scale linearly with host galaxy luminosity. A linear fit yields L$_{nuc} \sim $L$_{gal}^{0.57\pm0.05}$. The nuclear star cluster-host galaxy luminosity scaling relation for low-mass early-type dwarf galaxies is consistent with formation by globular cluster accretion. We find that at similar luminosities, galaxies with higher S\'ersic indices have slightly more luminous nuclear star clusters. Rounder galaxies have on average more luminous clusters. 

Some of the nuclear star clusters are resolved, despite the distance of Coma. We argue that the relation between nuclear star cluster mass and size is consistent with both formation by globular cluster accretion and in situ formation. 

Our data are consistent with GC inspiraling being the dominant mechanism at low masses, although the observed trend with S\'ersic index suggests that in situ star formation is an important second order effect.
\end{abstract}
\begin{keywords}
galaxies: dwarf --- galaxies: nuclei --- galaxies: star clusters
\end{keywords}
\section{Introduction}
Nuclear star clusters (NSCs) are dense concentrations of stars in galaxy centres. Although their sizes, as measured in nearby spiral galaxies, are found to be comparable to globular clusters \citep{BokSarMcL04}, their luminosities can exceed the brightest Milky Way globular cluster by orders of magnitude. NSCs have been found in galaxies of all Hubble types, from sub-L$_\star$ early-type galaxies \citep{CotPiaFer06}, to late-type spiral galaxies \citep{CarStideZ97,BokvanVac99,MatGalKri99,BokvanMaz01,BokLaivan02}, early-type spiral bulges \citep{BalGraDom03, BalGraPel07} and in early-type dwarf galaxies \citep[dEs\footnote{We use the term dE both for dwarf elliptical as well as dwarf lenticular galaxies}:][]{GraGuz03,LotMilFer04,GraKuiPhi05,CotPiaFer06}. Also the Milky Way is known to have a NSC \citep{LauZylMez02,SchEckAle07,SchMerEck09}.

 There are several formation scenarios for NSCs, which can essentially be divided into two main scenarios: build-up through the accretion of (globular) star clusters, or by star formation {\it in situ}. 

The globular cluster (GC) accretion scenario was first developed by \citet{TreOstSpi75}. In this scenario, GCs gradually approach the galaxy centre as angular momentum is slowly removed by dynamical friction with stars. At the centre, the GCs merge to form a NSC. This model was later refined to account for the long dynamical friction time-scales \citep[e.g.][]{Mil04}. Because of their low velocity dispersion, dynamical friction of star clusters should be most efficient in low-mass dwarf galaxies \citep[e.g.][]{LotTelFer01}.

In the {\it in situ} formation scenario, gas is transported to the centre of a galaxy, where it cools and forms a nuclear star cluster. Several mechanisms to transport gas to the inner parsecs of gas-rich galaxies have been proposed \citep[e.g.][]{MihHer94,Mil04,HopQua11}. Since simulations and calculations involving hydrodynamics are more complex than pure gravitational computations, the theoretical framework for the dissipational formation of NSCs provides less quantitative predictions than the inspiraling cluster scenario.
 
There is evidence that both the {\it in situ} as well as the GC inspiraling formation happen in galaxies. Evidence for the GC inspiral scenario is seen in the lack of globular clusters in the inner parts of dwarf elliptical galaxies \citep{LotMilFer04}. However, the colours of some of the NSCs studied by these authors are inconsistent with the colours of the old populations typically seen in GCs. Similarly, nuclear star clusters in spiral galaxies show often evidence for recent star formation \citep{WalvanMcL05}. Evidently, there is no concensus on the dominant formation channel of NSCs in dEs. And if such a dominant formation channel exists, it may well be dependent on the morphological type or mass of the host galaxy or indirectly on the environment.

For the spiral galaxy NGC4244, \citet{HarDebSet11} infer from a detailed dynamical study that at least 50\% of the mass of the NSC was formed by gas accretion, but also that at least 10\% of the stars have to have formed outside the cluster. Such detailed studies are currently only possible for nearby galaxies. However, at larger distances, the study of scaling relations of NSCs may provide insights into the formation mechanisms. 

The luminosities of NSCs in dEs are known to scale with the host galaxy luminosity \citep{GraGuz03}. This relation was also analyzed by \citet{GraKuiPhi05,CotPiaFer06,TurCotFer12,ScoGra13}, although, due to sample selection, only for bright dEs ($M_B \lesssim -15$). The same groups have studied the sizes of NSCs in dEs as a function of NSC luminosity. It is unclear what these scaling relations look like for NSCs in low-mass dEs. In this paper, we address the scaling relations of NSC in low-mass early-type galaxies through the analysis of the light distribution of 200 dwarf ellipticals in the Coma cluster and try to infer the dominant formation mechanism for the NSCs. A majority of these dEs are located (in projection) close to the core of the cluster.

The GC inspiral scenario provides testable predictions for the mass-ratio of NSCs and host galaxies, as well as for the geometrical sizes of the NSCs, although not without some uncertainty, since also a widely accepted theory for the formation of GCs is still lacking. \citet{BekCouDri04} present scaling relations for the sizes of NSCs based on N-body simulations of the merging of equal-mass star clusters, however they do not predict a relation between host galaxy mass and NSC mass. \citet{Ant13} on the other hand contains a comprehensive overview of scaling relations for both formation mechanisms, most of which are analytically derived. 

One of the few qualitative predictions for the relation between NSC mass and host galaxy velocity dispersion for {\it in situ} formation is given by \citet{McLKinNay06}, where it is assumed that the winds from giant stars and supernovae in the star cluster provide feedback to the galaxy, in a way that is similar to supermassive black hole (SMBH) feedback. There is however direct evidence for episodic gas accretion into NSCs in  spiral galaxies. \citet{Emsvan08} analyze the tidal forces in S\'ersic galaxies, and find that in galaxy centres on the scale of the size of typical NSCs, tidal forces become disruptive in galaxies that are too concentrated (i.e. S\'ersic index $n>3.5$). They derive that the mass of a central massive object that would remove all compressive tidal forces scales linearly with host-galaxy mass. Although reaching this maximum CMO mass depends on the amount of gas available for star formation, if indeed tidal forces are a key ingredient in the formation of NSCs, the formation efficiency should depend negatively on the S\'ersic index.

It has been suggested that NSCs and central supermassive black-holes (SMBHs) follow a M$_{\mbox{cmo}}$-$\sigma$ scaling relation with the same slope \citep{FerCotDal06,WehHar06}, although this has been challenged \citep[see for example][]{BalGraPel07,Gra12, ErwGad12, NeuWal12,LeiBokKni12,ScoGra13}. The existence of the M$_{\mbox{cmo}}$-$\sigma$ relation implies that the evolution of nuclear star clusters and SMBHs may possibly somehow be linked. Likely there is some interaction between nuclear star clusters and SMBHs, such that one may prevent the growth of the other or even destroy it \citep{McLKinNay06,Mer09,NayWilKin09}. The Milky Way was the first galaxy known to host both a NSC and a SMBH. Although this could be interpreted as a peculiar case, it has been shown that also in other galaxies NSCs and SMBHs coexist \citep{SetAguLee08,GraSpi09}. 

Very massive ellipticals have been found to lack nuclear star clusters. \citet{Mer09} provides a mechanism to dissolve NSCs by absorbtion of energy from the surrounding galaxy, and shows that the addition of a SMBH to the NSC always leads to expansion of the NSC. \citet{BekGra10} suggest that a merger of intermediate-mass elliptical galaxies in which nuclear star clusters and black holes co-exist will result in an elliptical galaxy in which the newly-formed star cluster is structurally and dynamically altered so that it may easily get destroyed. However, \citet{Ant13} provides a different explanation for the lack of NSCs in galaxies with SMBHs (the argument was derived for dissipationless formation): the tidal forces from the SMBH rip inspiraling clusters apart at large radii from the centre, so that the NSC ends up having a much lower density than in the absence of a SMBH and may therefore not be properly disentangled from the host galaxy.

Whatever physical interaction is going on between NSCs and SMBHs, both trace the mass accretion to the central parsecs of a galaxy, and the study of NSCs can therefore provide insights in the formation of SMBHs and their host galaxies, as well as their haloes. 

Due to its exquisite resolution, the Hubble Space Telescope (HST) is a particularly good instrument for studying NSCs. The work of \citet{LotMilFer04} and \citet{CotPiaFer06} has shown that ground-based observations of nuclear clusters are often of too low resolution for studying remote nuclear star clusters, because a significant fraction of the clusters is missed. So far, only a small set of NSCs in dEs in Coma has been studied \citep{GraGuz03}. Coma is the {\it most convenient} very rich (Abell class 2) and dense cluster to study in the local Universe. In the centre, the projected density of faint galaxies is almost a magnitude higher than in Virgo \citep{WeiLisGuo11}. If the environment plays a role in shaping NSCs, Coma is the place where this could be measured best.

For this paper, we focus on the scaling relations of a sample of dwarf early-type galaxies in the Coma ACS survey. Despite the distance to Coma, we manage to double the number of know NSCs in dEs observed with HST.  

In Section \ref{sec:data} and \ref{sec:analysis}, we briefly summarize the Coma ACS survey, data reduction and analysis. Section \ref{sec:results} presents some basic results of this analysis. We discuss these results in Sec. \ref{sec:discussion}.

\section{Observations and Data reduction}\label{sec:data}
The Coma ACS Survey \citep[][henceforth C08]{CarGouMob08} provides data in two passbands for 25 fields pointed at the core of the Coma cluster and at the outskirts. The exposure times in the two passbands, F814W and F475W (which are roughly equivalent to the I$_C$ and $g$ band) were $\sim$1400 and $\sim$2600s. The original envisaged coverage of the cluster was much larger than 25 fields, but due to the ACS failure in January 2008 the survey was not completed. Of the 25 observed fields, five fields have data of lower sensitivity  (visits 1, 3, 12, 13 and 14, see also C08) because of a lack of data or poor quality. 

The pixel scale of the ACS data reduced with the drizzling scheme as described in C08 is 0.05'' per pixel. Throughout this paper we assume a distance to Coma of 100Mpc (see C08). This corresponds to $\sim25$ pc per pixel at the distance of Coma. The 10$\sigma$ detection limit of point sources is 26.8 in F814W(AB) and 27.6 in F475W(AB), although this is in optimal conditions.  

\subsection{Sample}
An advantage of observing the Coma cluster compared to nearby clusters such as the Virgo cluster, is that the relative distance uncertainty between galaxies in the cluster is low. The disadvantage is that membership of galaxies is not always well determined. 

In \citet{HamVerHoy10} more than 70,000 sources were detected in the F814W passband along the line of side of Coma. Most of these sources are either globular clusters in Coma or background galaxies. The best way to establish cluster membership is by spectroscopically confirming that a galaxy has a redshift consistent with the velocities of other cluster members. However, spectroscopic observations have targeted mainly bright galaxies. Luckily, faint dE cluster members stand out against the background because of their low surface brightness. 

We use a subset of the sample defined in Trentham et al. (in preparation, Paper XI), which is used for determining the luminosity function of galaxies in Coma. The selection of this sample was based on by-eye identification of galaxies on the ACS frames. The catalogue is complete, except at faint magnitudes. Faint galaxies ($M_{F814W} = -10$) with similar structural properties as the Local Group galaxy Draco should be detected in the ACS frames. However, we exclude very faint dEs to avoid the introduction of strong biases in the sample of nuclear clusters: it has been found that nuclear clusters are generally 5 to 6 magnitudes fainter than the host galaxy (see also Section \ref{sec:cor_mag}) so that detection of nuclear star clusters is probably not possible in the faintest galaxies, whether or not these galaxies are really nucleated.

At faint magnitudes, the dominant noise is a combination of the readnoise and the sky background noise with a small contribution of Poisson noise from the host galaxy (see also Sec. \ref{sec:cor_mag} and  Fig. \ref{fig:maghost_magnuc}). For a host galaxy absolute magnitude M$_{F814W} = -13$ mag, the difference between the host galaxy magnitude and the magnitude of a point source that could be detected at 3-$\sigma$ level is expected to be $\sim 4.7$ mag (and lower than 4 for  M$_{F814W} = -11$).

We decided therefore to clip the faint part of our sample at $m_{F814W}$(AB) = 22.6 ($M_{F814W}$(AB) = $-12.6$, $M_B \approx -11.0$), also because obtaining reliable structural parameter fits of the host galaxy becomes difficult for galaxies that are fainter than this magnitude. We use the F814W magnitudes for clipping since it is a better proxy of stellar mass than the F475W magnitudes. We also introduce a limiting magnitude on the bright side of the magnitude range since for this paper we focus exclusively on nuclear star clusters in dwarf elliptical galaxies: it has been found that low-mass (non-dwarf) ellipticals often have a central light excess, but lack the distinct surface brightness bump and often distinct colours seen in lower-luminosity galaxies \citep[e.g.][]{CotPiaFer06,KorFisCor09}. Although scaling relations of the luminosity and size of this nuclear excess light and nuclear star clusters with host galaxy properties may be continuous, the formation mechanisms are not necessarily the same. The other reason for discarding high-mass galaxies is that they often require more than two components for the fitting \citep[e.g.][]{AguIglVil04,HuaHoPen13, WeiJogNei14,LasFervan14,JanLauLis14, DulGra13, HeaLucHud14}. We therefore only use galaxies fainter than $M_{F814W}$(AB) = $-19.0$.

\begin{figure}
\begin{minipage}{82mm}
\center
\scalebox{0.45}{\includegraphics{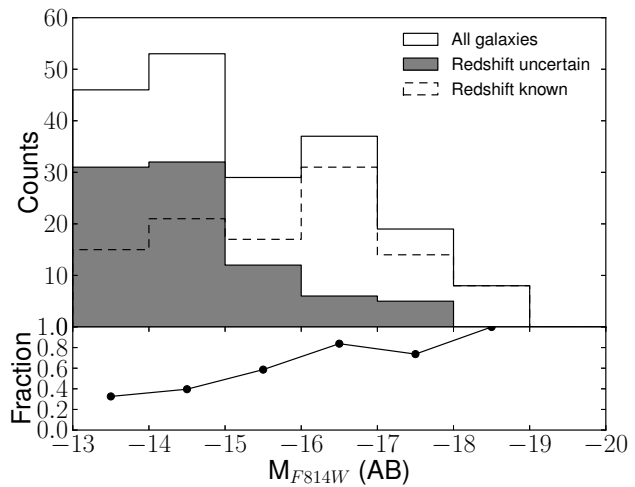}} 
\caption{Counts of galaxies with known and unknown redshift the fraction of those as a function of absolute magnitude of the host galaxy (for likely members galaxies). The black histogram shows the distribution of sources without redshift information as a function of host galaxy magnitude. The dashed histogram shows the sources which are confirmed member galaxies.}\label{fig:hist_z}
\end{minipage}
\end{figure}

Our sample contains both spectroscopically confirmed members and possible members, which were identified by eye (by NT and HF). In paper XI, these possible members are categorized into 3 classes: almost certainly members (class 1), likely members (class 2) and possible members (class 3). Spectroscopic follow-up of a subset of possible members in the paper XI sample has shown that the contamination of non-members in class 3 is around 50\%, and approximately 10\% for sources in class 2 \citep{ChiTulMar10}. A distribution of confirmed and non-confirmed sources over magnitude can be found in Fig. \ref{fig:hist_z}. The number of sources in our magnitude-limited sample with successful structural parameter fits that are in class 0 (confirmed members) is 104, in class 1 is 13, in class 2 is 58 and in class 3 is 23. In this paper, we analyze galaxies in all classes, but since membership is uncertain for class 3 objects, we exclude them from our measurements of scaling relations. For clarification, we note that whenever we refer to the 'full sample', we mean the magnitude-limited sample containing objects from all 4 classes, whereas with 'likely members' and 'probable members' we denote the magnitude-limited sample of galaxies in classes 2 and lower. We note that a number of galaxies were excluded from our analysis, because they either had late-type or irregular features such as ISM or significant spiral arms, were compact ellipticals, or were very difficult to fit, usually because they had a close bright companion. These sources are listed in Apx. \ref{apx:excl}. Similarly, due to the chosen dithering scheme, galaxies close to the edges of the frame suffer from lower signal-to-noise and cosmic rays. A handful of these galaxies also ended up in Apx. \ref{apx:excl}. For sources that appeared in multiple frames, we chose to fit the source in the frame with the best signal-to-noise. We note that, primarily due to the curtailment of the ACS survey, 90\% of our sample galaxies are at a projected distance of 500kpc from the core, whereas the virial radius of Coma is about 3Mpc \citep{okaMam03}.

\section{Analysis}\label{sec:analysis}
\subsection{Choice and justification of surface brightness profile parametrization}
The stellar populations of NSCs are distinct from those of the host galaxy \citep[e.g.][]{LotMilFer04}, so that treating the host galaxy and the NSC as distinct morphological components is justified. We use the S\'ersic model \citep{Ser68} to describe the light profile of the host galaxy. This profile is a generalization of the exponential profile, which has been used to fit disc galaxies \citep{Pat40,Fre70}, and the $R^{1/4}$-profile, commonly used for elliptical galaxies \citep{deV48}.  The S\'ersic profile is given by  
\begin{eqnarray*}
I(R) = I_e \exp\left\{{-b_n\left[\left(R/R_e\right)^{\frac{1}{n}}-1\right]}\right\},\label{eq:sersic}
\end{eqnarray*}
and provides an accurate description of the light profile of low-mass early-type galaxies outside the centre \citep{CaoCapDOn93,DOnCapCao94}.  $I_e$ is the surface brightness at the effective radius. The S\'ersic index $n$ defines the curvature of the profile; profiles with high $n$ are more cuspy. $b_n$ plays the r\^ole of a normalisation constant, and can be approximated by $b_n \approx 0.19992n-0.3271$ for $0.5 < n < 10$ \citep{Cap89}. A comprehensive review of the S\'ersic model can be found in \citet{GraDri05}

If NSCs in Coma follow the same size distribution as those in Virgo \citep{CotPiaFer06}, we expect most NSCs to be unresolved. Our first choice of modelling is therefore to fit them as a Point Spread Function (PSF) convolved point source. However, since some of them may be marginally resolved, we also fit them with a Gaussian, similar to \citet{GraGuz03} and \citet{KouKhoCar12}. Although there is no physical justification for using a Gaussian profile, the use of a more complicated profile is not warranted by our data, and the Gaussian profile has one fewer free parameters than a King model. 

\subsection{Fitting code}
We make use of \texttt{Bagatelle} (see Appendix \ref{apx:bag}), a 2-dimensional Bayesian fitting code for modelling surface brightness images of galaxies. The advantage of this code over other codes is that it allows for i) a full exploration of the posterior distribution, which can help quantify any degeneracies, and ii) a quantitave way of deciding which profile fits a galaxy best; given two model profiles with priors on their parameters, the code calculates and compares the marginalization over all variables, known as the Bayesian {\it evidence}, of the profiles. We describe the fitting code in more detail in Apx. \ref{apx:bag} and give a comparison between the results of fits carried out by \citet{WeiJogNei14} with \galfit\ and our fits in Apx. \ref{apx:comp_tw}.

\subsection{Masking}
Neighbouring stars and galaxies were masked according to the \sex segmentation maps. As masking in this way was generally not sufficient, the \sex mask apertures were enlarged to an ellipse with major and minor axes of $4\times$\texttt{A\_IMAGE} and $4\times$\texttt{B\_IMAGE}. As close neighbours are sometimes missed by \sex, we manually checked the masks and tweaked them where necessary.

\subsection{Point Spread Function}
When fitting sources which are unresolved or marginally resolved, detailed knowledge of the Point Spread Function (PSF) is crucial. For our analysis, we have chosen to use the artificial PSFs from {\sc TinyTim}, drizzled in the same way as our HST/ACS data using a script called {\sc DrizzlyTim}\footnote{DrizzlyTim is written by Luc Simard} \cite[for details see][]{HoyDenVer11}. Given the importance of the PSF, we have also determined empirical PSFs. The PSF of the ACS camera varies with position on the chip because the camera is mounted off-axis, and varies also as a function of time, since thermal fluctuations change the focus of the telescope \citep[e.g.][]{RhoMasAlb07}.

To construct the empirical PSFs, we used archival data of the Galactic globular cluster 47 Tuc. The exposure time of both the Coma and 47 Tuc observations are shorter than one HST orbit, which is the main time-scale of variation \citep{RhoMasAlb07}. Moreover, the time between the observations of the two targets is longer than 6 months. Although the empirical PSF will not be the optimal representation of the real PSF, it will at least show us how the uncertainty in the shape of the PSF affects our analysis.

\begin{figure}
\begin{minipage}{82mm}
\center
\scalebox{0.4}[0.4]{\includegraphics{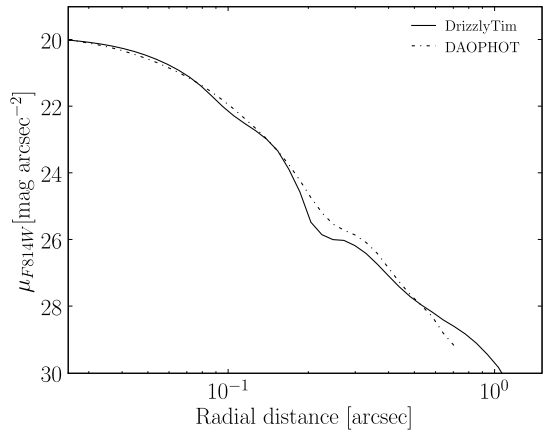}}
\caption{Normalised profiles of the F814W psf at the location $x,y=(3300,2400)$ The solid line is the \textsc{DrizzlyTim} PSF, the dash-dotted line is the empirical PSF.}\label{fig:psfcomplog}
\end{minipage}
\end{figure}

The drizzled 47 Tuc data were processed using the DAOPHOT II package in IRAF \citep[e.g.][]{Ste87}. In overview and summary, we detected stars with the task \texttt{daofind} and calculated a PSF using 669 stars which were relatively bright, isolated and not saturated, with the task \texttt{psf}. After calculation of a PSF, stars near the PSF stars were subtracted from the image and the PSF was recalculated. We used a few such iterations, where during the last iterations we let the PSF vary quadratically as a function of position. We assumed a Moffat profile with $\beta=1.5$ for the analytic component of the PSF, although other analytic profiles fitted equally well. A comparison between radial surface brightness profiles of the empirical PSF and the {\sc DrizzlyTim} PSF is given in Fig. \ref{fig:psfcomplog}. The empirical PSF is slightly broader than the theoretical PSF. The mismatch between the profiles of the two PSFs in the outer parts is a consequence of estimating the sky close to PSF stars because of crowding in the 47 Tuc observations. For detecting and measuring sizes of star clusters, and for the photometry of the main body of the galaxy, these wings are not relevant. However, it may cause us to underestimate the flux of a nuclear star cluster. Based on the differences between the empirical and TinyTim PSF in both F475W and F814W, we expect a systematic flux error of 2\% in magnitude, and less so in the colour, since it is a differential measurement.

\begin{figure*}
\begin{minipage}{172mm}
\center
\scalebox{0.42}[0.42]{\includegraphics{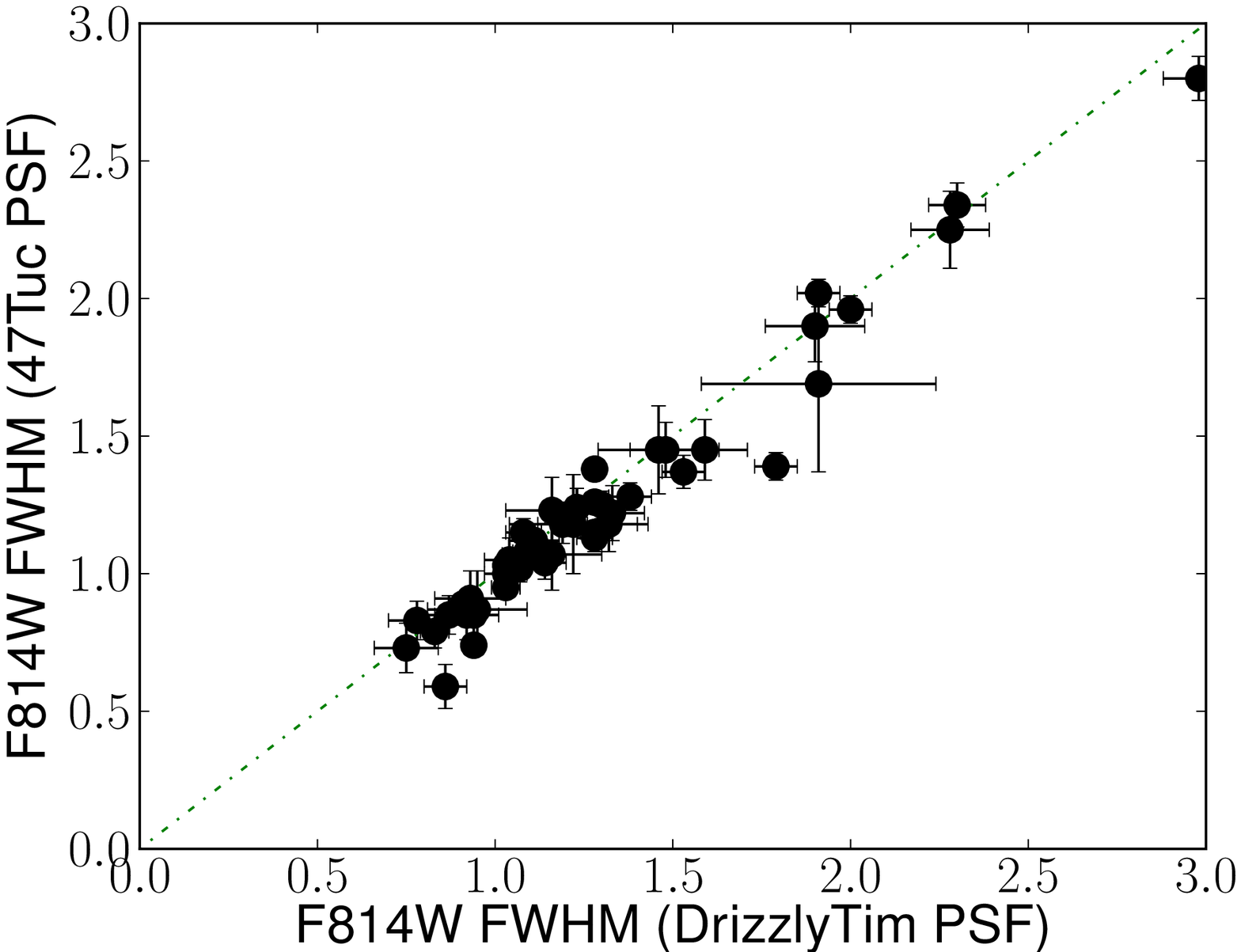}}
\scalebox{0.42}[0.42]{\includegraphics{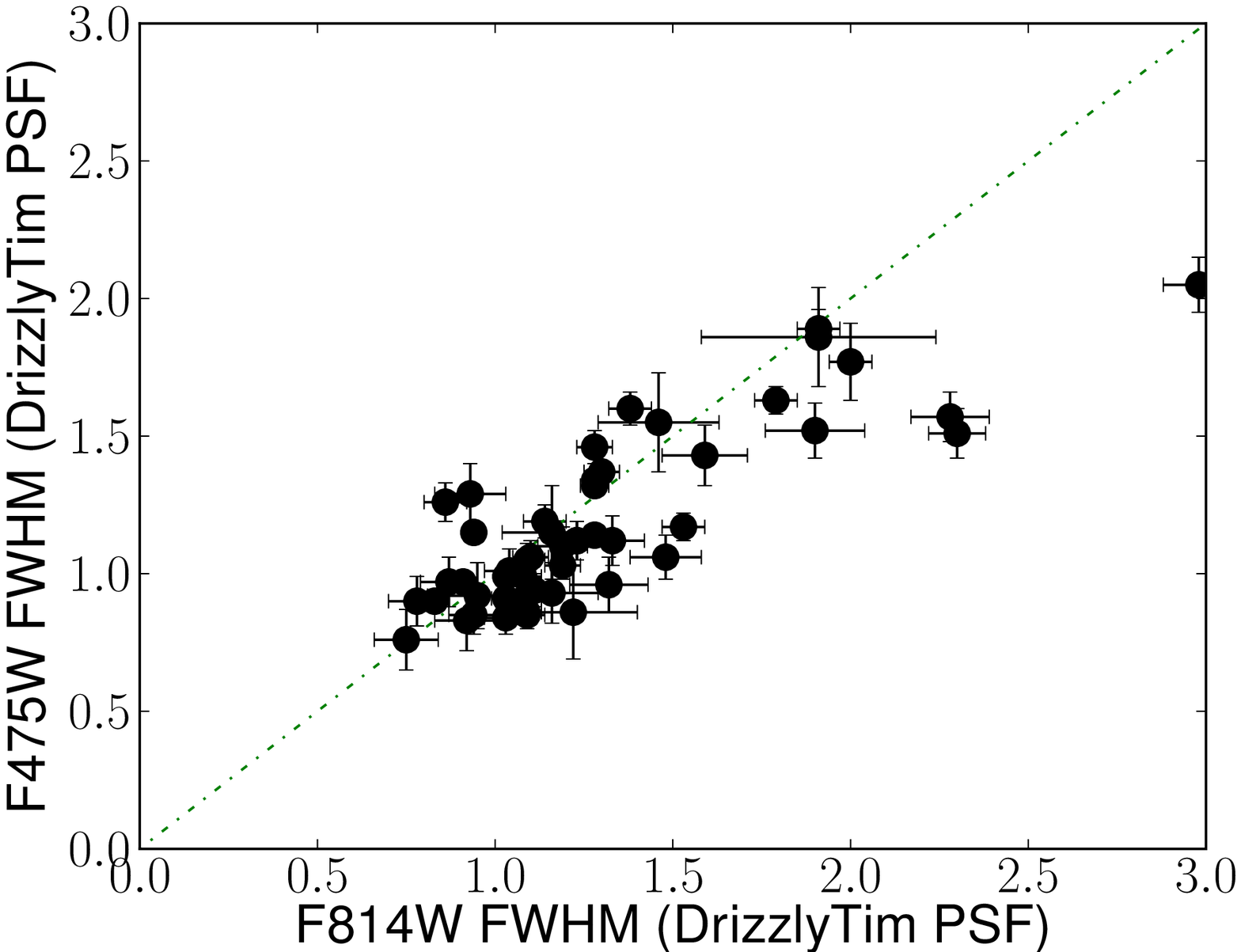}}
\caption{{\bf Left:} F814W sizes of NSCs (in pixels) as measured with the empirical PSF vs F814W sizes of NSCs as determined with a theoretical PSF. {\bf Right:} F475W sizes of NSCs (in pixels) vs F814W sizes of NSCs. Both sizes were measured using a {\sc DrizzlyTim} PSF convolved with a gaussian, while in parallel fitting the host galaxy with a S\'ersic profile. }\label{fig:sizes_814_tuc_dt} \label{fig:sizes_475_814}
\end{minipage}
\end{figure*}

During the course of this work, we found that several NSCs were resolved. To ensure that the sizes of the NSCs were not due to problems with the PSF, we measured sizes with both the empirical and theoretical PSFs. The left panel of Fig. \ref{fig:sizes_814_tuc_dt} shows that the measured sizes are not dependent on which PSF is used. To check that the sizes are not due to data-related issues, we show in the right panel of Fig. \ref{fig:sizes_475_814} the sizes of resolved star clusters in the F814W band versus those in the F475W band. Although there is considerable scatter, the correlation suggests that the clusters are indeed resolved.


\section{Results}\label{sec:results}
\subsection{Detection fraction}
\begin{figure}
\begin{minipage}{82mm}
\center
\scalebox{0.45}{\includegraphics{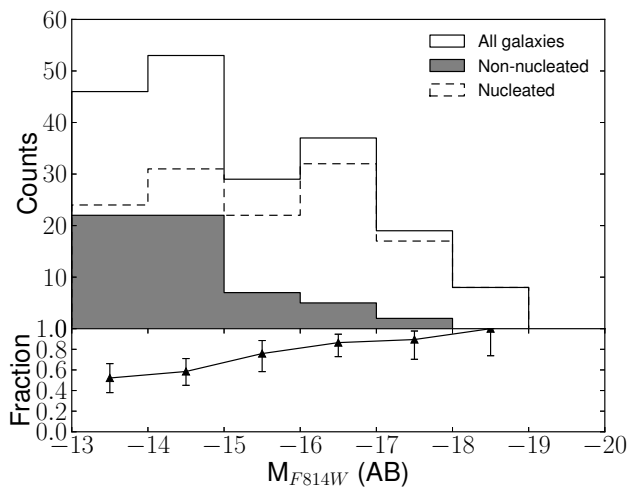}} 
\caption{Counts of nucleated and non-nucleated galaxies and frequency of nucleation as a function of absolute magnitude of the host galaxy (for likely members galaxies). The black histogram shows the distribution of non-nucleated sources as a function of host galaxy magnitude. The dashed histogram shows the nucleated galaxies. Since faint nuclear star clusters may have been missed, the nucleation fraction is denoted as a lower limit for the points in the lower panel. The error bars denote the 95\% Jeffreys interval.}\label{fig:hist_nucl}
\end{minipage}
\end{figure}

 Fig. \ref{fig:hist_nucl} shows a histogram of galaxy counts as a function of absolute F814W magnitude, for nucleated and non-nucleated galaxies. We confirm the result of \citet{GraGuz03} and \citet{CotPiaFer06} that the nuclear star cluster fraction in bright dwarf galaxies is close to unity. Faint dEs show a lower fraction of nucleation, qualitatively consistent with results of \citet{SanBinTam85,van86}. However, it is not inconceivable that we have missed faint nuclei; the nucleation fractions in faint dEs are therefore likely lower-limits. In fact, injecting the faintest genuine nucleated galaxy (VCC1895) from the ACS Virgo Cluster Survey with updated structural parameters and noise added in a Coma image, we do not recover the nucleus. Over the full luminosity range, the fraction of non-nucleated galaxies is 16\%, similar to the 13\% found by \citet{GraGuz03} for luminous dEs in Coma.  

Even though this number is in agreement with the number found by \citet{CotPiaFer06}, we note that the average galaxy luminosity of the Virgo sample is higher. The samples have mean and median magnitude difference of 4.0 and 3.8 mags (2.8 and 2.7 for the nucleated galaxies in both samples). Given that our sample is volume and luminosity-complete, it is representative of the luminosity function of dwarf ellipticals in the core of the Coma cluster. 

As a possible caveat, we note that the detection fraction of NSCs in visits 3 and 13, both with only half the planned exposure time, is lower than for the visits with complete exposure times in the core of the cluster. The total detection fraction of NSCs between $-16 >$ M$_{F814W} > -18 = 92 \pm 7$\% is consistent with the detection fraction of $85 \pm 5$\% in the z-band for galaxies in the same magnitude range in C\^ot\'e et al. The small difference in detection fraction can probably be explained by different sample properties, since we have excluded galaxies with late-type features.

\subsection{Luminosity function}
\begin{figure}
\begin{minipage}{82mm}
\center
\scalebox{0.4}[0.4]{\includegraphics{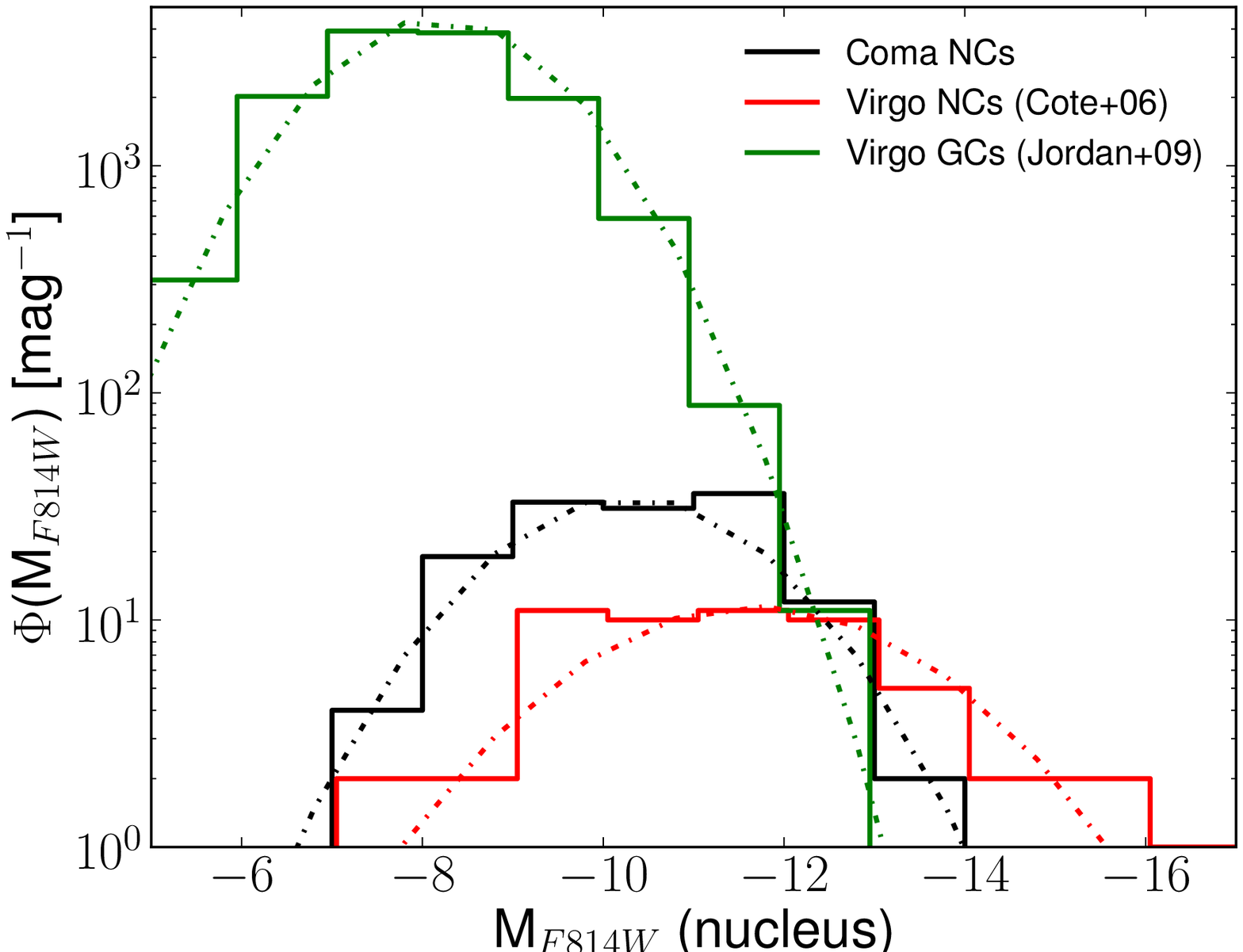}}
\caption{Luminosity function of NSCs in the F814W band. Also shown are the NSCs of galaxies in Virgo \citep[][red]{CotPiaFer06} and GCs in Virgo \citep[][green]{JorPenBla09}, with the best-fitting Gaussians shown as dash-dotted lines.}\label{fig:magplumfunc}
\end{minipage}
\end{figure}

In Fig. \ref{fig:magplumfunc} we show the luminosity function of the nuclear star clusters, together with the luminosity function of NSCs in the Virgo cluster from Cot\'e et al. and the luminosity function of GCs in Virgo \citep{JorPenBla09}. The luminosity function is well-described by a Gaussian, although we note that this functional form for the luminosity function is somewhat ad hoc and depends on survey depth \citep{TurCotFer12}. 

The nuclei in our sample are slighly fainter than the NSCs in the ACS Virgo Cluster Survey sample (by about 1.5 mag. in the F814W band), which is a consequence of the Coma sample containing more low-mass galaxies.

The faintest NSCs in our sample coincide with GCs at the turnover magnitude. For both formation scenarios there are qualitative arguments why there could be a lower limit to the NSC mass: in the accretion scenario, low-mass GCs have much longer dynamical friction time-scales, thus the formation of low-mass NSCs is suppressed; in the {\it in situ} scenario, infant mortality may be higher for low-mass star clusters in addition to feedback effects related to putative central black holes. The fact that the faintest NSCs coincide with the turnover magnitude of GCs in Virgo may also be coincidence. We note however that, except for the very brightest nuclei (F814W $\approx -14$), most NSC magnitudes are consistent with being drawn from the massive end of the GC luminosity function. 

\subsection{Luminosity scaling relations}\label{sec:cor_mag}
\begin{figure*}
\begin{minipage}{\textwidth}
\center
\scalebox{0.42}[0.42]{\includegraphics{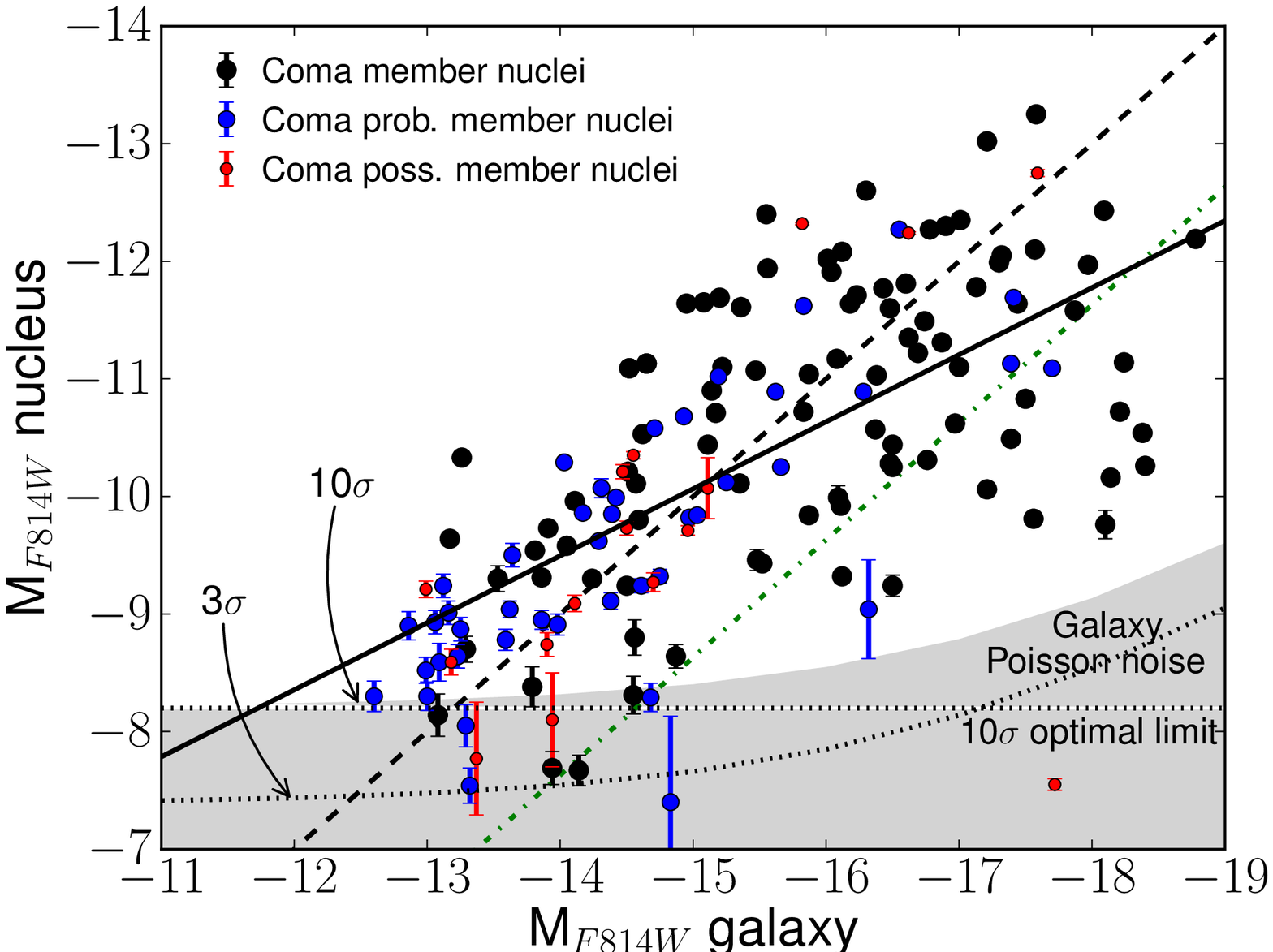}}
\scalebox{0.42}[0.42]{\includegraphics{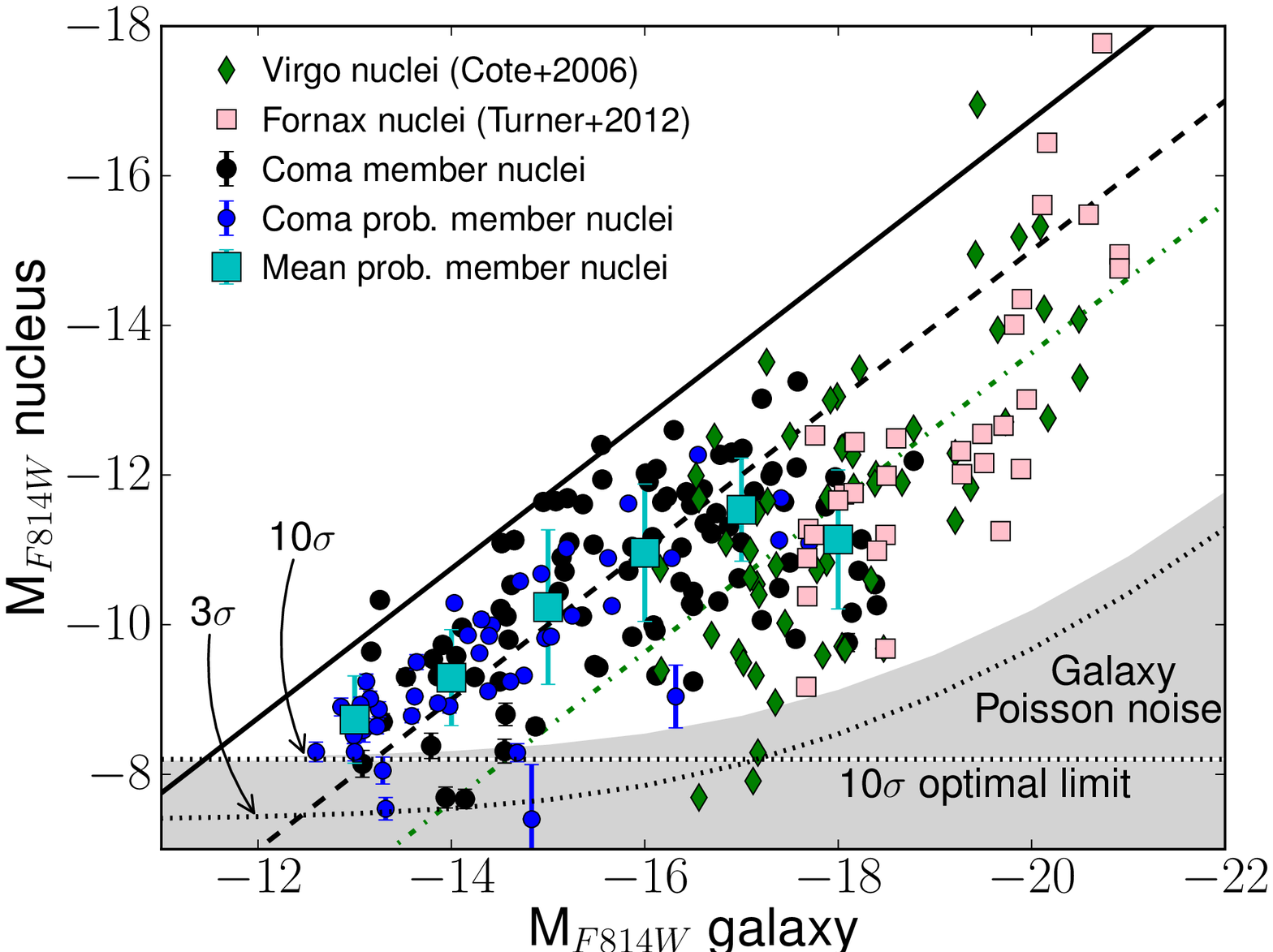}}
\caption{{\bf Left:} Absolute F814W(AB) magnitudes of nuclear star clusters versus the absolute F814W(AB) magnitudes of the host galaxy for Coma members. The magnitudes were obtained by profile fitting. The host galaxy luminosity does not include the star cluster luminosity. The dashed line denotes our best fitting line with unit slope. The green dash-dotted line represents a similar fit from Cot\'e et al. The black solid line shows our best fitting linear relation with non-unity slope. The grey areas approximate the regions where sky and galactic Poisson noise become dominant.  {\bf Right:} Same plot, but now only containing members and probable members of the Coma cluster. Overplotted in green diamonds and pink squares are the uncorrected (see text) F850LP(AB) measurements from \citet{CotPiaFer06} and \citet{TurCotFer12}. We also overplot in big cyan squares the mean NSC magnitude and error on the mean by binning the likely members of Coma by host galaxy magnitude. The black solid line denotes the line where the NSC would contain 5\% of the light of the host galaxy.}\label{fig:maghost_magnuc}
\end{minipage}
\end{figure*}

Several authors have pointed out a relationship between NSC luminosity and host galaxy luminosity in dEs \citep[e.g.][]{GraGuz03,LotMilFer04,GraKuiPhi05,CotPiaFer06}. With our data set, we double the number of measurements for this relationship. Fig.\ \ref{fig:maghost_magnuc} shows the nuclear star cluster magnitude as a function of host galaxy magnitude in the F814W band. The host galaxy magnitude and NSC magnitude were determined as described in Section \ref{sec:analysis}. The magnitude of the host galaxy was determined by extrapolating the S\'ersic profile to infinity, and does not include the nuclear star cluster. Because the nuclear star cluster is on average 5-6 magnitudes fainter, this should only make a marginal difference ($\sim0.01$ mag) in the magnitude of the host galaxy. The region where the detection limit for point sources is below 10$\sigma$ is indicated by the grey area below the dotted line. At low magnitudes, the sky dominates in the detection efficiency, whereas at high magnitudes, the Poisson noise of the centre of the host galaxy is dominant. In the same plot we also show data from nuclear star clusters in the ACS Virgo Cluster Survey \citep{CotPiaFer06} and the ACS Fornax Cluster Survey \citep{TurCotFer12}. The conversion between F850LP and F814W is small (generally less than 0.2 mag) and holds for both host galaxy and star cluster -- it is therefore unlikely that increased scatter is due to this conversion. 

The magnitude-magnitude relation of Coma cluster NSCs and their host galaxies actually extends the relation found for Virgo and Fornax toward fainter magnitudes, although with a change of slope. Interestingly, the nucleus-galaxy luminosity ratio is higher than the one found by C\^ot\'e et al. Performing a linear fit with fixed slope of one, we find (for the F814W band), for weighted and unweighted fits:
\begin{eqnarray}
M_{\mbox{nuc}} = M_{\mbox{gal}} + (5.13  \pm 0.10) \\
M_{\mbox{nuc}} = M_{\mbox{gal}} + (5.54  \pm 0.10)
\end{eqnarray}
This is in agreement with earlier values \citep{GraGuz03,GraKuiPhi05,LotMilFer04} (Table\ \ref{tab:scaling}). Cot\'e et al. speculate that the difference they find with earlier findings may be due to the use of different fitting functions or the higher sensitivity of their ACS observations, which allows them to detect fainter nuclei. Performing a fit with a free slope, we find
\begin{eqnarray}
M_{\mbox{nuc}} = (0.57 \pm 0.05)(M_{\mbox{gal}} + 17.5) - (11.49  \pm 0.14),
\end{eqnarray}
where we assumed uniform errorbars for all data points, since otherwise the fit was strongly biased by the few brightest NSCs in our sample. In fact, fitting the relation with a constant intrinsic Gaussian scatter gives an intrinsic dispersion of 0.9 mag, not very different from unit errorbars. Our slope is marginally consistent (overlapping uncertainties) with the slope found by Graham and Guzman but less steep then what was found by Grant et al. The intersect is consistent with the value given by \citet{GraGuz03} but is about 0.5 mag off from the fitted value from Cot\'e et al. Our slope agrees with those of \citet{BalGraPel07} ($0.76 \pm 0.17$)  and \citet{ScoGra13} ($0.60 \pm 0.10$).

\begin{table*}
\begin{tabular}{llcl}
Source & Relation & Band & Comments \\
\hline
\citet{GraGuz03} & M$_\mathrm{nuc}$ = ($0.87 \pm 0.26$) (M$_\mathrm{gal}$ + 17.5) - ($11.90 \pm 0.25$)& F606W & outliers clipped\\
\citet{GraKuiPhi05} &  M$_\mathrm{nuc}$ =($0.74 \pm 0.06$) (M$_\mathrm{gal}$ + 17.5) - ($12.59 $)& I-band & zeropoint estimated from plot\\
\citet{CotPiaFer06} & M$_\mathrm{nuc}$ =($1.0$) (M$_\mathrm{gal}$ + 17.5) - ($11.13 \pm 0.22 $)& F850LP &  unit slope\\
& M$_\mathrm{nuc}$ =($1.05 \pm 0.18$) (M$_\mathrm{gal}$ + 17.5) - ($11.05 \pm 2.19 $) & F850LP &\\
\citet{BalGraPel07} & M$_\mathrm{nuc}$ =($0.76 \pm 0.17$) (M$_\mathrm{bulge}$ + 25.0) - ($15.5 \pm 0.45 $)& K-band &  for bulges\\
\citet{ScoGra13} & M$_\mathrm{nuc}$ =($0.60 \pm 0.10$) (M$_\mathrm{gal}$ + 20.4) - ($16.57 \pm 0.175 $)& K-band &  \\
This work & M$_\mathrm{nuc}$ = ($0.57 \pm 0.05$)(M$_\mathrm{gal}$ + 17.5) - ($11.49  \pm 0.14$) & F814W & uniform errors\\
& M$_\mathrm{nuc}$ = ($1.0$)(M$_\mathrm{gal}$ + 17.5) - ($12.36  \pm 0.10$) & F814W & 
\end{tabular}
\caption{Luminosity scaling relations between NSCs and their host galaxies from this work and the literature.}\label{tab:scaling}
\end{table*}

\subsection{Resolved star clusters}
We find that a non-negligible number of star clusters ($\sim$29\% for the full sample, 25\% of all likely members) are resolved. 

Although large (sizes up to 30 pc), we note that the sizes are not excessive: for example, the Milky Way globular cluster NGC2419 has a half-light radius of around 20 pc \citep{Har96,McLvan05}, and also \citet{MacHuxFer06} find globular clusters around Andromeda with similar sizes, showing that these sizes are not uncommon for star clusters.
We note that in principle it is possible that the objects with large sizes are not NSCs but nuclear discs. However, local NSCs are known to be disky \citep[e.g.][]{SetBluBas08}. Since the difference between NSCs and nuclear discs is diffuse, we treat both resolved and unresolved sources in the same way when analyzing scaling relations. In other words, we define a NSC to be the central excess light above a S\'ersic profile.

\subsection{Correlation with S\'ersic index}

\begin{figure}
\begin{minipage}{82mm}
\center
\scalebox{0.4}[0.4]{\includegraphics{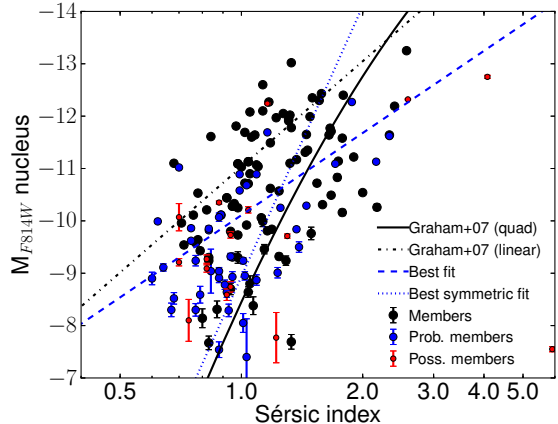}}
\caption{The luminosity of the NSC as a function of host galaxy S\'ersic index. Overplotted are the curved and linear relations from \citet{GraDri07} in black, which were converted to luminosity by assuming a constant mass-to-light ratio. Our own best-fit relation $M_{F814W} =  (-5.2 \pm 0.7) \log_{10}(n/3) - (12.6 \pm 0.3)$ is given by the blue dashed line. The blue dotted line shows the result of a symmetrical regression fit, which miminizes the offsets perpendicular to the best-fit line.}\label{fig:mag_n}
\end{minipage}
\end{figure}

In Fig. \ref{fig:mag_n}, we show the luminosity of the NSC as a function of the S\'ersic index of the host galaxy. We note that we did not separate out different components (discs, bulges) so that the S\'ersic index used here provides a measure of the overall concentration of the host galaxy. A comparison between the scatter on this relation with the scatter on the NSC/host galaxy luminosity relation (0.9 mag, see also Fig. \ref{fig:maghost_magnuc}) implies that the S\'ersic index is not the main driver of NSC formation. \citet{GraDri07} found a relation between the logarithm of the mass of a SMBH and the S\'ersic index of the host galaxy with remarkably low scatter. They showed that this relation continues for low S\'ersic index galaxies, when one replaces the mass of the SMBH with the NSC mass. They provide both a linear and curved CMO-S\'ersic index relation. \citet{Gra12} argues that this relation should in fact be a broken relation. We convert the masses in the CMO-relations from \citet{GraDri07} to luminosities and show these relations in Fig. \ref{fig:mag_n}, where we assumed a constant mass-to-light ratio in the I-band of 2.0, based on Miles models \citep{VazSanFal10} for a 4 Gyr old stellar population with solar metallicity. A lower metallicity, which is likely for the fainter NSCs, will push these lines upwards, so that also the faintest NSC points will follow the curve on average. At a fixed S\'ersic index, the points show significantly higher scatter than the 0.9 mag in the nucleus-host galaxy luminosity scaling relation. If indeed NSCs form the counterpart of SMBHs at low galaxy masses -- and this is debated \citep[e.g.][]{Gra12} -- it remains to be explained why the scatter in NSC magnitudes is so much higher than for the M$_{BH}$-$n$ relation. For completeness, we note that our best fit relation between magnitude and S\'ersic index is $M_{F814W} =  (-5.2 \pm 0.7) \log_{10}(n/3) - (12.6 \pm 0.3)$. A symmetric fit that minimizes the distances perpendicular to the best-fit line yields $M_{F814W} =  (-17.2 \pm 4.7) \log_{10}(n/3) - (17.1 \pm 3.9)$, which has a slope that is similar to the log-quadratic fit of Graham at low S\'ersic indices.  A log-quadratic fit is not warranted by our data.

\citet{Emsvan08} argue that the tidal field in galaxy centres becomes disruptive for steep inner profiles, and that one therefore should not expect to see many NSCs in galaxies with S\'ersic index $n \gtrsim 3.5$. Since we selected dEs, which preferentially have low S\'ersic indices, we are not able to check this prediction. However, trends with S\'ersic index are expected, and we will discuss this more in Section \ref{sec:orig_lum_lum}.

\section{Discussion}\label{sec:discussion}
\subsection{Sizes of nuclear clusters}
\begin{figure*}
\begin{minipage}{\textwidth}
\center
\scalebox{0.7}[0.7]{\includegraphics{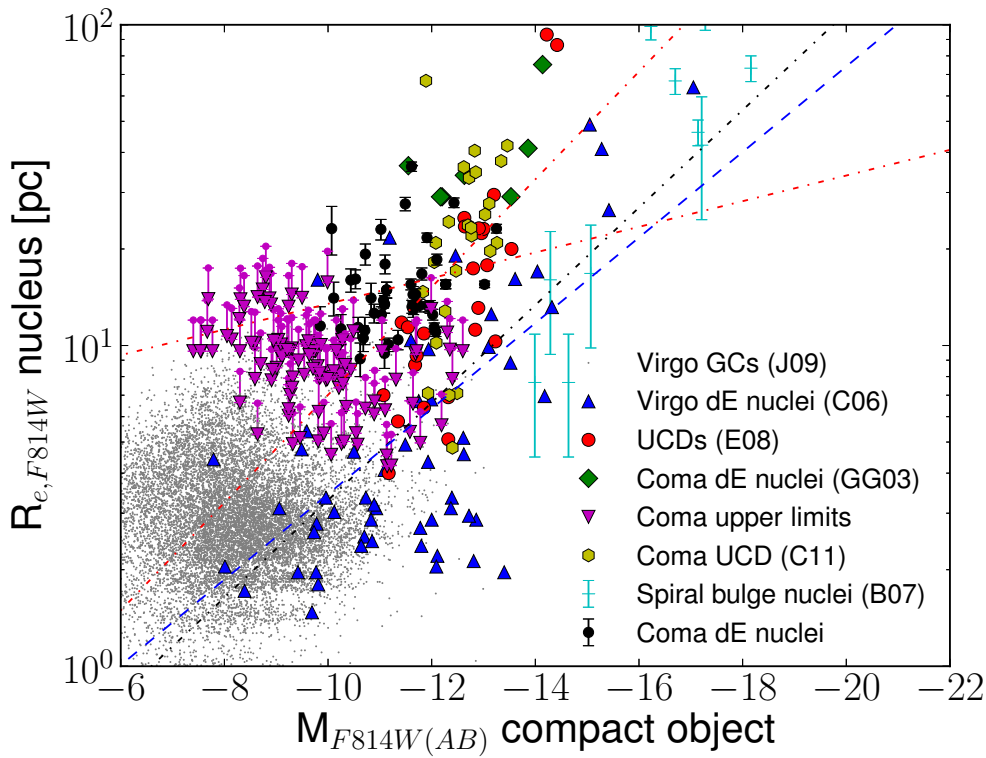}}
\caption{Effective radii of NSCs versus absolute NSC magnitudes in the F814W band (black cirles) and upper limits (purple arrows, see text). Overplotted in blue triangles are the F850LP(AB) measurements from \citet{CotPiaFer06}, derived from Michie-King model fits to their data. Red circles are half-light radii of UCDs, from \citet{EvsDriPen08}, derived from fitting King models or combinations of King and S\'ersic models. The green diamonds are NSCs in dEs in Coma fitted with a gaussian component by \citet{GraGuz03}. The grey small dots are the GC parameters from \citet{JorPenBla09}. The yellow hexagons are UCDs in the Coma cluster from \citet{ChiTulMar11}, and the cyan pluses are the resolved nuclear components in spiral bulges from near-infrared photometry \citep{BalGraPel07}. The dash-dotted (black) line is the relation of \citet{BekCouDri04} and the double-dashed (blue) line shows the expected growth of NSCs by disposition of stars/gas at a radius determined by tidal forces. We fitted a linear relation to the sizes and magnitudes of resolved NSCs in Coma, both with and without taking into account upper limits on the unresolved sources (red dashed lines).}\label{fig:magsize}
\end{minipage}
\end{figure*}
In Fig. \ref{fig:magsize} we show sizes of NSCs as a function of the NSC's absolute magnitude, together with similar measurements of NSCs (from the ACS VCS and from near-infrared photometry of nearby spiral bulges), UCDs and GCs. The measured sizes of Coma NSCs are similar to sizes of the other compact systems. Unresolved NSCs are shown as upper limits. Since our abilility to resolve NSCs depends on S/N, we use upper limits from the \bagatelle fits of resolved clusters to the data, where we define an upper limit as that size for which we know with 67\% certainty that the cluster is smaller than that. We have limited our analysis to dEs and have not analysed the bright nuclear excess light typically seen in $L_\star$-ellipticals, as was done for the ACS Virgo Cluster Survey. The Coma NSC points cover a different locus of the luminosity-size diagram than the NSCs from the ACSVCS, in the sense that at a fixed luminosity NSCs in Coma are somewhat larger. We find that $\sim 70\%$ of star clusters with F814W absolute magnitude between -14 and -11 have halflight radii $>$ 9 pc, whereas for the Virgo ACS Survey this percentage is closer to 25\%. 

A possible reason for this may be the use of different functional forms for fitting the NSC (we use a gaussian profile; Cot\'e et al. use a King profile) as well as the use of different PSFs. We have tested the influence of a different functional form by fitting a source with an empirical King profile and find that the effective radius is consistent with the one inferred from a Gaussian, unless we put strong constraints on the concentration. More likely is that, despite all our tests to confirm the sizes of the NSCs, we may be overestimating the quality of our data or our knowledge about the PSF, or, alternatively, the structural parameters fits may be biased to resolved sources.

Both the upper right part of the diagram and the lower left part are uninhabited: there are no luminous NSCs with small sizes, or faint NSCs with large sizes. Although the latter could be caused by selection effects, the first relation probably reflects the formation of bright NSCs.

If the sizes of the Coma NSCs are indeed larger than the sizes of nuclear clusters of Virgo dwarfs, how can this be explained? Given the compactness of NSCs in general, it is hard to believe that tidal interactions with other galaxies in the cluster have modified the structure of the NSC. However, the NSCs in Coma and Virgo may be at different evolutionary stages; it is possible that some of the secular processes mentioned in the introduction (absorption of kinetic energy from the host galaxy, puffing up by a  central BH) are responsible for the larger sizes.

The globular cluster merging scenario for NSCs gives predictions for the sizes of NSCs as a function of mass \citep[e.g.][]{BekCouDri04}. We can infer the slope of the mass-size relation based on just our Coma data. We fit a linear relation between magnitude and size to the resolved clusters in our sample. Under the assumption of a constant mass to light ratio, this relation is:
\begin{eqnarray}
\mbox{Size} \propto \mbox{Mass}^{0.1 \pm 0.05}.
\end{eqnarray}
This relation does not take into account that for most of the sources in our sample we only have upper limits on the size. We know the luminosity of these sources. However, the size can be anything between roughly 1pc and 8pc. Adding the unresolved sources to the likelihood, we find that the sizes of the Coma NSCs follow a slightly steeper relation:
\begin{eqnarray}
\mbox{Size} \propto \mbox{Mass}^{0.42 \pm 0.04}.
\end{eqnarray}  
In Fig. \ref{fig:magsize} we also show the size-luminosity relation predicted by  \citet{BekCouDri04} for nuclear star clusters formed from merging GCs. The slope of this relation is consistent with the one of the Coma NSCs including upperlimits. 

For the dissipational formation of NSCs, there exist no predictions for a mass--radius relation. However, \citet{SetDalHod06} discuss the formation of a stellar ring around the NSC in IC5052 from the tidal disruption of a molecular cloud by the already existing NSC. If this mechanism is universal and NSCs grow inside-out with new material deposited at a radius determined by tidal forces, we expect a mass-size relation with the same slope as the mass-size relation predicted by \citet{Ant13} for GC accretion ($\mbox{Size} \propto \mbox{Mass}^{\frac{1}{3}}$).

Several authors have compared sizes and luminosities of NSCs with those of UCDs. We note that in the same luminosity range as the UCDs of \citet{ChiTulMar11}, the NSCs and UCDs in Coma dEs have similar sizes. If UCDs are the high-mass end of the GC luminosity function, this similarity in size is not unexpected, since the high-mass GCs are the ones that most easily reach the centre. Similarly, if UCDs are stripped dEs, the sizes of UCDs and dE NSCs should be similar. Given the overlap in luminosity between NSCs and GCs (Fig. \ref{fig:magplumfunc}) a part of the GC cluster population \changed{brighter than F814W $=-10$ mag} likely consists partly of stripped NSCs. 

\subsection{Origin of the L$_{\mathrm{nuc}}$-L$_\mathrm{gal}$ relation}
\label{sec:orig_lum_lum}

There exist predictions for the scaling relation between the mass of the host galaxy and the mass of the nuclear star cluster under the assumption that it formed entirely from globular cluster accretion. \citet{Ant13} provides predictions and derivations for both. For GC accretion, the mass of the star cluster is:
\begin{eqnarray*}
M_{NSC} &=& 3 \times 10^7 \left(\frac{f}{10^{-6}} \right) \left(\frac{\ln \Lambda}{3} \right)  \left(\frac{m_\star}{M_\odot} \right) \\
 & & \left(\frac{<m_{cl}>}{10^5 M_\odot} \right)^{3/2}  \left(\frac{t}{10^{10}} \right)^{1/2} \left(\frac{\sigma}{50 \kms} \right)^{3/2}, 
\end{eqnarray*}
with $f$ the initial number fraction of GCs and $\ln \Lambda$ the Coulomb logarithm, which is defined as $\Lambda = \frac{b_{max} v_{\star}^2}{G(M_{GC}+m_{\star})}$, with $b_{max}$ the largest possible impact parameter and $m_{\star}$ and $v_{\star}$ the typical masses and velocities of the stars.

Since for low mass galaxies the Faber-Jackson relation scales as $L \propto \sigma^2$ \citep{DavEfsFal83,HeldeZMou92,MatGuz05}, this yields a prediction between the magnitude of the star cluster and the host galaxy with slope 0.75 (with maybe a small additional dependence on the globular cluster fraction with host galaxy magnitude). For a more complex model of GC accretion, \citet{GneOstTre14} derive that the mass of the NSC and the mass of the galactic spheroid scale as $M_{NSC}/{M_{star}}\approx 0.0025 \, {M_{star}}_{,11}^{-0.5}$, i.e. with a slope of 0.5. Here ${M_{star}}_{,11}$ denotes the stellar mass of the spheroid divided by $10^{11} \msun$.

For in situ formation of the cluster, the power law slope between $M_{NSC}$ and $\sigma$ is predicted to be 4.0 for a constant momentum feedback model \citep{McLKinNay06,Ant13}, leading to a significantly steeper slope between nuclear star cluster magnitude and host galaxy magnitude of 2.0. This slope seems to be consistent with the the Virgo and Fornax data in Fig. \ref{fig:maghost_magnuc}. However, it is significantly steeper than our fitted slope. 

Although a correlation between $\sigma$ and NSC mass with slope 4.0 was reported by \citet{FerCotDal06}, following up on this work, \citet{ScoGra13} found a much shallower slope with $M_{NSC} \propto \sigma^{2.11}$. Our slope seems to be consistent with that. Although the coincidence of the slopes for the GC accretion scenario and the slope found above are striking, it remains to be seen if dissipational formation of NSCs is ruled out by this. Since the momentum feedback in the McLaughlin model may not be important at later times, when the winds from supernovae and young stars have faded, the predicted slope of 4 may be overestimated compared to reality. As an example, \citet{SetDalHod06} came up with a model in which gas is accreted episodically in the centre of the host galaxy. Although it depends critically on the gas physics, it is not unthinkable that if a fixed fraction of the gas in the galaxy ends up in the nucleus, the NSC mass grows proportionally to the mass of the host galaxy. We conclude that, although the slope of the host galaxy--NSC luminosity relation is in good agreement with the GC inspiral scenario, the lack of detailed model predictions for the formation and evolution of NCS through dissipational collapse make the exclusion of the latter scenario premature. 

\begin{figure}
\center
\scalebox{0.42}[0.42]{\includegraphics{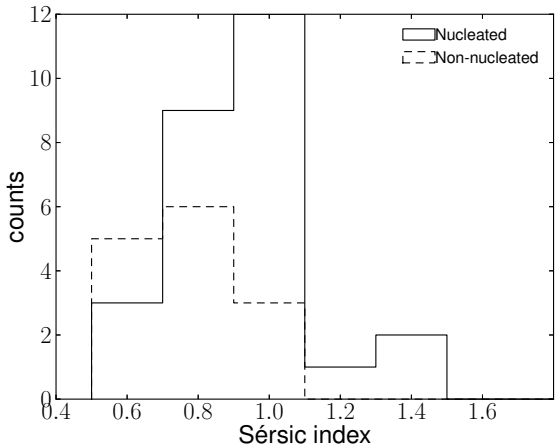}}
\caption{Histograms of the distribution of S\'ersic indices for nucleated (solid line) and non-nucleated (dashed line) galaxies fainter than $M_{F814W} = -14$. The mean S\'ersic index of non-nucleated galaxies is slightly, but not significantly, lower than for nucleated galaxies ($0.76\pm 0.17$ vs. $0.92\pm0.18$), which implies that the missing central star cluster in non-nucleated galaxies is not mistaken for a galaxy with a more cusped profile. \label{fig:hist_n}}
\end{figure}
It may be possible that non-nucleated galaxies do contain a nuclear star cluster which is to faint to be detected in our data. As a sanity check, we have compared for the lowest luminosity galaxies (M$_{F814W} < -14$) the S\'ersic indices of non-nucleated galaxies with those of nucleated galaxies. The S\'ersic indices of non-nucleated galaxies are on average lower than those of nucleated galaxies, suggesting that at least the nucleus is not mistaken for a central cusp (Fig. \ref{fig:hist_n}). 
We also checked by how much the scaling relations would change if we would replace the non-detections by upper limits. The results suggest that the main change happens in the intersect of the relation and the intrinsic scatter, and not so much in the slope (which changes by less than 0.1). If we allow for a variable nucleation fraction among the non-detections, 10\% of the non-detections would be consistent with our fitted relation. We note that although this is comforting, it is by no means solid evidence that these galaxies are not nucleated.

In general, we find that NSCs are almost never brighter than 5\% of the host galaxy. This is slightly different for the bright galaxies in the Virgo and Fornax samples, however, we will argue that structurally (in this paper) and in terms of stellar populations (in a follow-up paper) the NSCs in those galaxies have problaby formed or evolved through a different channel. Galaxies with F814W magnitude $\sim -18$ appear to not reach this maximum NSC formation efficiency. It is well known that the specific frequency of globular clusters is curved in this magnitude range \citep[e.g.][]{PenJorCot08} -- as the formation of globular clusters is happening mainly in the very early Universe before feedback shuts off the star formation in the host galaxy, the globular cluster specific frequency can be high for low-mass galaxies. This may be a possible explanation for the curvature in the NSC-host galaxy magnitude relation.

\subsubsection{Dependence on S\'ersic index}
Both formation scenarios predict that the mass of the star cluster is proportional to the mass of the host galaxy. For dissipational collapse, a larger gas reservoir has been available for formation of the nuclear cluster. For the GC accretion scenario, the number of GCs is proportional (although not linearly) to the mass of the host galaxy. We now try to distinguish between these two scenarios by looking at the scaling relations of NSCs and host galaxies.

\begin{figure}
\center
\scalebox{0.42}[0.42]{\includegraphics{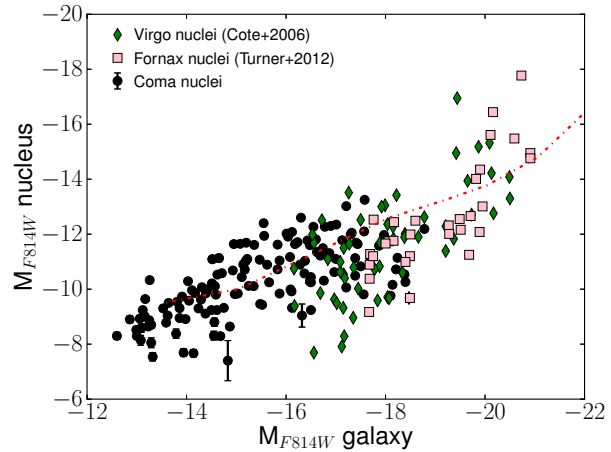}}
\caption{Magnitude of NSCs versus host galaxy magnitude, with the average luminosity in globular clusters (red dash-dotted line), based on measurements in the Virgo cluster.}\label{fig:mag_vs_mag_globular}
\end{figure}

Fig. \ref{fig:mag_vs_mag_globular} shows the luminosity-luminosity relation with superimposed the average luminosity of GC systems of dEs in Virgo, based on the data of Peng et al., and Lotz et al. There is a strange coincidence, which has been noted before \citep[e.g.][]{CotPiaFer06}, that the luminosity of the NSCs is similar to the combined luminosity of all GCs in the galaxy. If NSCs form purely from accretion of GCs, this means that GC inspiraling in low-mass galaxies is less efficient. The NSCs in high-mass galaxies stand out as well, as they do not follow the GC luminosity line.

\begin{figure*}
\scalebox{0.43}[0.43]{\includegraphics{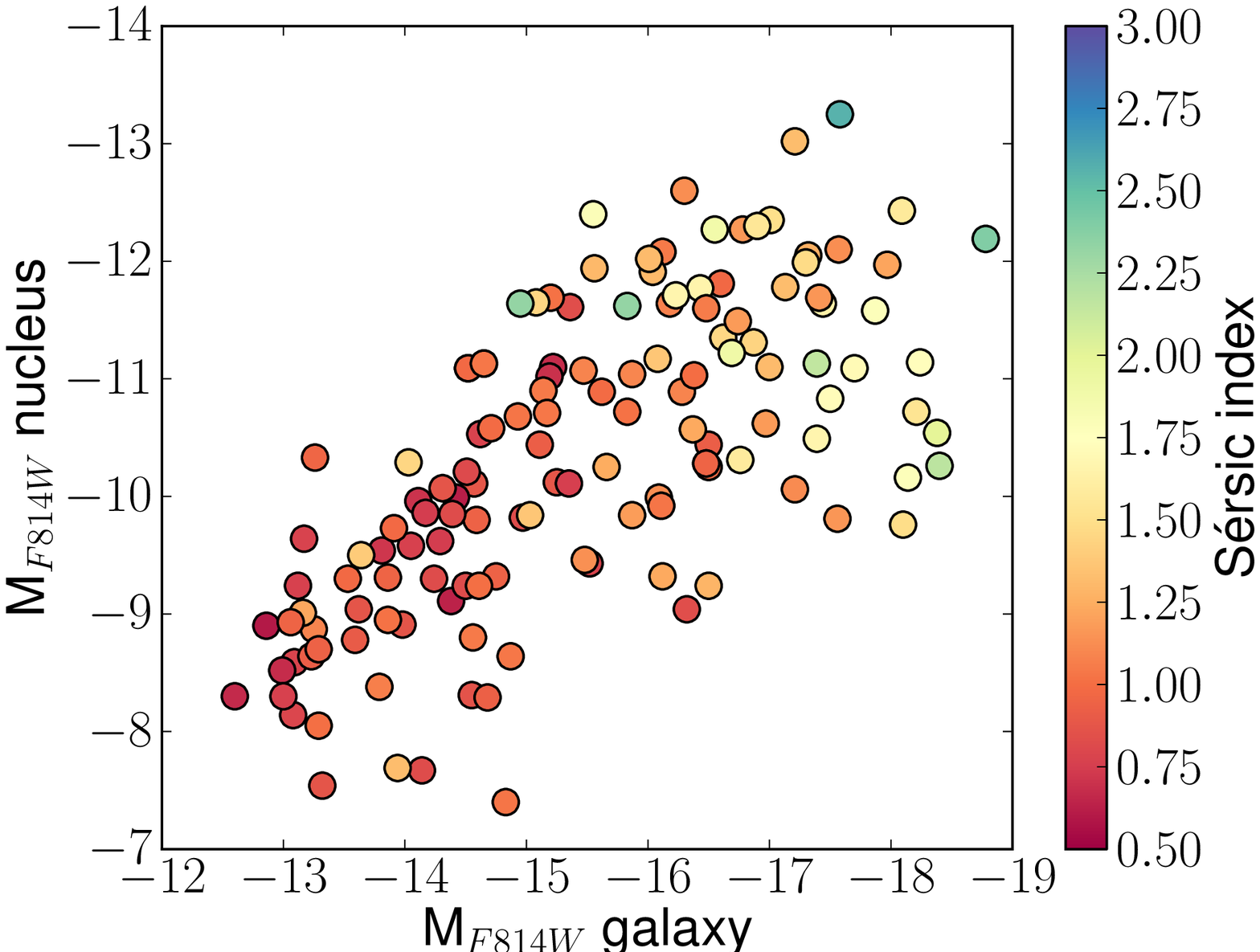}}
\scalebox{0.43}[0.43]{\includegraphics{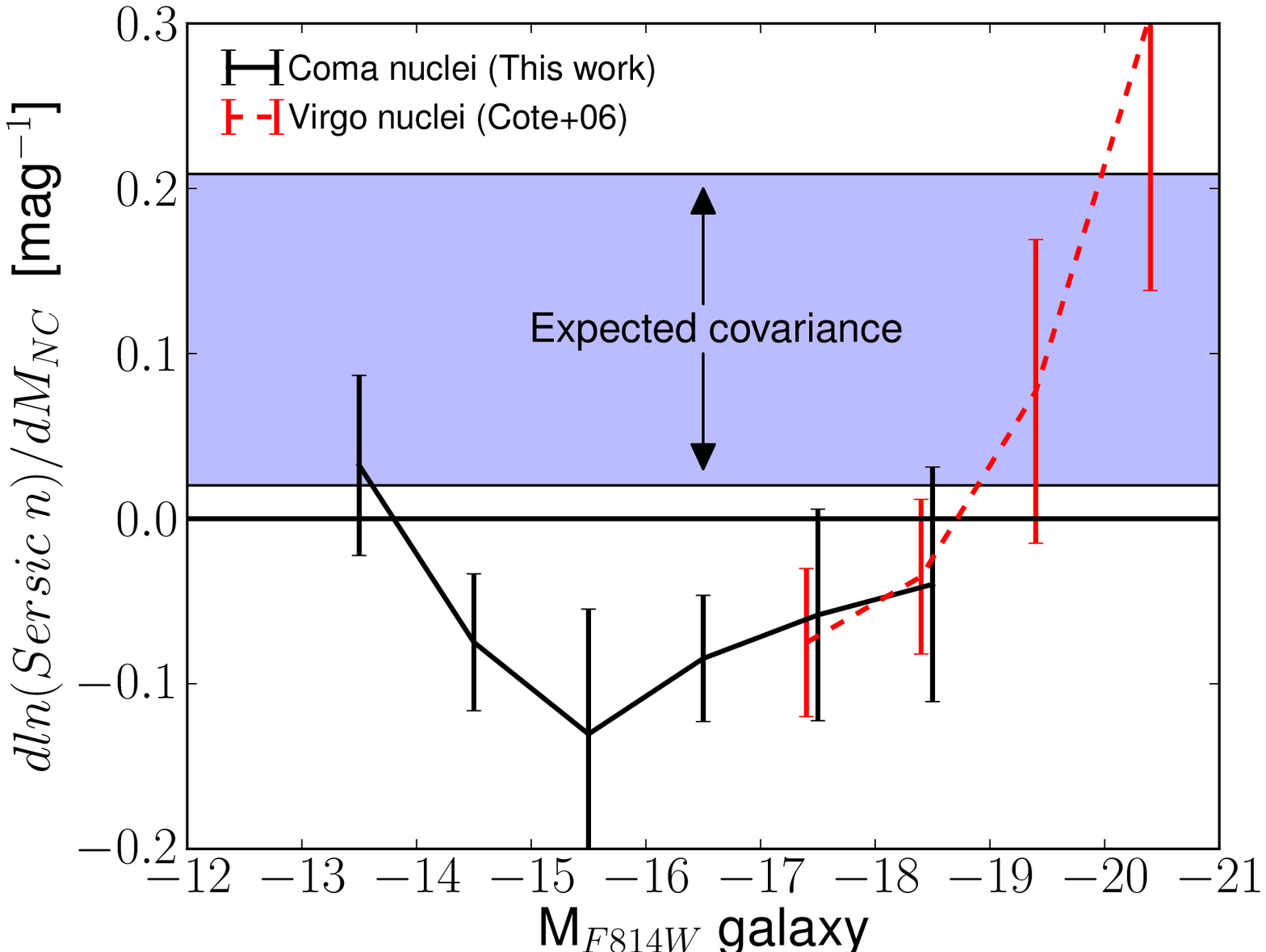}}
\caption{{\bf Left:} Magnitude of NSCs versus host galaxy magnitude, colour-coded by host galaxy S\'ersic index. {\bf Right:} The dependence of nuclear star cluster luminosity on S\'ersic index, for different host galaxy magnitudes, for both the Coma likely members sample (solid black line) and the ACS Virgo sample (dashed red line). Negative values mean that a higher S\'ersic index implies a more luminous NSC, contrary to the covariance between NSC luminosity and S\'ersic index determined from the MCMC output (purple region, see text).}\label{fig:mag_vs_mag_sersic_2}\label{fig:mag_vs_mag_sersic_1}
\end{figure*}

In Fig. \ref{fig:mag_vs_mag_sersic_2} we show how the NSC luminosity vs. host galaxy luminosity depends on the S\'ersic index. In this plot we show, for a given magnitude bin, the slope of a linear relation between NSC luminosities and S\'ersic indices (assuming unit error bars while fitting). For the low-mass end (M$_{F814W} \sim -13$), this relation may be influenced by the low signal-to-noise of the data, however, for intermediate-mass galaxies, we find that {\it in a given magnitude bin, galaxies with high S\'ersic indices have on average higher NSC luminosities.}. We also determine this dependence for galaxies in the Virgo ACS survey, and find that the relations overlap, in the region where the samples overlap.

The S\'ersic index dependence at the low-mass end was found to be different from the slope found for intermediate-luminosity dEs. However, the S\'ersic index and NSC luminosity of a galaxy are expected to be covariant: if we fit to a galaxy a profile with a S\'ersic index that is slightly lower than the actual S\'ersic index ({\sl ceteris paribus}), thereby lowering $\mu_0$, we may still be able to obtain a good fit if we compensate the missing central light by increasing the NSC luminosity. As our fitting code explores different parameter configurations and calculates a posterior probability distribution for each of them, we can infer the magnitude of this effect by diagonalising the covariance matrix for S\'ersic index and NSC magnitude. Since it is difficult to calculate the uncertainty on the covariance from the data of a single galaxy (as this would involve the calculation of an additional covariance matrix for the elements of the covariance matrix), we take the standard deviation of the values found for several galaxies. From the MCMC output of our fitting code, we thus determine that the average covariance between S\'ersic $n$ and M$_{\mbox{nuc}}$ translates into $d\ln{n}/dmag = 0.114 \pm 0.094$. This is close to the value that we find for the magnitude -13 bin, suggesting that that particular value may be due to low signal-to-noise. More importantly however, the sign of the slope inferred from the covariance is opposite the sign of the result for intermediate-luminosity dEs, as already qualitatively argued above, suggesting that this result is solid.

We have thus found that galaxies with a relatively high S\'ersic index have on average a more luminous NSC. The explanation may be two-fold: galaxies with high S\'ersic index may produce more gas in the centre from stellar outflows, and may also be better at retaining gas since the potential is deeper, which may also affect the ambient pressure. On the other hand, GC accretion may be more efficient in galaxies with higher S\'ersic index. A test of the first scenario would require hydrodynamic simulations, which go beyond the scope of this paper. We do note that this trend is opposite what is expected from tidal forces in galaxy centres \citep{Emsvan08}. In order to see if we can reproduce the S\'ersic index dependence with the GC accretion scenario, we make a simple model for the production of NSCs.

\subsubsection{A toy model for GC accretion}
This model\footnote{We note that a similar calculation can be found in Section 8.2.2 of \citet{Mer13}} for the accretion of GCs is based on the dynamical friction formula of \citet{Cha43} for a Gaussian velocity distribution (note that we make strong assumptions on the geometry here, namely sphericity and the size scales which are taken into account by the Coulomb logarithm):
\begin{eqnarray}
\mathbf{f} = -4 \pi G^2 \mglob \rho(r)  \ln \Lambda \left[ \mbox{erf}(X) - \frac
{2X}{\sqrt{\pi}} e^{-X^2} \right] \frac{\mathbf{v}}{v^3}
\end{eqnarray}
where $\mathbf{v}$ is the velocity of the inspiraling GC and $X = v/(\sqrt{2}\sigma)$, with $\sigma$ the local velocity dispersion. Since we have no handle on the actual value of $\ln \Lambda$, the Coulomb logarithm, we make the somewhat {\it ad hoc} assumption that $\ln \Lambda = 3$ for all systems. \citet{SpiPorFel03} find that  $\ln \Lambda = 2.9$ for dense star clusters, and 6.6 for massive black holes. We note however that even assuming a value for the Coulomb logarithm as high as $\ln \Lambda = 10$ does not change the conclusions.
We assume that all matter follows visible matter, and that the projected mass-density is therefore described by a S\'ersic profile. In the model, GCs are orbiting the galaxy centre with the circular velocity at each galactocentric radius, which we calculate from the total enclosed mass by integrating the deprojected S\'ersic profile. The velocity dispersion at each radius is calculated from the spherical Jeans equation, assuming isotropy throughout the galaxy. We thus calculate how much time it takes for a GC with mass \mglob, starting out at radius $r$, to reach the centre of the galaxy.

This allows us to calculate the maximum radius for which a GC with mass \mglob can reach the galaxy centre in less than a Hubble time.
Given this radius, we assume that a fixed fraction of the enclosed mass in the galaxy formed in GCs and produces the NSC. Leaving all parameters (mass, effective radius) for the model galaxy fixed, except the S\'ersic index, we then determine how a change in S\'ersic index changes the NSC mass. 

\begin{figure}
\center
\scalebox{0.42}[0.42]{\includegraphics{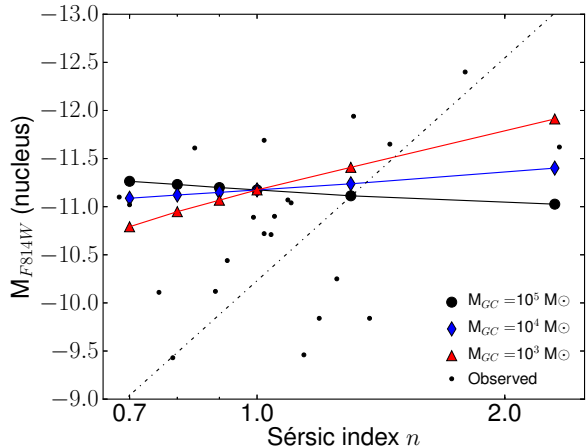}}
\caption{The final magnitude of the NSC after inspiraling of different mass GCs for a Hubble time for different S\'ersic indices of the host galaxy. The host galaxy is assumed to be a F814W=-15 spherical dE with M/L=2.3. The low-mass GCs have to form close to the centre to build up the NSC in a Hubble time and therefore show a strong dependence on S\'ersic index. The high-mass GCs can form farther out for higher S\'ersic indices, but in these galaxies the mass is also distributed in such a way that S\'ersic index dependence actually changes sign. The small dots show the observed probable member data in the $-16 <$ F814W $< -15$ bin, together with the fit assuming uniform errors on $\log{n}$. }\label{fig:modelling_ndependence}
\end{figure}

Our model suggests that if a NSC forms from low-mass GCs ($10^3 M_\odot$), then only GCs that are already close to the galaxy centre manage to reach the centre in a Hubble time, since the dynamical friction times for low-mass GCs in the outer part of the galaxy are too long. On the other hand, high-mass GCs ($M = 10^5-10^6 M_\odot$) may form farther out than low-mass GCs and still reach the centre, since the dynamical friction time is proportional $M^{-2}$.
Massive clusters, which start their journey to the galaxy centre at a distance of a few effective radii, in a galaxy for which we have increased the S\'ersic index slightly, reach the centre somewhat quicker indeed. However, this shorter inspiraling time and therefore larger initial radius does not lead to a larger NSC mass, since the increased S\'ersic index redistributes the mass in the galaxy in such a way that there is more mass in the outer parts, and therefore the increase in $n$ does not lead to an increase in the enclosed mass within the maximum inspiral radius of the GC. This is illustrated in Fig. \ref{fig:modelling_ndependence} for a $10^5$\msun GC, where we assumed a model dwarf galaxy with M$_{F814W}$=-15, an effective radius of 700pc and M/L=2.3 in the I-band and calculate for 5 different S\'ersic indices the expected NSC mass, which we normalized at $n=1$. Not only is the dependence on S\'ersic index much weaker than expected, also the sign of the relation is wrong for high-mass GCs. For low-mass GCs ($10^3$\msun), we find that the trend with S\'ersic index has the right sign, but is not strong enough to explain the observed trend.

Our model is based on strong assumptions (S\'ersic index constant over time, fixed fraction of stars form in clusters of similar mass) but at face value suggests that we cannot reproduce the S\'ersic index-luminosity trend unless we build up our star clusters from low-mass building blocks close to the galaxy centre. A more natural explanation may however be a small amount of residual dissipational star formation. 

\subsubsection{Dependence on host galaxy flattening}
We note that other authors \citep{van86,RydTer94} have already found that non-nucleated galaxies are generally more elliptical than nucleated galaxies. Our data confirm, that, for galaxies with similar luminosities, non-nucleated galaxies tend to have a higher axis-ratio than nucleated galaxies.

\begin{figure}
\center
\scalebox{0.42}[0.42]{\includegraphics{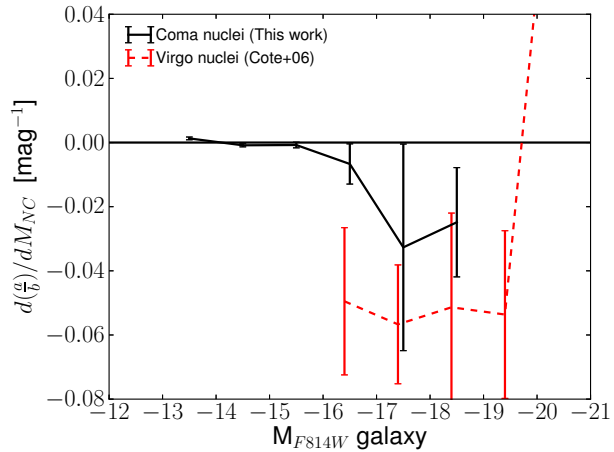}}
\caption{The slope of the nuclear star cluster luminosity--host galaxy axis ratio relation as a function of host galaxy magnitude.}\label{fig:mag_vs_mag_sersic_3}
\end{figure}

However, not only the nucleation fraction, but also the host galaxy--NSC luminosity appears to depend on the ellipticity of the host galaxy. In Fig. \ref{fig:mag_vs_mag_sersic_3}, we show a relation for the slope of the axis ratio--NSC luminosity as a function of host galaxy magnitude. For luminous dEs, the rounder galaxies tend to have brighter NSCs. A simple explanation for this may be that dEs that are rounder have been in the cluster for a longer time and hence have lost some of their mass, that is, instead of the NSC becoming brighter over time, the host galaxy has become fainter. It is unclear if the GC accretion scenario predicts this scaling with axis ratio, because, although dynamical friction may be less efficient outside the plane of the dEs, most globulars actually seem to lie in the plane of the galaxy \citep{BeaCenStr09,WanPenBla13}. For the dissipational model, it is possible that galaxies that are rounder have less rotational support, making it easier for the gas to reach the centre. 

Interestingly, \citet{SetAguLee08} find that late-type spiral galaxies have  NSCs that are on average one order of magnitude less massive than those found in elliptical galaxies of the same mass. An obvious explanation is that late-type spirals are the progenitors of early-type dwarf ellipticals, which may have lost part of their outer parts due to tidal stripping, or alternatively, have grown their nuclear star cluster disproportionally compared to their host galaxy after entering the cluster.
It may thus be the case that this relation tells us more about the evolution of the host galaxy in the cluster environment than about the formation of the nuclear star cluster.

\subsection{Influence of environment}
\begin{figure}
\center
\scalebox{0.42}[0.42]{\includegraphics{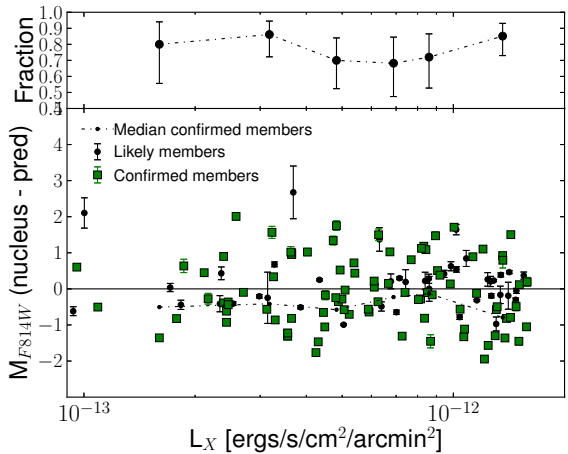}}
\caption{{\bf Upper panel:} Dependence of nucleation fraction on environment, as parametrized by the 0.5-20 keV X-ray flux. The error bars denote the 95\% Jeffreys interval. {\bf Lower panel}: Difference between NSC magnitude and average NSC magnitude for a fixed host galaxy magnitude bin as calculated in Fig. 5, shown as a function of 0.5-20 keV X-ray flux.}\label{fig:diff_xray}
\end{figure}
If nuclei are formed or still evolve after a galaxy falls into the cluster, one might expect that properties of the nuclei depend on environment. As the pressure of the intra-cluster medium (ICM) increases toward the centre of the cluster, dwarf galaxies near the cluster centre may have had additional bursts of star formation, because the ICM confines the gas \citep{BabRee92}. 

In Fig. \ref{fig:diff_xray} we show the difference in nuclear star cluster magnitude and host galaxy magnitude versus the X-ray flux at the centre of the host galaxy. For this, we used the XMM 0.5-20 keV map of \citet{FinBriHen03}, but note that using clustercentric distance does not change the results. If the hot cluster gas plays any role in enhancing the evolution of NSCs, the difference in magnitude between NSC and host galaxy should be higher at high X-ray flux. We find no indication that this is the case. We bin together datapoints in bin width of $2\times10^{-13}$ ergs/s/cm$^2$/arcmin$^{2}$ and calculate for each bin the median difference between the NSC and the host galaxy. If anything, the magnitude difference is slightly higher in high-density regions. An analysis for different host galaxy magnitude bins, similar to what we did in Section \ref{sec:orig_lum_lum} for S\'ersic index and ellipticity, does not reveal any environmental trend either. 

The virial radius of the Coma cluster is approximately 2.9 Mpc \citep{okaMam03}. Taking $\left<v_{\mbox{rms}}\right> \approx 1000$ km/s as a typical velocity for a dwarf galaxy in the cluster \citep{StrRoo99}, a crossing time is approximately 3 Gyr. If we assume that dwarf galaxies have ages in the range 4--7 Gyr \citep{SmiLucHud09,KolDeRPru09} and that the cluster has been responsible for shutting off star formation as soon as a dwarf passed the virial radius, a typical dwarf elliptical has already passed the cluster once or twice, and (depending on the orbit) may therefore have seen different environments of the cluster. It is therefore possible that the influence that environment may have had on the formation or evolution of the NSC is not reflected by the current environment of the dwarf galaxy.

\citet{LisGreBin07} analysed the clustering and flattening of nucleated dEs in the Virgo cluster, and found that these galaxies, compared to non-nucleated and disky dEs, formed a more relaxed population. This is consistent with the lack of trend with environment that we see in Coma. The nucleation fraction does suggest a change with density -- although the statistics are poor -- but not in a monotonic way. It is possible that nucleation is a secular process and that instead the formation/destruction of non-nucleated galaxies is dependent on environment. As an example we mention the possibility that some of the non-nucleated galaxies form as gas-poor tidal galaxies, and therefore should show a dependence on environment \citep[e.g.][]{OkaTan00,ChaDebKar14}. Since the non-nucleated galaxies have on average lower S\'ersic indices, they may be more easily destroyed in the cluster centre by tidal forces than the more centrally concentrated nucleated galaxies.

\subsection{Non-nucleated galaxies}
In the left panel of Fig. \ref{fig:hist_nucl} we showed how the detection fraction of NSCs depends on magnitude. The absence of NSCs in low-mass galaxies has been known since a long time \citep[][]{van86}. \citet{TurCotFer12}  suggest that galaxies may not be able to form nuclei when the NSC magnitude would fall below the turnover magnitude of the GC luminosity function.
 
\begin{figure}
\begin{minipage}{82mm}
\center
\scalebox{0.52}[0.52]{\includegraphics{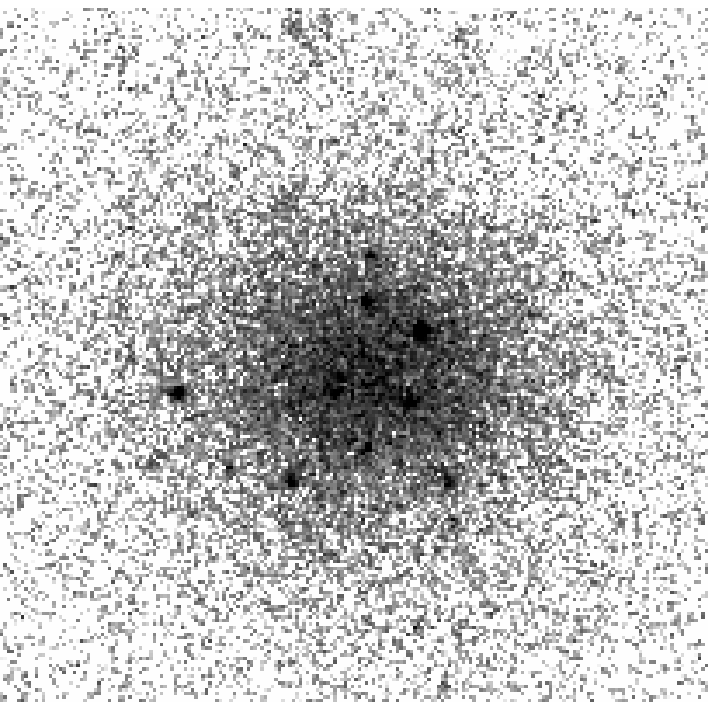}}
\scalebox{0.52}[0.52]{\includegraphics{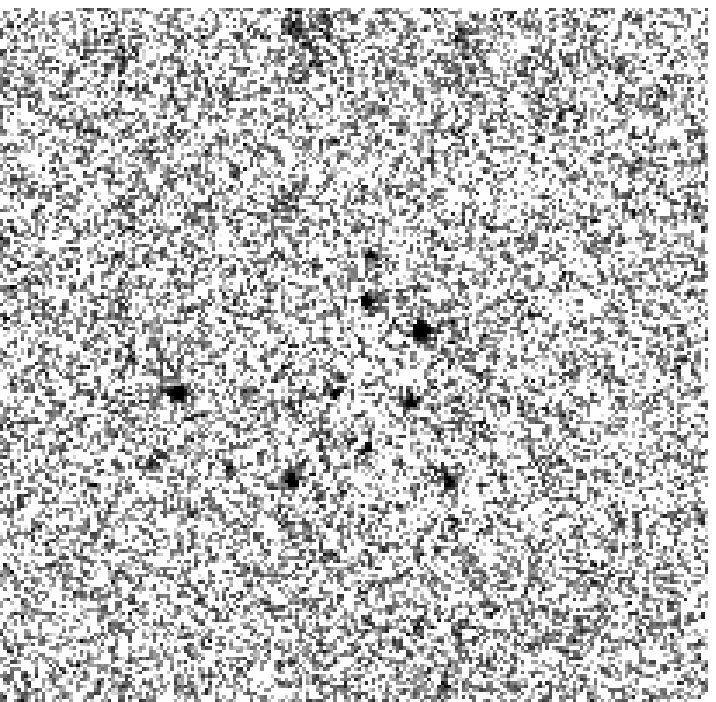}}
\caption{The non-nucleated galaxy SDSSJ125636.63\_271503.6 and the residual from the \bagatelle\ fit. Several star clusters are visible, but none of them is at the photometric centre of the galaxy.}\label{fig:example_nonnuc}
\end{minipage}
\end{figure}

Several galaxies do not show a NSC, although in a few cases, there are globular clusters surrounding the galaxy. As an example, we show in Fig \ref{fig:example_nonnuc} the galaxy SDSSJ125636.63\_271503.6, which, since it is in class 2, belongs to Coma with an 90\% probability. The galaxy is not nucleated, but shows several clusters within one effective radius, though all of them several pixels away from the photometric centre. The spatial coincidence of these clusters and the galaxy implies that they are physically associated. 

The galaxy, found in one of the more remote tiles (visit 63), is rather faint (F814W = -15.9), and the colour, which compared to other galaxies in the sample is typical for a galaxy of this magnitude, implies an SSP age (assuming [Z/H]=- 0.3, half the solar metallicity) of more than 3 Gyr. The core of this galaxy appears to be slightly redder than the outer parts, with no indication of the bluer central colour often found for NSCs. The brightest 6 clusters all have F814W magnitudes between 24.4 and 25.9. If we exclude the brightest and the faintest cluster, the 4 remaining clusters have a combined absolute magnitude of -11.4, consistent with the galaxy luminosity -- NSC luminosity relation. 

This galaxy poses a problem for both formation scenarios: if GC inspiralling is the dominant mechanism for NSC formation, why is there no NSC in the centre of this galaxy, since the most massive clusters should have had plenty of time to reach the centre? And if a galaxy is able to form GCs in its outer parts, why is it not able to form a NSC in its centre, where the high pressure is probably favorable for the formation of a huge star cluster? This galaxy shows that, even though the ability to form GC-mass objects may be a necessity to form a NSC, it is apparently not always sufficient.

\section{Conclusions}
We have analysed data of a sample of 200 dwarf elliptical galaxies in the Coma cluster, to study the scaling relations of their nuclear star clusters, and in doing so, double the sample of HST-observed nuclear star clusters in dEs. Our conclusions are the following.

\begin{itemize}
\item Nuclear star clusters are present in almost all low-mass galaxies. The detection fraction is consistent with what has been found for Virgo by C\^ot\'e et al. and previously for Coma by Graham \& Guzman. 
\item The luminosity function of NSCs peaks 1.5 mag fainter than the luminosity function of the Virgo sample of Cot\'e et al. A natural explanation is the difference in host galaxy magnitudes between the samples. This also means that the luminosity function is closer to the peak of the GC luminosity function.
\item The magnitudes of NSCs follow a curved relation w.r.t. host galaxy magnitude. A possible explanation for the shape at the bright end is that the excess light is due to wet or moist mergers. At the faint end star cluster formation may have been less efficient due to feedback or long inspiraling time-scales. A linear fit between the nuclear star cluster and host galaxy magnitude gives M$_\mathrm{nuc}$ = ($0.57 \pm 0.05$)(M$_\mathrm{gal}$ + 17.5) - ($11.49  \pm 0.14$).
\item At fixed magnitude, galaxies with higher S\'ersic indices tend to have brighter nuclear star clusters. A toy model for the formation of NSCs by the inspiraling of GCs does not predict the right trend or order of magnitude, although we cannot exclude that more sophisticated modelling will do so. A plausible explanation is, that additional dissipational star formation in the centre of galaxies with high S\'ersic indices is more efficient since the gas is more easily retained.
\item Rounder galaxies have slightly more luminous NSCs. The most obvious explanation is that these galaxies have been in the cluster for longer time and have lost part of their mass. On the other hand, it is possible that dynamical friction is less efficient in flattened galaxies or that angular momentum prevents gas to reach the centre.
\end{itemize}
From the scaling relations of nuclear star clusters a picture appears in which their formation is consistent with the GC accretion scenario. Despite this, the GC accretion model still fails to explain why the nucleation fraction goes down at the faint end where dynamical friction should be more efficient, and similarly does not explain the trend with S\'ersic index. 

\section*{Acknowledgements} 
Based on observations made with the NASA/ESA {\it Hubble Space Telescope} obtained at the Space Telescope Science Institute, which is operated by the association of Universities for Research in Astronomy, Inc., under NASA contract NAS 5-26555. This research is primarily associated with program GO-10861. AWG was supported by Australian Research Council funding through grants DP110103509 and FT110100263.

We are grateful to Alexis Finoguenov for providing the X-ray map of Coma. We thank the referee and David Merritt for valuable comments. MdB thanks Marc Verheijen for providing comments on an early draft of this paper.

\bibliographystyle{mn2e}
\bibliography{NUC}

\begin{thebibliography}{}

\bibitem[\protect\citeauthoryear{{Aguerri}, {Iglesias-Paramo}, {Vilchez} \&
  {Mu{\~n}oz-Tu{\~n}{\'o}n}}{{Aguerri} et~al.}{2004}]{AguIglVil04}
{Aguerri} J.~A.~L.,  {Iglesias-Paramo} J.,  {Vilchez} J.~M.,
  {Mu{\~n}oz-Tu{\~n}{\'o}n} C.,  2004, \aj, 127, 1344

\bibitem[\protect\citeauthoryear{{Antonini}}{{Antonini}}{2013}]{Ant13}
{Antonini} F.,  2013, \apj, 763, 62

\bibitem[\protect\citeauthoryear{{Babul} \& {Rees}}{{Babul} \&
  {Rees}}{1992}]{BabRee92}
{Babul} A.,  {Rees} M.~J.,  1992, \mnras, 255, 346

\bibitem[\protect\citeauthoryear{{Balcells}, {Graham},
  {Dom{\'{\i}}nguez-Palmero} \& {Peletier}}{{Balcells}
  et~al.}{2003}]{BalGraDom03}
{Balcells} M.,  {Graham} A.~W.,  {Dom{\'{\i}}nguez-Palmero} L.,    {Peletier}
  R.~F.,  2003, \apjl, 582, L79

\bibitem[\protect\citeauthoryear{{Balcells}, {Graham} \& {Peletier}}{{Balcells}
  et~al.}{2007}]{BalGraPel07}
{Balcells} M.,  {Graham} A.~W.,    {Peletier} R.~F.,  2007, \apj, 665, 1084

\bibitem[\protect\citeauthoryear{{Beasley}, {Cenarro}, {Strader} \&
  {Brodie}}{{Beasley} et~al.}{2009}]{BeaCenStr09}
{Beasley} M.~A.,  {Cenarro} A.~J.,  {Strader} J.,    {Brodie} J.~P.,  2009,
  \aj, 137, 5146

\bibitem[\protect\citeauthoryear{{Bekki}, {Couch}, {Drinkwater} \&
  {Shioya}}{{Bekki} et~al.}{2004}]{BekCouDri04}
{Bekki} K.,  {Couch} W.~J.,  {Drinkwater} M.~J.,    {Shioya} Y.,  2004, \apjl,
  610, L13

\bibitem[\protect\citeauthoryear{{Bekki} \& {Graham}}{{Bekki} \&
  {Graham}}{2010}]{BekGra10}
{Bekki} K.,  {Graham} A.~W.,  2010, \apjl, 714, L313

\bibitem[\protect\citeauthoryear{{B{\"o}ker}, {Laine}, {van der Marel},
  {Sarzi}, {Rix}, {Ho} \& {Shields}}{{B{\"o}ker} et~al.}{2002}]{BokLaivan02}
{B{\"o}ker} T.,  {Laine} S.,  {van der Marel} R.~P.,  {Sarzi} M.,  {Rix} H.,
  {Ho} L.~C.,    {Shields} J.~C.,  2002, \aj, 123, 1389

\bibitem[\protect\citeauthoryear{{B{\"o}ker}, {Sarzi}, {McLaughlin}, {van der
  Marel}, {Rix}, {Ho} \& {Shields}}{{B{\"o}ker} et~al.}{2004}]{BokSarMcL04}
{B{\"o}ker} T.,  {Sarzi} M.,  {McLaughlin} D.~E.,  {van der Marel} R.~P.,
  {Rix} H.,  {Ho} L.~C.,    {Shields} J.~C.,  2004, \aj, 127, 105

\bibitem[\protect\citeauthoryear{{B{\"o}ker}, {van der Marel}, {Mazzuca},
  {Rix}, {Rudnick}, {Ho} \& {Shields}}{{B{\"o}ker} et~al.}{2001}]{BokvanMaz01}
{B{\"o}ker} T.,  {van der Marel} R.~P.,  {Mazzuca} L.,  {Rix} H.,  {Rudnick}
  G.,  {Ho} L.~C.,    {Shields} J.~C.,  2001, \aj, 121, 1473

\bibitem[\protect\citeauthoryear{{B{\"o}ker}, {van der Marel} \&
  {Vacca}}{{B{\"o}ker} et~al.}{1999}]{BokvanVac99}
{B{\"o}ker} T.,  {van der Marel} R.~P.,    {Vacca} W.~D.,  1999, \aj, 118, 831

\bibitem[\protect\citeauthoryear{{Caon}, {Capaccioli} \& {D'Onofrio}}{{Caon}
  et~al.}{1993}]{CaoCapDOn93}
{Caon} N.,  {Capaccioli} M.,    {D'Onofrio} M.,  1993, \mnras, 265, 1013

\bibitem[\protect\citeauthoryear{{Capaccioli}}{{Capaccioli}}{1989}]{Cap89}
{Capaccioli} M.,  1989, in {H.~G.~Corwin Jr.~\& L.~Bottinelli} ed., World of
  Galaxies (Le Monde des Galaxies) {Photometry of early-type galaxies and the R
  exp 1/4 law}.
pp 208--227

\bibitem[\protect\citeauthoryear{{Carollo}, {Stiavelli}, {de Zeeuw} \&
  {Mack}}{{Carollo} et~al.}{1997}]{CarStideZ97}
{Carollo} C.~M.,  {Stiavelli} M.,  {de Zeeuw} P.~T.,    {Mack} J.,  1997, \aj,
  114, 2366

\bibitem[\protect\citeauthoryear{{Carter}, {Goudfrooij}, {Mobasher} \&
  {Ferguson} H.~C.}{{Carter} et~al.}{2008}]{CarGouMob08}
{Carter} D.,  {Goudfrooij} P.,  {Mobasher} B.,    {Ferguson} H.~C. e.~a.,
  2008, \apjs, 176, 424

\bibitem[\protect\citeauthoryear{{Chandrasekhar}}{{Chandrasekhar}}{1943}]{Cha43}
{Chandrasekhar} S.,  1943, \apj, 97, 255

\bibitem[\protect\citeauthoryear{{Chattopadhyay}, {Debsarma}, {Karmakar} \&
  {Davoust}}{{Chattopadhyay} et~al.}{2014}]{ChaDebKar14}
{Chattopadhyay} T.,  {Debsarma} S.,  {Karmakar} P.,    {Davoust} E.,  2014,
  ArXiv e-prints

\bibitem[\protect\citeauthoryear{{Chiboucas}, {Tully}, {Marzke}, {Phillipps},
  {Price}, {Peng}, {Trentham}, {Carter} \& {Hammer}}{{Chiboucas}
  et~al.}{2011}]{ChiTulMar11}
{Chiboucas} K.,  {Tully} R.~B.,  {Marzke} R.~O.,  {Phillipps} S.,  {Price} J.,
  {Peng} E.~W.,  {Trentham} N.,  {Carter} D.,    {Hammer} D.,  2011, \apj, 737,
  86

\bibitem[\protect\citeauthoryear{{Chiboucas}, {Tully}, {Marzke}, {Trentham},
  {Ferguson}, {Hammer}, {Carter} \& {Khosroshahi}}{{Chiboucas}
  et~al.}{2010}]{ChiTulMar10}
{Chiboucas} K.,  {Tully} R.~B.,  {Marzke} R.~O.,  {Trentham} N.,  {Ferguson}
  H.~C.,  {Hammer} D.,  {Carter} D.,    {Khosroshahi} H.,  2010, \apj, 723, 251

\bibitem[\protect\citeauthoryear{{C{\^o}t{\'e}}, {Piatek}, {Ferrarese},
  {Jord{\'a}n}, {Merritt}, {Peng}, {Ha{\c s}egan}, {Blakeslee}, {Mei}, {West},
  {Milosavljevi{\'c}} \& {Tonry}}{{C{\^o}t{\'e}} et~al.}{2006}]{CotPiaFer06}
{C{\^o}t{\'e}} P.,  {Piatek} S.,  {Ferrarese} L.,  {Jord{\'a}n} A.,  {Merritt}
  D.,  {Peng} E.~W.,  {Ha{\c s}egan} M.,  {Blakeslee} J.~P.,  {Mei} S.,  {West}
  M.~J.,  {Milosavljevi{\'c}} M.,    {Tonry} J.~L.,  2006, \apjs, 165, 57

\bibitem[\protect\citeauthoryear{{Davies}, {Efstathiou}, {Fall}, {Illingworth}
  \& {Schechter}}{{Davies} et~al.}{1983}]{DavEfsFal83}
{Davies} R.~L.,  {Efstathiou} G.,  {Fall} S.~M.,  {Illingworth} G.,
  {Schechter} P.~L.,  1983, \apj, 266, 41

\bibitem[\protect\citeauthoryear{{de Vaucouleurs}}{{de
  Vaucouleurs}}{1948}]{deV48}
{de Vaucouleurs} G.,  1948, Annales d'Astrophysique, 11, 247

\bibitem[\protect\citeauthoryear{{D'Onofrio}, {Capaccioli} \&
  {Caon}}{{D'Onofrio} et~al.}{1994}]{DOnCapCao94}
{D'Onofrio} M.,  {Capaccioli} M.,    {Caon} N.,  1994, \mnras, 271, 523

\bibitem[\protect\citeauthoryear{{Dullo} \& {Graham}}{{Dullo} \&
  {Graham}}{2013}]{DulGra13}
{Dullo} B.~T.,  {Graham} A.~W.,  2013, ArXiv e-prints

\bibitem[\protect\citeauthoryear{{Emsellem} \& {van de Ven}}{{Emsellem} \& {van
  de Ven}}{2008}]{Emsvan08}
{Emsellem} E.,  {van de Ven} G.,  2008, \apj, 674, 653

\bibitem[\protect\citeauthoryear{{Erwin} \& {Gadotti}}{{Erwin} \&
  {Gadotti}}{2012}]{ErwGad12}
{Erwin} P.,  {Gadotti} D.~A.,  2012, Advances in Astronomy, 2012

\bibitem[\protect\citeauthoryear{{Evstigneeva}, {Drinkwater}, {Peng}, {Hilker},
  {De Propris}, {Jones}, {Phillipps}, {Gregg} \& {Karick}}{{Evstigneeva}
  et~al.}{2008}]{EvsDriPen08}
{Evstigneeva} E.~A.,  {Drinkwater} M.~J.,  {Peng} C.~Y.,  {Hilker} M.,  {De
  Propris} R.,  {Jones} J.~B.,  {Phillipps} S.,  {Gregg} M.~D.,    {Karick}
  A.~M.,  2008, \aj, 136, 461

\bibitem[\protect\citeauthoryear{{Ferrarese}, {C{\^o}t{\'e}}, {Dalla
  Bont{\`a}}, {Peng}, {Merritt}, {Jord{\'a}n}, {Blakeslee}, {Ha{\c s}egan},
  {Mei}, {Piatek}, {Tonry} \& {West}}{{Ferrarese} et~al.}{2006}]{FerCotDal06}
{Ferrarese} L.,  {C{\^o}t{\'e}} P.,  {Dalla Bont{\`a}} E.,  {Peng} E.~W.,
  {Merritt} D.,  {Jord{\'a}n} A.,  {Blakeslee} J.~P.,  {Ha{\c s}egan} M.,
  {Mei} S.,  {Piatek} S.,  {Tonry} J.~L.,    {West} M.~J.,  2006, \apjl, 644,
  L21

\bibitem[\protect\citeauthoryear{{Finoguenov}, {Briel} \& {Henry}}{{Finoguenov}
  et~al.}{2003}]{FinBriHen03}
{Finoguenov} A.,  {Briel} U.~G.,    {Henry} J.~P.,  2003, \aap, 410, 777

\bibitem[\protect\citeauthoryear{{Freeman}}{{Freeman}}{1970}]{Fre70}
{Freeman} K.~C.,  1970, \apj, 160, 811

\bibitem[\protect\citeauthoryear{{Gnedin}, {Ostriker} \& {Tremaine}}{{Gnedin}
  et~al.}{2014}]{GneOstTre14}
{Gnedin} O.~Y.,  {Ostriker} J.~P.,    {Tremaine} S.,  2014, \apj, 785, 71

\bibitem[\protect\citeauthoryear{{Graham}}{{Graham}}{2012}]{Gra12}
{Graham} A.~W.,  2012, \mnras, 422, 1586

\bibitem[\protect\citeauthoryear{{Graham} \& {Driver}}{{Graham} \&
  {Driver}}{2005}]{GraDri05}
{Graham} A.~W.,  {Driver} S.~P.,  2005, PASA, 22, 118

\bibitem[\protect\citeauthoryear{{Graham} \& {Driver}}{{Graham} \&
  {Driver}}{2007}]{GraDri07}
{Graham} A.~W.,  {Driver} S.~P.,  2007, \apj, 655, 77

\bibitem[\protect\citeauthoryear{{Graham} \& {Guzm{\'a}n}}{{Graham} \&
  {Guzm{\'a}n}}{2003}]{GraGuz03}
{Graham} A.~W.,  {Guzm{\'a}n} R.,  2003, \aj, 125, 2936

\bibitem[\protect\citeauthoryear{{Graham} \& {Spitler}}{{Graham} \&
  {Spitler}}{2009}]{GraSpi09}
{Graham} A.~W.,  {Spitler} L.~R.,  2009, \mnras, 397, 2148

\bibitem[\protect\citeauthoryear{{Grant}, {Kuipers} \& {Phillipps}}{{Grant}
  et~al.}{2005}]{GraKuiPhi05}
{Grant} N.~I.,  {Kuipers} J.~A.,    {Phillipps} S.,  2005, \mnras, 363, 1019

\bibitem[\protect\citeauthoryear{{Hammer}, {Verdoes Kleijn}, {Hoyos}, {den
  Brok}, {Balcells}, {Ferguson}, {Goudfrooij}, {Carter}, {Guzm{\'a}n},
  {Peletier}, {Smith}, {Graham}, {Trentham}, {Peng}, {Puzia}, {Lucey}, {Jogee}
  \& {et al.}}{{Hammer} et~al.}{2010}]{HamVerHoy10}
{Hammer} D.,  {Verdoes Kleijn} G.,  {Hoyos} C.,  {den Brok} M.,  {Balcells} M.,
   {Ferguson} H.~C.,  {Goudfrooij} P.,  {Carter} D.,  {Guzm{\'a}n} R.,
  {Peletier} R.~F.,  {Smith} R.~J.,  {Graham} A.~W.,  {Trentham} N.,  {Peng}
  E.,  {Puzia} T.~H.,  {Lucey} J.~R.,  {Jogee} S.,    {et al.} 2010, \apjs,
  191, 143

\bibitem[\protect\citeauthoryear{{Harris}}{{Harris}}{1996}]{Har96}
{Harris} W.~E.,  1996, \aj, 112, 1487

\bibitem[\protect\citeauthoryear{{Hartmann}, {Debattista}, {Seth}, {Cappellari}
  \& {Quinn}}{{Hartmann} et~al.}{2011}]{HarDebSet11}
{Hartmann} M.,  {Debattista} V.~P.,  {Seth} A.,  {Cappellari} M.,    {Quinn}
  T.~R.,  2011, \mnras, 418, 2697

\bibitem[\protect\citeauthoryear{{Head}, {Lucey}, {Hudson} \& {Smith}}{{Head}
  et~al.}{2014}]{HeaLucHud14}
{Head} J.~T.~C.~G.,  {Lucey} J.~R.,  {Hudson} M.~J.,    {Smith} R.~J.,  2014,
  \mnras, 440, 1690

\bibitem[\protect\citeauthoryear{{Held}, {de Zeeuw}, {Mould} \&
  {Picard}}{{Held} et~al.}{1992}]{HeldeZMou92}
{Held} E.~V.,  {de Zeeuw} T.,  {Mould} J.,    {Picard} A.,  1992, \aj, 103, 851

\bibitem[\protect\citeauthoryear{{Hopkins} \& {Quataert}}{{Hopkins} \&
  {Quataert}}{2011}]{HopQua11}
{Hopkins} P.~F.,  {Quataert} E.,  2011, \mnras, 411, L61

\bibitem[\protect\citeauthoryear{{Hoyos}, {den Brok}, {Verdoes Kleijn},
  {Carter}, {Balcells}, {Guzm{\'a}n}, {Peletier} \& {et al.}}{{Hoyos}
  et~al.}{2011}]{HoyDenVer11}
{Hoyos} C.,  {den Brok} M.,  {Verdoes Kleijn} G.,  {Carter} D.,  {Balcells} M.,
   {Guzm{\'a}n} R.,  {Peletier} R.,    {et al.} 2011, \mnras, 411, 2439

\bibitem[\protect\citeauthoryear{{Huang}, {Ho}, {Peng}, {Li} \&
  {Barth}}{{Huang} et~al.}{2013}]{HuaHoPen13}
{Huang} S.,  {Ho} L.~C.,  {Peng} C.~Y.,  {Li} Z.-Y.,    {Barth} A.~J.,  2013,
  \apj, 766, 47

\bibitem[\protect\citeauthoryear{{Janz}, {Laurikainen}, {Lisker}, {Salo},
  {Peletier}, {Niemi}, {Toloba}, {Hensler}, {Falc{\'o}n-Barroso}, {Boselli},
  {den Brok}, {Hansson}, {Meyer} \& {Ry{\'s}} A.~and{Paudel}}{{Janz}
  et~al.}{2014}]{JanLauLis14}
{Janz} J.,  {Laurikainen} E.,  {Lisker} T.,  {Salo} H.,  {Peletier} R.~F.,
  {Niemi} S.-M.,  {Toloba} E.,  {Hensler} G.,  {Falc{\'o}n-Barroso} J.,
  {Boselli} A.,  {den Brok} M.,  {Hansson} K.~S.~A.,  {Meyer} H.~T.,
  {Ry{\'s}} A.~and{Paudel} P.,  2014, \apj, 786, 105

\bibitem[\protect\citeauthoryear{{Jord{\'a}n}, {Peng}, {Blakeslee},
  {C{\^o}t{\'e}}, {Eyheramendy}, {Ferrarese}, {Mei}, {Tonry} \&
  {West}}{{Jord{\'a}n} et~al.}{2009}]{JorPenBla09}
{Jord{\'a}n} A.,  {Peng} E.~W.,  {Blakeslee} J.~P.,  {C{\^o}t{\'e}} P.,
  {Eyheramendy} S.,  {Ferrarese} L.,  {Mei} S.,  {Tonry} J.~L.,    {West}
  M.~J.,  2009, \apjs, 180, 54

\bibitem[\protect\citeauthoryear{{Koleva}, {de Rijcke}, {Prugniel}, {Zeilinger}
  \& {Michielsen}}{{Koleva} et~al.}{2009}]{KolDeRPru09}
{Koleva} M.,  {de Rijcke} S.,  {Prugniel} P.,  {Zeilinger} W.~W.,
  {Michielsen} D.,  2009, \mnras, 396, 2133

\bibitem[\protect\citeauthoryear{{Kormendy}, {Fisher}, {Cornell} \&
  {Bender}}{{Kormendy} et~al.}{2009}]{KorFisCor09}
{Kormendy} J.,  {Fisher} D.~B.,  {Cornell} M.~E.,    {Bender} R.,  2009, \apjs,
  182, 216

\bibitem[\protect\citeauthoryear{{Kourkchi}, {Khosroshahi}, {Carter} \&
  {Mobasher}}{{Kourkchi} et~al.}{2012}]{KouKhoCar12}
{Kourkchi} E.,  {Khosroshahi} H.~G.,  {Carter} D.,    {Mobasher} B.,  2012,
  \mnras, 420, 2835

\bibitem[\protect\citeauthoryear{{L{\"a}sker}, {Ferrarese} \& {van de
  Ven}}{{L{\"a}sker} et~al.}{2014}]{LasFervan14}
{L{\"a}sker} R.,  {Ferrarese} L.,    {van de Ven} G.,  2014, \apj, 780, 69

\bibitem[\protect\citeauthoryear{{Launhardt}, {Zylka} \& {Mezger}}{{Launhardt}
  et~al.}{2002}]{LauZylMez02}
{Launhardt} R.,  {Zylka} R.,    {Mezger} P.~G.,  2002, \aap, 384, 112

\bibitem[\protect\citeauthoryear{{Leigh}, {B{\"o}ker} \& {Knigge}}{{Leigh}
  et~al.}{2012}]{LeiBokKni12}
{Leigh} N.,  {B{\"o}ker} T.,    {Knigge} C.,  2012, \mnras, 424, 2130

\bibitem[\protect\citeauthoryear{{Lisker}, {Grebel}, {Binggeli} \&
  {Glatt}}{{Lisker} et~al.}{2007}]{LisGreBin07}
{Lisker} T.,  {Grebel} E.~K.,  {Binggeli} B.,    {Glatt} K.,  2007, \apj, 660,
  1186

\bibitem[\protect\citeauthoryear{{{\L}okas} \& {Mamon}}{{{\L}okas} \&
  {Mamon}}{2003}]{okaMam03}
{{\L}okas} E.~L.,  {Mamon} G.~A.,  2003, \mnras, 343, 401

\bibitem[\protect\citeauthoryear{{Lotz}, {Miller} \& {Ferguson}}{{Lotz}
  et~al.}{2004}]{LotMilFer04}
{Lotz} J.~M.,  {Miller} B.~W.,    {Ferguson} H.~C.,  2004, \apj, 613, 262

\bibitem[\protect\citeauthoryear{{Lotz}, {Telford}, {Ferguson}, {Miller},
  {Stiavelli} \& {Mack}}{{Lotz} et~al.}{2001}]{LotTelFer01}
{Lotz} J.~M.,  {Telford} R.,  {Ferguson} H.~C.,  {Miller} B.~W.,  {Stiavelli}
  M.,    {Mack} J.,  2001, \apj, 552, 572

\bibitem[\protect\citeauthoryear{{Mackey}, {Huxor}, {Ferguson}, {Tanvir},
  {Irwin}, {Ibata}, {Bridges}, {Johnson} \& {Lewis}}{{Mackey}
  et~al.}{2006}]{MacHuxFer06}
{Mackey} A.~D.,  {Huxor} A.,  {Ferguson} A.~M.~N.,  {Tanvir} N.~R.,  {Irwin}
  M.,  {Ibata} R.,  {Bridges} T.,  {Johnson} R.~A.,    {Lewis} G.,  2006,
  \apjl, 653, L105

\bibitem[\protect\citeauthoryear{{Matkovi{\'c}} \& {Guzm{\'a}n}}{{Matkovi{\'c}}
  \& {Guzm{\'a}n}}{2005}]{MatGuz05}
{Matkovi{\'c}} A.,  {Guzm{\'a}n} R.,  2005, \mnras, 362, 289

\bibitem[\protect\citeauthoryear{{Matthews}, {Gallagher} III, {Krist},
  {Watson}, {Burrows}, {Griffiths}, {Hester}, {Trauger} \& {et al.}}{{Matthews}
  et~al.}{1999}]{MatGalKri99}
{Matthews} L.~D.,  {Gallagher} III J.~S.,  {Krist} J.~E.,  {Watson} A.~M.,
  {Burrows} C.~J.,  {Griffiths} R.~E.,  {Hester} J.~J.,  {Trauger} J.~T.,
  {et al.} 1999, \aj, 118, 208

\bibitem[\protect\citeauthoryear{{McLaughlin}, {King} \&
  {Nayakshin}}{{McLaughlin} et~al.}{2006}]{McLKinNay06}
{McLaughlin} D.~E.,  {King} A.~R.,    {Nayakshin} S.,  2006, \apjl, 650, L37

\bibitem[\protect\citeauthoryear{{McLaughlin} \& {van der Marel}}{{McLaughlin}
  \& {van der Marel}}{2005}]{McLvan05}
{McLaughlin} D.~E.,  {van der Marel} R.~P.,  2005, \apjs, 161, 304

\bibitem[\protect\citeauthoryear{{Merritt}}{{Merritt}}{2009}]{Mer09}
{Merritt} D.,  2009, \apj, 694, 959

\bibitem[\protect\citeauthoryear{{Merritt}}{{Merritt}}{2013}]{Mer13}
{Merritt} D.,  2013, {Dynamics and Evolution of Galactic Nuclei}

\bibitem[\protect\citeauthoryear{{Mihos} \& {Hernquist}}{{Mihos} \&
  {Hernquist}}{1994}]{MihHer94}
{Mihos} J.~C.,  {Hernquist} L.,  1994, \apjl, 437, L47

\bibitem[\protect\citeauthoryear{{Milosavljevi{\'c}}}{{Milosavljevi{\'c}}}{2004}]{Mil04}
{Milosavljevi{\'c}} M.,  2004, \apjl, 605, L13

\bibitem[\protect\citeauthoryear{{Mukherjee}, {Parkinson} \&
  {Liddle}}{{Mukherjee} et~al.}{2006}]{MukParLid06}
{Mukherjee} P.,  {Parkinson} D.,    {Liddle} A.~R.,  2006, \apjl, 638, L51

\bibitem[\protect\citeauthoryear{{Nayakshin}, {Wilkinson} \&
  {King}}{{Nayakshin} et~al.}{2009}]{NayWilKin09}
{Nayakshin} S.,  {Wilkinson} M.~I.,    {King} A.,  2009, \mnras, 398, L54

\bibitem[\protect\citeauthoryear{{Neumayer} \& {Walcher}}{{Neumayer} \&
  {Walcher}}{2012}]{NeuWal12}
{Neumayer} N.,  {Walcher} C.~J.,  2012, Advances in Astronomy, 2012

\bibitem[\protect\citeauthoryear{{Okazaki} \& {Taniguchi}}{{Okazaki} \&
  {Taniguchi}}{2000}]{OkaTan00}
{Okazaki} T.,  {Taniguchi} Y.,  2000, \apj, 543, 149

\bibitem[\protect\citeauthoryear{{Patterson}}{{Patterson}}{1940}]{Pat40}
{Patterson} F.~S.,  1940, Harvard College Observatory Bulletin, 914, 9

\bibitem[\protect\citeauthoryear{{Peng}, {Ho}, {Impey} \& {Rix}}{{Peng}
  et~al.}{2002}]{PenHoImp02}
{Peng} C.~Y.,  {Ho} L.~C.,  {Impey} C.~D.,    {Rix} H.-W.,  2002, \aj, 124, 266

\bibitem[\protect\citeauthoryear{{Peng}, {Jord{\'a}n}, {C{\^o}t{\'e}},
  {Takamiya}, {West}, {Blakeslee}, {Chen}, {Ferrarese}, {Mei}, {Tonry} \&
  {West}}{{Peng} et~al.}{2008}]{PenJorCot08}
{Peng} E.~W.,  {Jord{\'a}n} A.,  {C{\^o}t{\'e}} P.,  {Takamiya} M.,  {West}
  M.~J.,  {Blakeslee} J.~P.,  {Chen} C.-W.,  {Ferrarese} L.,  {Mei} S.,
  {Tonry} J.~L.,    {West} A.~A.,  2008, \apj, 681, 197

\bibitem[\protect\citeauthoryear{{Rhodes}, {Massey}, {Albert}, {Collins},
  {Ellis}, {Heymans}, {Gardner}, {Kneib}, {Koekemoer}, {Leauthaud}, {Mellier},
  {Refregier}, {Taylor} \& {Van Waerbeke}}{{Rhodes} et~al.}{2007}]{RhoMasAlb07}
{Rhodes} J.~D.,  {Massey} R.~J.,  {Albert} J.,  {Collins} N.,  {Ellis} R.~S.,
  {Heymans} C.,  {Gardner} J.~P.,  {Kneib} J.,  {Koekemoer} A.,  {Leauthaud}
  A.,  {Mellier} Y.,  {Refregier} A.,  {Taylor} J.~E.,    {Van Waerbeke} L.,
  2007, \apjs, 172, 203

\bibitem[\protect\citeauthoryear{{Ryden} \& {Terndrup}}{{Ryden} \&
  {Terndrup}}{1994}]{RydTer94}
{Ryden} B.~S.,  {Terndrup} D.~M.,  1994, \apj, 425, 43

\bibitem[\protect\citeauthoryear{{Sandage}, {Binggeli} \& {Tammann}}{{Sandage}
  et~al.}{1985}]{SanBinTam85}
{Sandage} A.,  {Binggeli} B.,    {Tammann} G.~A.,  1985, \aj, 90, 1759

\bibitem[\protect\citeauthoryear{{Sch{\"o}del}, {Eckart}, {Alexander},
  {Merritt}, {Genzel}, {Sternberg}, {Meyer}, {Kul}, {Moultaka}, {Ott} \&
  {Straubmeier}}{{Sch{\"o}del} et~al.}{2007}]{SchEckAle07}
{Sch{\"o}del} R.,  {Eckart} A.,  {Alexander} T.,  {Merritt} D.,  {Genzel} R.,
  {Sternberg} A.,  {Meyer} L.,  {Kul} F.,  {Moultaka} J.,  {Ott} T.,
  {Straubmeier} C.,  2007, \aap, 469, 125

\bibitem[\protect\citeauthoryear{{Sch{\"o}del}, {Merritt} \&
  {Eckart}}{{Sch{\"o}del} et~al.}{2009}]{SchMerEck09}
{Sch{\"o}del} R.,  {Merritt} D.,    {Eckart} A.,  2009, \aap, 502, 91

\bibitem[\protect\citeauthoryear{{Scott} \& {Graham}}{{Scott} \&
  {Graham}}{2013}]{ScoGra13}
{Scott} N.,  {Graham} A.~W.,  2013, \apj, 763, 76

\bibitem[\protect\citeauthoryear{{Sersic}}{{Sersic}}{1968}]{Ser68}
{Sersic} J.~L.,  1968, {Atlas de galaxias australes}

\bibitem[\protect\citeauthoryear{{Seth}, {Ag{\"u}eros}, {Lee} \&
  {Basu-Zych}}{{Seth} et~al.}{2008}]{SetAguLee08}
{Seth} A.,  {Ag{\"u}eros} M.,  {Lee} D.,    {Basu-Zych} A.,  2008, \apj, 678,
  116

\bibitem[\protect\citeauthoryear{{Seth}, {Blum}, {Bastian}, {Caldwell} \&
  {Debattista}}{{Seth} et~al.}{2008}]{SetBluBas08}
{Seth} A.~C.,  {Blum} R.~D.,  {Bastian} N.,  {Caldwell} N.,    {Debattista}
  V.~P.,  2008, \apj, 687, 997

\bibitem[\protect\citeauthoryear{{Seth}, {Dalcanton}, {Hodge} \&
  {Debattista}}{{Seth} et~al.}{2006}]{SetDalHod06}
{Seth} A.~C.,  {Dalcanton} J.~J.,  {Hodge} P.~W.,    {Debattista} V.~P.,  2006,
  \aj, 132, 2539

\bibitem[\protect\citeauthoryear{{Skilling}}{{Skilling}}{2004}]{Ski04}
{Skilling} J.,  2004, in {Fischer} R.,  {Preuss} R.,   {Toussaint} U.~V.,  eds,
  American Institute of Physics Conference Series Vol.~735 of American
  Institute of Physics Conference Series, {Nested Sampling}.
pp 395--405

\bibitem[\protect\citeauthoryear{{Smith}, {Lucey}, {Hudson}, {Allanson},
  {Bridges}, {Hornschemeier}, {Marzke} \& {Miller}}{{Smith}
  et~al.}{2009}]{SmiLucHud09}
{Smith} R.~J.,  {Lucey} J.~R.,  {Hudson} M.~J.,  {Allanson} S.~P.,  {Bridges}
  T.~J.,  {Hornschemeier} A.~E.,  {Marzke} R.~O.,    {Miller} N.~A.,  2009,
  \mnras, 392, 1265

\bibitem[\protect\citeauthoryear{{Spinnato}, {Portegies Zwart}, {Fellhauer},
  {van Albada} \& {Sloot}}{{Spinnato} et~al.}{2003}]{SpiPorFel03}
{Spinnato} P.~F.,  {Portegies Zwart} S.~F.,  {Fellhauer} M.,  {van Albada}
  G.~D.,    {Sloot} P.~M.~A.,  2003, Memorie della Societa Astronomica Italiana
  Supplementi, 1, 54

\bibitem[\protect\citeauthoryear{{Stetson}}{{Stetson}}{1987}]{Ste87}
{Stetson} P.~B.,  1987, \pasp, 99, 191

\bibitem[\protect\citeauthoryear{{Struble} \& {Rood}}{{Struble} \&
  {Rood}}{1999}]{StrRoo99}
{Struble} M.~F.,  {Rood} H.~J.,  1999, \apjs, 125, 35

\bibitem[\protect\citeauthoryear{{Tremaine}, {Ostriker} \& {Spitzer}
  Jr.}{{Tremaine} et~al.}{1975}]{TreOstSpi75}
{Tremaine} S.~D.,  {Ostriker} J.~P.,    {Spitzer} Jr. L.,  1975, \apj, 196, 407

\bibitem[\protect\citeauthoryear{{Turner}, {C{\^o}t{\'e}}, {Ferrarese},
  {Jord{\'a}n}, {Blakeslee}, {Mei}, {Peng} \& {West}}{{Turner}
  et~al.}{2012}]{TurCotFer12}
{Turner} M.~L.,  {C{\^o}t{\'e}} P.,  {Ferrarese} L.,  {Jord{\'a}n} A.,
  {Blakeslee} J.~P.,  {Mei} S.,  {Peng} E.~W.,    {West} M.~J.,  2012, \apjs,
  203, 5

\bibitem[\protect\citeauthoryear{{van den Bergh}}{{van den
  Bergh}}{1986}]{van86}
{van den Bergh} S.,  1986, \aj, 91, 271

\bibitem[\protect\citeauthoryear{{Vazdekis}, {S{\'a}nchez-Bl{\'a}zquez},
  {Falc{\'o}n-Barroso}, {Cenarro}, {Beasley}, {Cardiel}, {Gorgas} \&
  {Peletier}}{{Vazdekis} et~al.}{2010}]{VazSanFal10}
{Vazdekis} A.,  {S{\'a}nchez-Bl{\'a}zquez} P.,  {Falc{\'o}n-Barroso} J.,
  {Cenarro} A.~J.,  {Beasley} M.~A.,  {Cardiel} N.,  {Gorgas} J.,    {Peletier}
  R.~F.,  2010, \mnras, 404, 1639

\bibitem[\protect\citeauthoryear{{Walcher}, {van der Marel}, {McLaughlin},
  {Rix}, {B{\"o}ker}, {H{\"a}ring}, {Ho}, {Sarzi} \& {Shields}}{{Walcher}
  et~al.}{2005}]{WalvanMcL05}
{Walcher} C.~J.,  {van der Marel} R.~P.,  {McLaughlin} D.,  {Rix} H.-W.,
  {B{\"o}ker} T.,  {H{\"a}ring} N.,  {Ho} L.~C.,  {Sarzi} M.,    {Shields}
  J.~C.,  2005, \apj, 618, 237

\bibitem[\protect\citeauthoryear{{Wang}, {Peng}, {Blakeslee}, {C{\^o}t{\'e}},
  {Ferrarese}, {Jord{\'a}n}, {Mei} \& {West}}{{Wang}
  et~al.}{2013}]{WanPenBla13}
{Wang} Q.,  {Peng} E.~W.,  {Blakeslee} J.~P.,  {C{\^o}t{\'e}} P.,  {Ferrarese}
  L.,  {Jord{\'a}n} A.,  {Mei} S.,    {West} M.~J.,  2013, \apj, 769, 145

\bibitem[\protect\citeauthoryear{{Wehner} \& {Harris}}{{Wehner} \&
  {Harris}}{2006}]{WehHar06}
{Wehner} E.~H.,  {Harris} W.~E.,  2006, \apjl, 644, L17

\bibitem[\protect\citeauthoryear{{Weinmann}, {Lisker}, {Guo}, {Meyer} \&
  {Janz}}{{Weinmann} et~al.}{2011}]{WeiLisGuo11}
{Weinmann} S.~M.,  {Lisker} T.,  {Guo} Q.,  {Meyer} H.~T.,    {Janz} J.,  2011,
  \mnras, 416, 1197

\bibitem[\protect\citeauthoryear{{Weinzirl}, {Jogee}, {Neistein}, {Khochfar},
  {Kormendy}, {Marinova}, {Hoyos}, {Balcells}, {den Brok}, {Hammer},
  {Peletier}, {Kleijn}, {Carter} \& {et al.}}{{Weinzirl}
  et~al.}{2014}]{WeiJogNei14}
{Weinzirl} T.,  {Jogee} S.,  {Neistein} E.,  {Khochfar} S.,  {Kormendy} J.,
  {Marinova} I.,  {Hoyos} C.,  {Balcells} M.,  {den Brok} M.,  {Hammer} D.,
  {Peletier} R.~F.,  {Kleijn} G.~V.,  {Carter} D.,    {et al.} 2014, \mnras,
  441, 3083

\bibitem[\protect\citeauthoryear{{Yoon}, {Weinberg} \& {Katz}}{{Yoon}
  et~al.}{2011}]{YooWeiKat11}
{Yoon} I.,  {Weinberg} M.~D.,    {Katz} N.,  2011, \mnras, 414, 1625

\end{thebibliography}

\begin{appendix}
\section{Tables and profiles}\label{apx:excl}
\onecolumn
\begin{longtable}{llll|llll}
\hline
Name & RA & DEC &  $P_m$ & F814W(gal)& S\'ersic $n$ & F814W(nuc) & FWHM  \\
\hline
SDSSJ130018.54\_280549.7 & 195.0773 & 28.0972 & 0 & 16.8 $\pm$ 0.01 & 1.5 $\pm$ 0.01 & 24.3 $\pm$ 0.05 & - \\ 
LEDA126789 & 194.8829 & 27.8613 & 0 & 16.9 $\pm$ 0.01 & 1.6 $\pm$ 0.01 & 22.6 $\pm$ 0.03 & 2.3 $\pm$ 0.1 \\ 
SDSSJ130041.19\_280242.4 & 195.1717 & 28.0451 & 0 & 16.8 $\pm$ 0.01 & 1.7 $\pm$ 0.01 & 23.9 $\pm$ 0.04 & - \\ 
SDSSJ130034.42\_275604.9 & 195.1435 & 27.9347 & 0 & 16.6 $\pm$ 0.01 & 2.2 $\pm$ 0.01 & 24.7 $\pm$ 0.06 & - \\ 
RB068 & 194.9978 & 27.9406 & 0 & 16.2 $\pm$ 0.01 & 2.4 $\pm$ 0.01 & 22.8 $\pm$ 0.01 & - \\ 
LEDA126815 &  194.6898 &  27.7539  & 0 &  16.62$\pm $ 0.01  &  2.0$\pm$ 0.01 & 24.46 $\pm$0.04  & -  \\
\vdots    & \vdots & \vdots & \vdots & \vdots & \vdots & \vdots & \vdots \\
\vdots    & \vdots & \vdots & \vdots & \vdots & \vdots & \vdots & \vdots \\
\vdots    & \vdots & \vdots & \vdots & \vdots & \vdots & \vdots & \vdots \\
COMAi125932.883p275800.05 & 194.8870 & 27.9666 & 2 & 21.84$\pm$0.36 & 1.2$\pm$0.3 & 26.0$\pm$0.1 & -\\ 
COMAi125929.995p275348.12 & 194.8748 & 27.8967 & 2 & 21.14$\pm$0.17 & 1.0$\pm$0.1 & 26.0$\pm$0.1 & -\\ 
COMAi125952.543p275824.21 & 194.9688 & 27.9734 & 2 & 21.94$\pm$0.22 & 0.9$\pm$0.2 & 26.1$\pm$0.1 & -\\ 
COMAi13021.712p275650.16 & 195.0904 & 27.9472 & 2 & 21.41$\pm$0.16 & 0.9$\pm$0.1 & 26.2$\pm$0.1 & -\\ 
COMAi125944.017p275615.29 & 194.9333 & 27.9375 & 0 & 21.71$\pm$0.06 & 0.9$\pm$0.1 & 26.3$\pm$0.1 & -\\ 
COMAi125924.938p275320.35 & 194.8538 & 27.8889 & 2 & 21.71$\pm$0.13 & 1.0$\pm$0.1 & 27.0$\pm$0.2 & -\\ 
\caption{Structural parameters of galaxies used in the paper. The complete table is available in the online version of this paper.}
\end{longtable}
\newpage
\begin{table}
\begin{tabular}{lllllllll}
Name & P$_M$ & Type & Tile & $z$ & RA & DEC & F814W & Notes\\
\hline
  SDSSJ130013.42\_280311.8 & 0 & S0 & 10 & 8124.0 & 195.0559 & 28.0533 & 17.3 & Bright and completely edge-on\\
  SDSSJ130035.42\_275633.9 & 0 & S0 & 22 & 6925.0 & 195.1458 & 27.9428 & 17.3 & This galaxy is not axisymmetric: \\ & & & & & & & & either barred or two spiral arms\\
  SDSSJ125937.00\_280106.9 & 0 & E & 12 & 7195.0 & 194.9042 & 28.0186 & 17.7 & The fit is degenerate -  \\ & & & & & & & & centre can be fit in many ways.\\
  SDSSJ125815.27\_272752.9 & 0 & S0 & 45 & 7615.0 & 194.5636 & 27.4647 & 17.8 & Edge-on galaxy\\
  SDSSJ130009.46\_275456.3 & 3 & E & 24 & - & 195.0398 & 27.9156 & 17.9 & Close to bright star\\
  SDSSJ130039.10\_280035.5 & 0 & dE & 8 & 5785.0 & 195.1627 & 28.0099 & 18.1 & The isophotes of this object are  \\ & & & & & & & &not ellipcical. Not clear if central object is NSC\\
  SDSSJ130017.64\_275915.1 & 0 & E & 16 & 5966.0 & 195.0735 & 27.9876 & 18.1 & Fit requires two outer components \\ & & & & & & & & + central S\'ersic.\\
  CcGV18 & 0 & E & 18 & 6535.0 & 194.9996 & 27.9894 & 20.1 & Compact\\
  CcGV9b & 0 & E & 9 & 6425.0 & 195.1137 & 28.0092 & 19.2 & Compact\\
  CcGV1 & 0 & E & 1 & 6775.0 & 195.1986 & 28.0927 & 19.3 & Compact\\
  CcGV19b & 0 & E & 19 & 7075.0 & 194.9133 & 27.9985 & 19.9 & Compact\\
  CcGV18 & 0 & E & 18 & 6535.0 & 194.9996 & 27.9894 & 20.1 & Compact\\
  CcGV12 & 0 & E & 12 & 7721.0 & 194.9263 & 28.0153 & 20.5 & Compact\\
  RB110 & 0 & dE,N & 8 & 7615.0 & 195.1607 & 28.016 & 18.3 & Nucleated. In halo of neighbouring galaxy\\
  -- & 0 & dE,N & 63 & 9054.0 & 194.1588 & 27.2178 & 18.6 & Galaxy shows spiral arms.\\
  -- & 1 & dE,N & 10 & - & 195.0238 & 28.0259 & 18.8 & In halo of huge galaxy.\\
  SDSSJ130023.47\_280301.9 & 0 & dE & 9 & 6925.0 & 195.0977 & 28.0505 & 19.8 & Strong gradient  \\ & & & & & & & &due to neighbouring galaxy\\
  -- & 1 & VLSB & 9 & - & 195.134 & 28.0376 & 19.9 & Even though this was identified as a galaxy,  \\ & & & & & & & & it looks more like a halo of another galaxy\\
  -- & 2 & dE,N & 25 & - & 194.9245 & 27.9262 & 20.1 & In halo/edge. Nucleated\\
  -- & 2 & dE & 19 & - & 194.9022 & 27.9608 & 20.1 & Inside halo\\
  -- & 2 & dE,N & 22 & - & 195.1479 & 27.9438 & 20.2 & In halo of another galaxy. Nucleation uncertain\\
  -- & 2 & dE,N & 33 & - & 194.8487 & 27.8451 & 20.3 & Galaxy is in edge - fit looks  \\ & & & & & & & & decent but nucleation is uncertain.\\
  SDSSJ125845.91\_274655.5 & 2 & dE,N & 75 & - & 194.6912 & 27.7823 & 20.3 & Galaxy is edge-on/irregular\\
  SDSSJ125942.92\_275954.6 & 0 & dE,N & 19 & 8274.0 & 194.9288 & 27.9984 & 20.4 & Strong gradient in the sky \\ & & & & & & & & due to nearby galaxy/Irregular\\
  -- & 2 & dE & 25 & - & 194.9535 & 27.9155 & 20.4 & In area of low S/N\\
  -- & 0 & S0 & 18 & 8814.0 & 194.9988 & 27.9983 & 20.5 & Either a bar or offset NSC\\
  -- & 1 & VLSB & 19 & - & 194.9158 & 27.9905 & 20.5 & VLSB - not nucleated\\
  -- & 3 & dE & 23 & - & 195.0918 & 27.8987 & 20.6 & In halo\\
  SDSSJ130039.76\_280601.9 & 1 & VLSB & 2 & - & 195.1658 & 28.1008 & 20.9 & Nucleated but difficult to fit.\\
  SDSSJ125850.42\_274445.7 & 2 & dE,N & 75 & - & 194.71 & 27.746 & 20.9 & The fit is dubious -  \\ & & & & & & & & asymmetric residuals around the NSC.\\
  -- & 2 & dE & 15 & - & 195.1595 & 28.0024 & 21.0 & Covered by a star\\
  -- & 3 & dE,N & 25 & - & 194.9599 & 27.9216 & 21.0 & Crowded area: difficult to obtain good fit\\
  SDSSJ130011.81\_280504.0 & 2 & dE,N & 3 & - & 195.049 & 28.0846 & 21.1 & Blended with neighbour. \\ & & & & & & & & Likely nucleated.\\
  -- & 2 & dE & 23 & - & 195.0737 & 27.9368 & 21.1 & In halo\\
  SDSSJ126944.76\_275807.1 & 0 & dE & 18,19 & 9623.0 & 194.9363 & 27.9685 & 21.2 & VLSB in edge of frame\\
  -- & 3 & dE,N & 13 & - & 194.8752 & 28.0439 & 21.6 & Extremely elongated -- difficult to fit \\ & & & & & & & & and probably background\\
  SDSSJ130000.97\_275929.5 & 1 & dE,N & 18 & - & 195.0042 & 27.9929 & 21.6 & In halo of elliptical galaxy - nucleated\\
  -- & 3 & dE,N & 13 & - & 194.8776 & 28.0223 & 21.7 & In halo - nucleated\\
  SDSSJ130037.83\_275840.9 & 0 & dE & 15 & 4684.0 & 195.1576 & 27.9779 & 21.7 & Compact source\\
  -- & 2 & dE & 33 & - & 194.9011 & 27.8853 & 21.7 & VLSB - nonnucleated\\
  -- & 2 & dE,N & 19 & - & 194.9108 & 27.9496 & 21.8 & Unfittable - in halo (nucleated)\\
  -- & 2 & dE & 19 & - & 194.8825 & 28.0015 & 22.0 & Irregular galaxy\\
  SDSSJ125832.93\_272406.5 & 3 & dE,N & 46 & - & 194.637 & 27.402 & 22.0 & Irregular galaxy probably background\\
  -- & 2 & VLSB & 19 & - & 194.9204 & 27.9548 & 22.1 & Very crowded area - nucleation and \\ & & & & & & & & centre uncertain.\\
  -- & 3 & dE & 22 & - & 195.1553 & 27.9236 & 22.1 & In the spokes of a bright foreground star\\
  -- & 3 & dE & 2 & - & 195.1355 & 28.0606 & 22.2 & Fit affected by nearby bright galaxy\\
  -- & 3 & dE,N & 12 & - & 194.9351 & 28.0413 & 22.2 & Poor data quality - possible NSC \\ & & & & & & & & is masked out as cosmic.\\
  -- & 2 & dE,N & 19 & - & 194.9068 & 27.9686 & 22.2 & In halo\\
  -- & 3 & dE,N & 10 & - & 195.0617 & 28.0085 & 22.4 & Centre uncertain\\
  -- & 3 & dE,N & 24 & - & 195.0102 & 27.9497 & 22.4 & This galaxy is misclassified as dE,N\\
  -- & 3 & dE & 24 & - & 195.0385 & 27.9076 & 22.4 & Difficult to fit\\
  -- & 2 & dE,N & 45 & - & 194.6363 & 27.4596 & 22.4 & In edge and halo\\
  -- & 2 & dE & 10 & - & 195.0524 & 28.0085 & 22.5 & Difficult to fit\\
  -- & 2 & dE,N & 19 & - & 194.9121 & 27.9789 & 22.5 & Very difficult to fit, in 4874 halo\\
  -- & 3 & dE & 22 & - & 195.1334 & 27.9519 & 22.5 & Difficult to fit\\
\end{tabular}\caption{Sources excluded from the analysis.}\label{tab:ex}
\end{table}
\twocolumn
\newpage

\onecolumn
%\newpage
\begin{figure}
\includegraphics[width=0.32\textwidth]{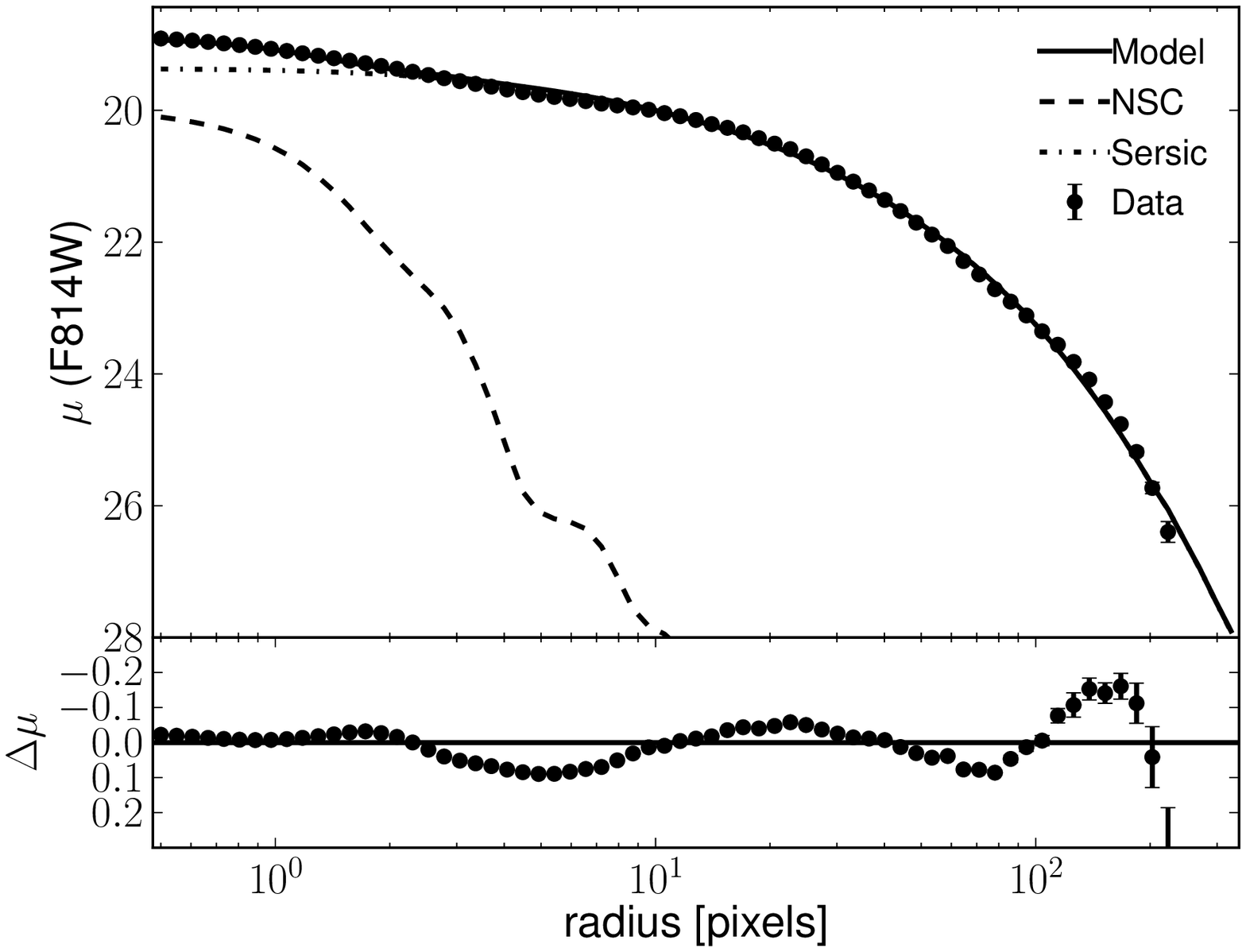}
\includegraphics[width=0.32\textwidth]{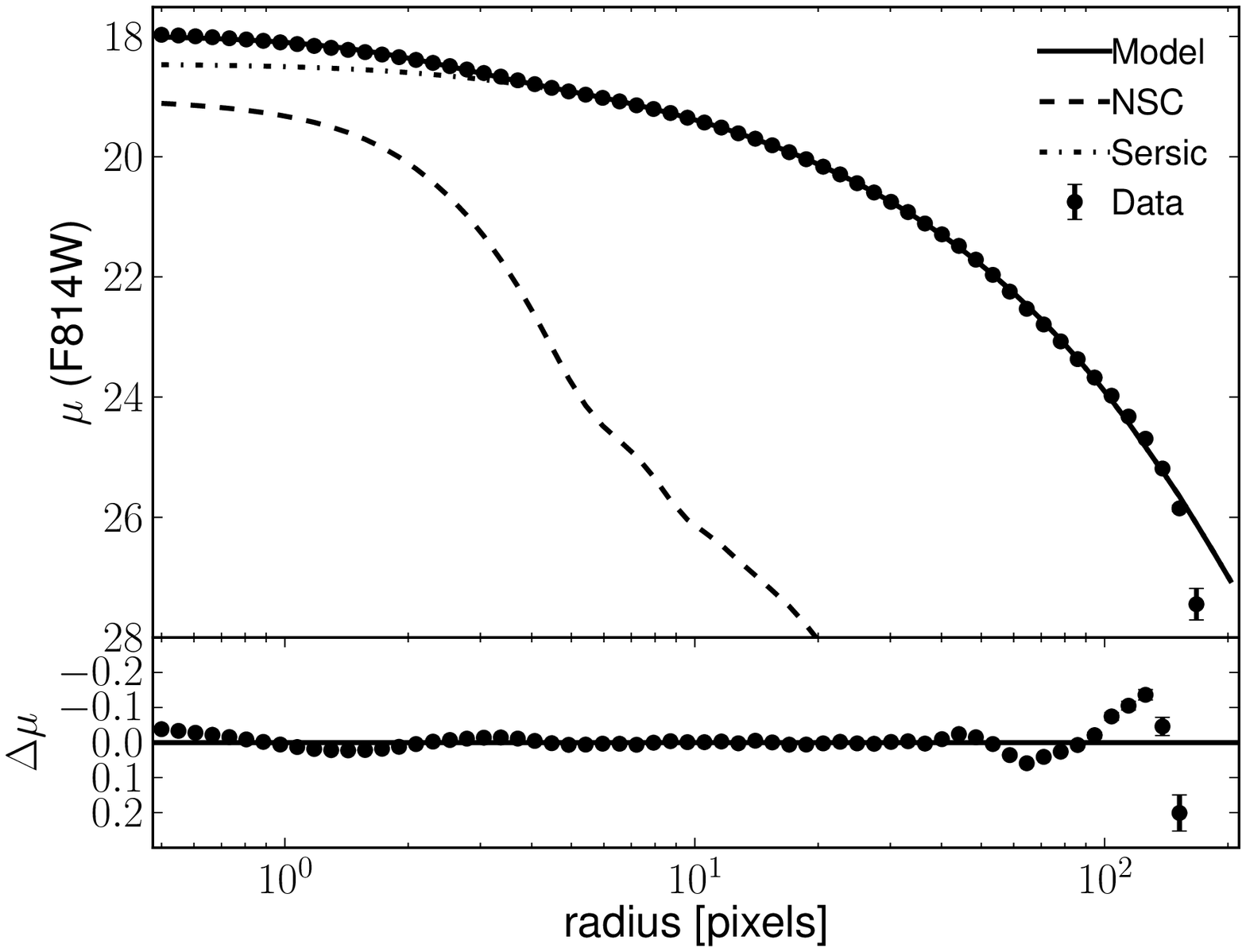}
\includegraphics[width=0.32\textwidth]{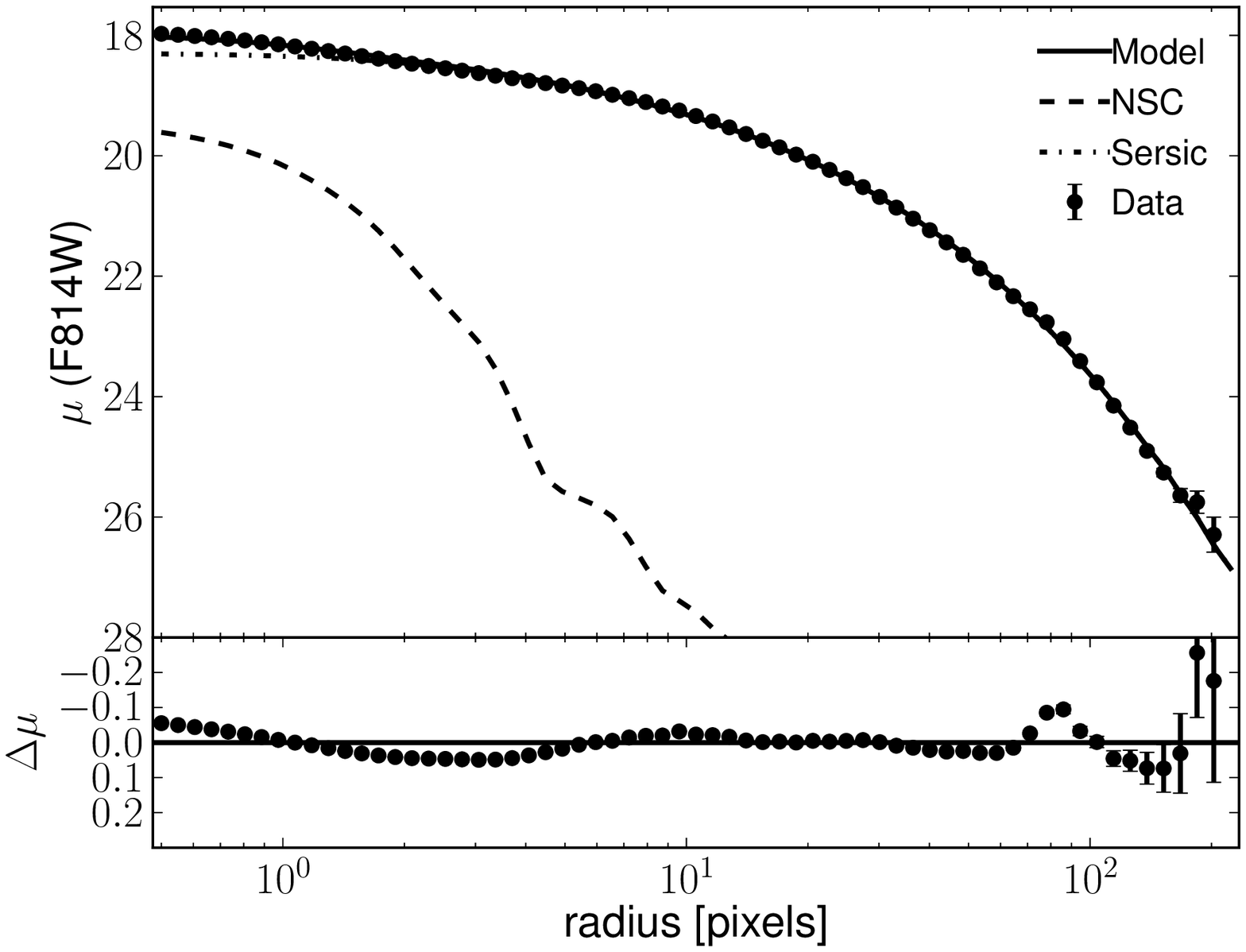}\\
%\end{figure}
%\begin{figure}
\includegraphics[width=0.32\textwidth]{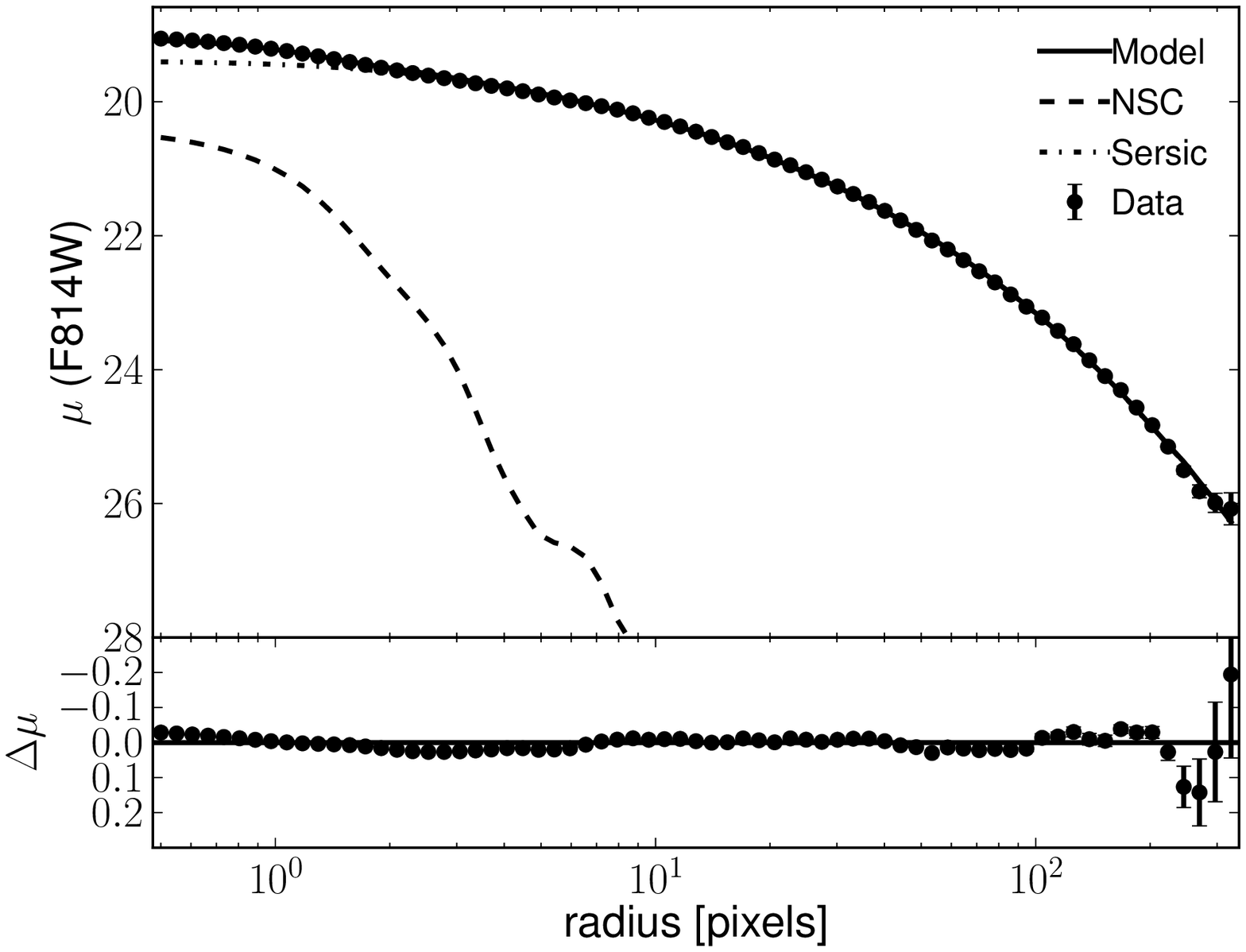}
\includegraphics[width=0.32\textwidth]{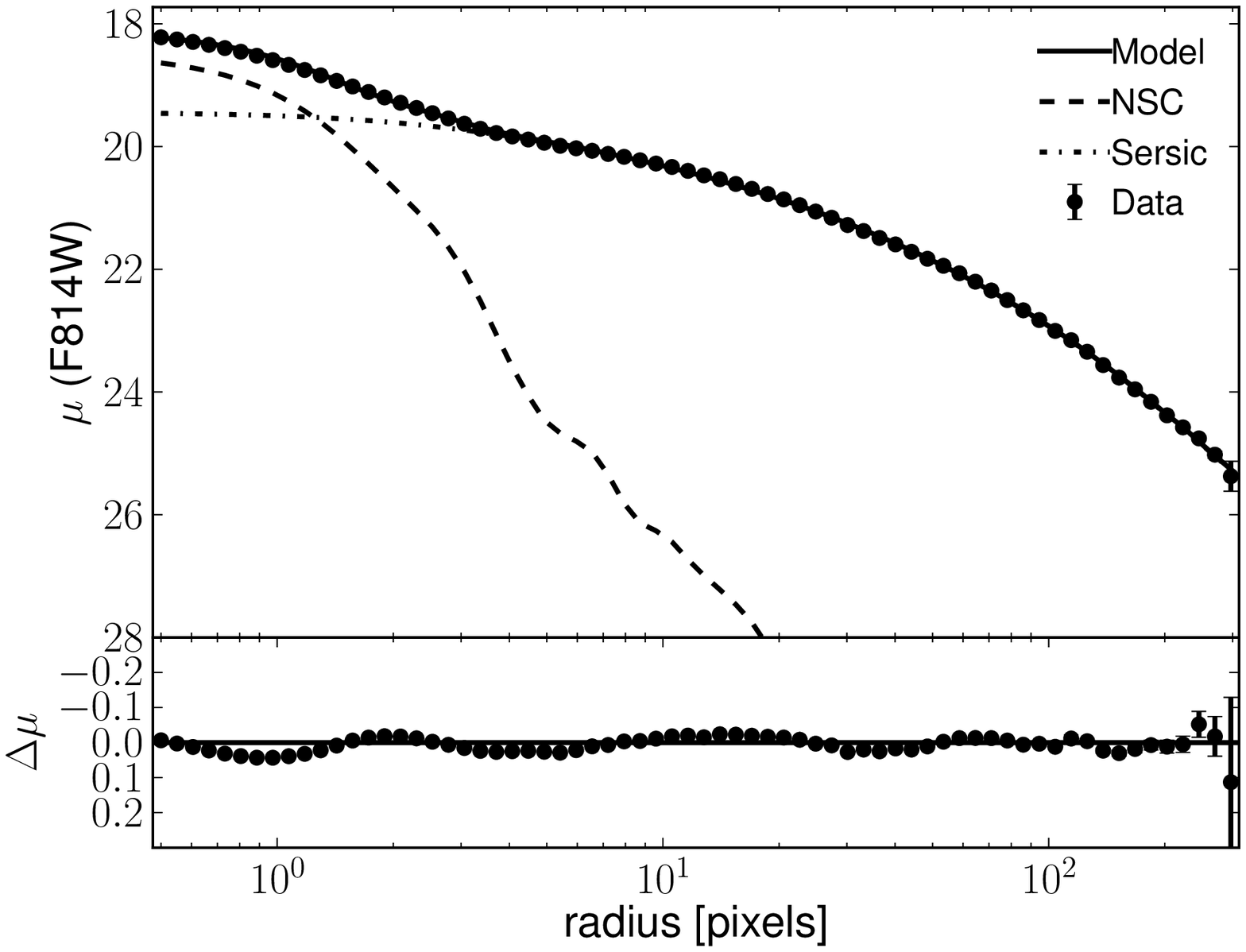}
\includegraphics[width=0.32\textwidth]{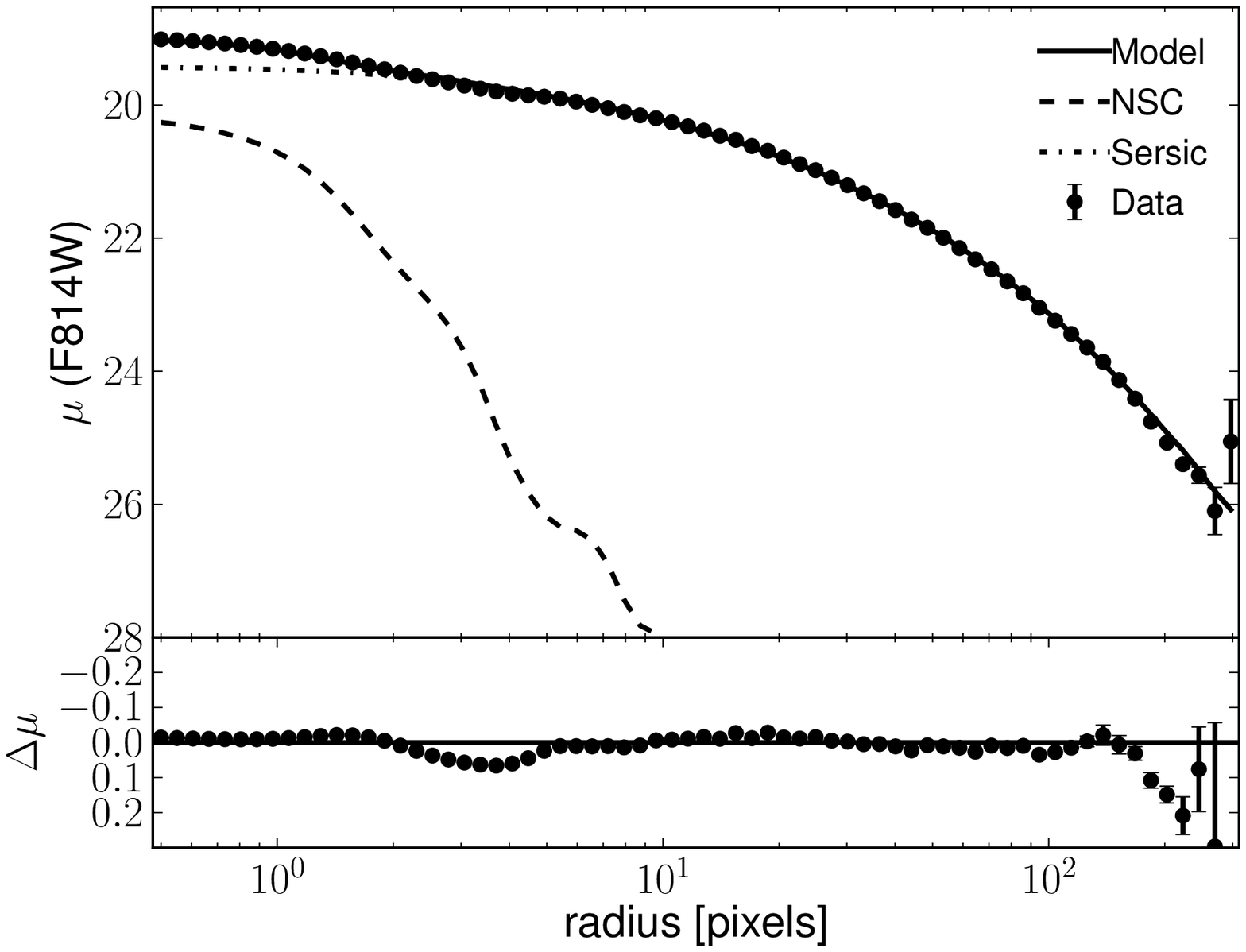}\\
%\end{figure}
%\begin{figure}
\includegraphics[width=0.32\textwidth]{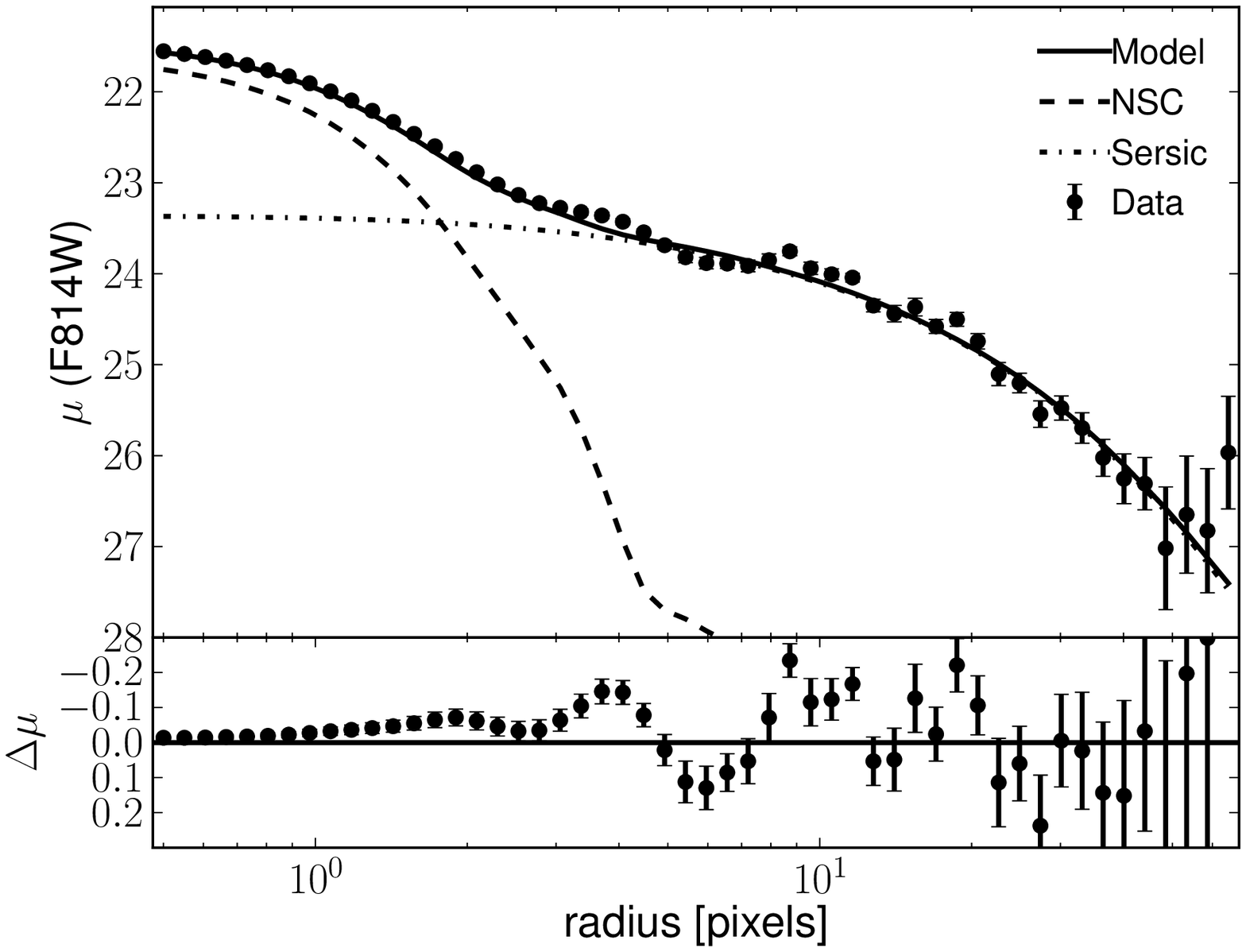}
\includegraphics[width=0.32\textwidth]{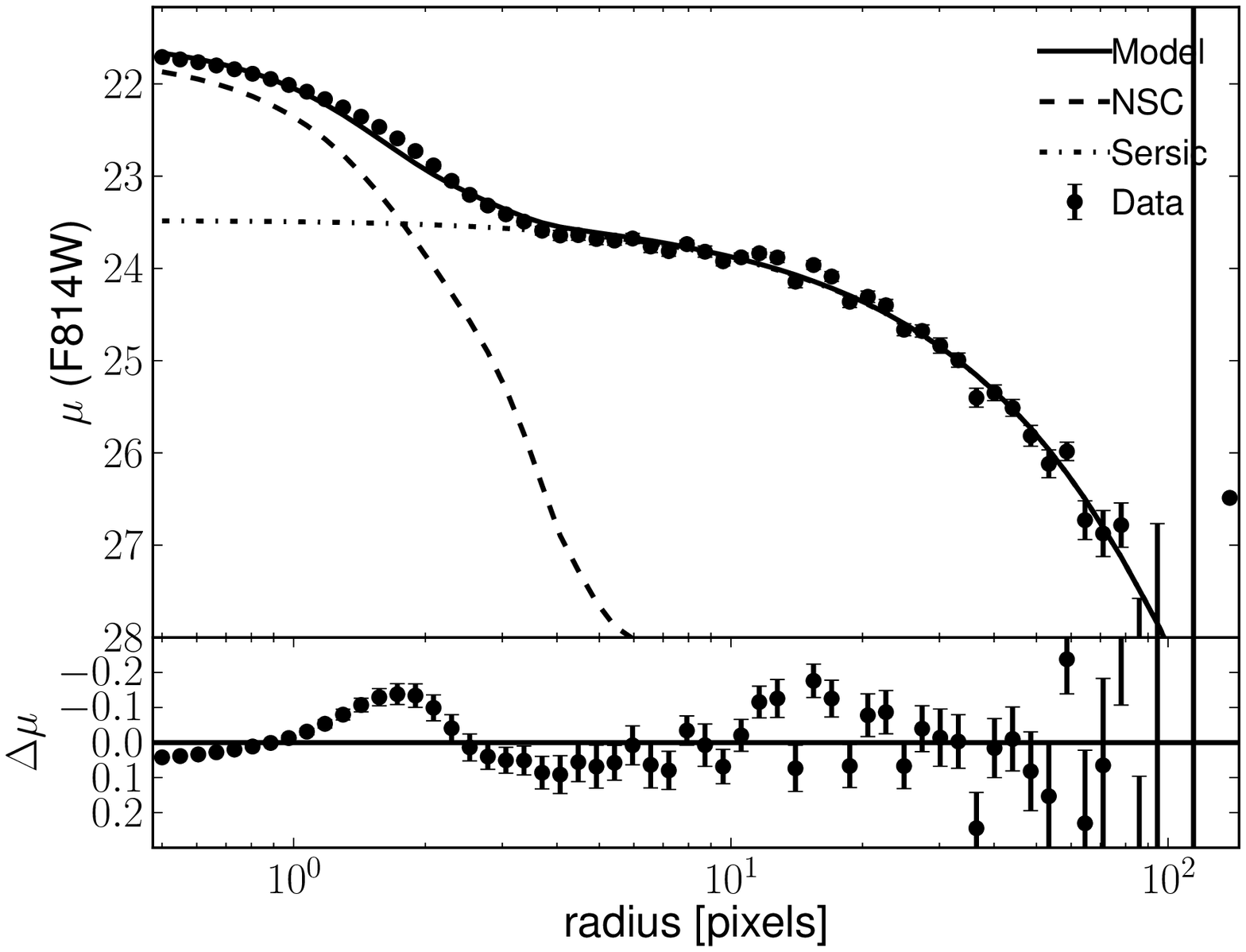}
\includegraphics[width=0.32\textwidth]{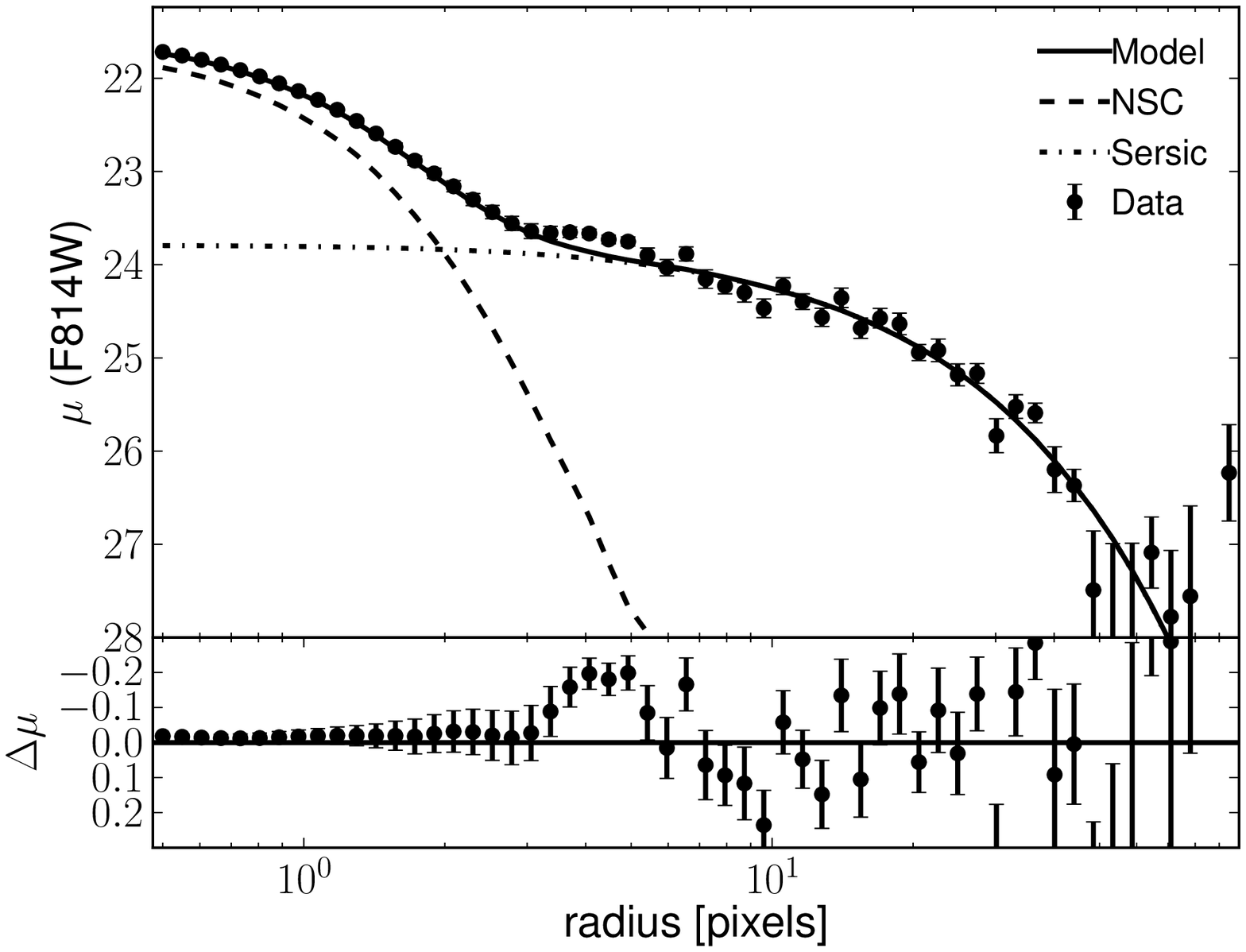}\\
%\end{figure}
%\begin{figure}
\includegraphics[width=0.32\textwidth]{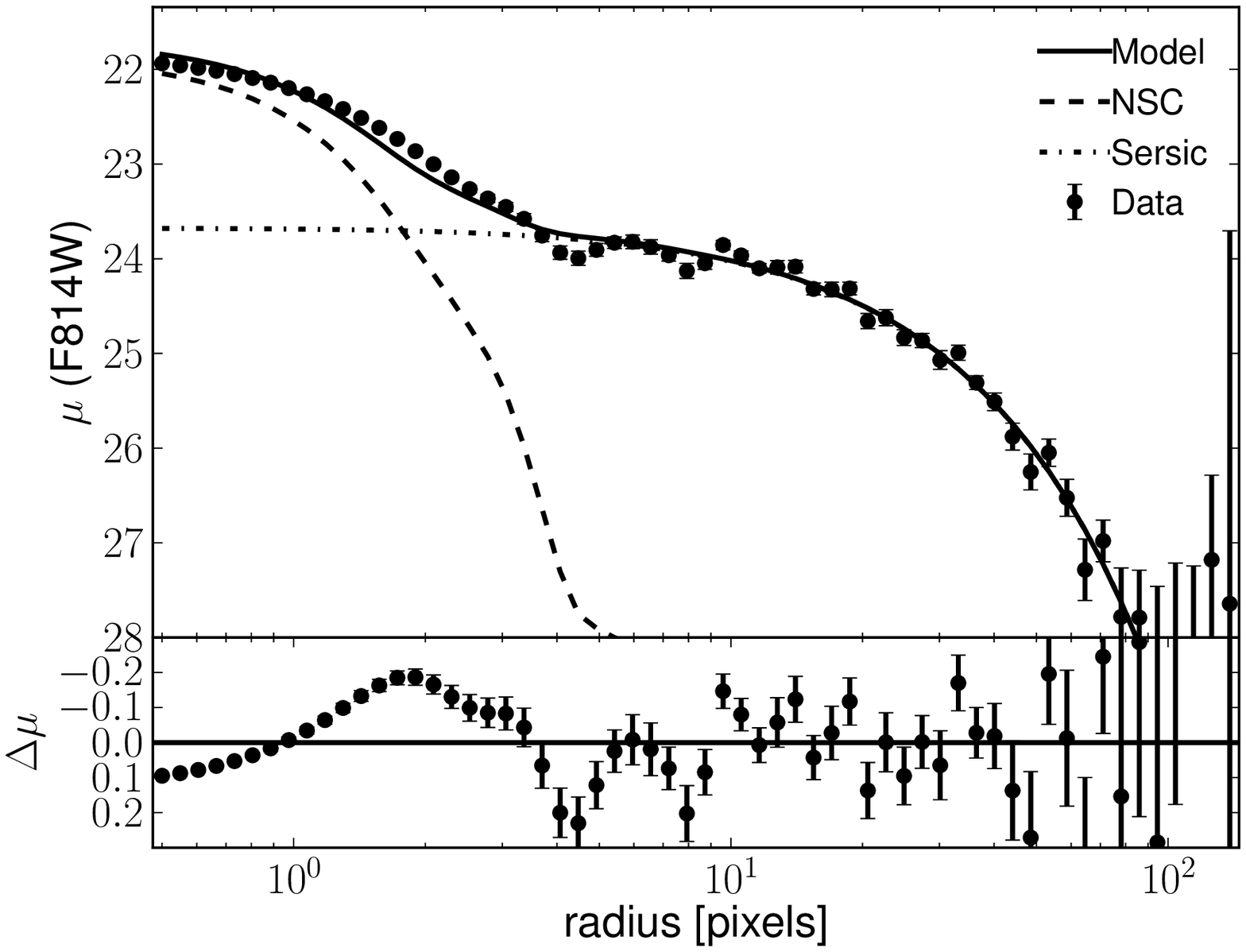}
\includegraphics[width=0.32\textwidth]{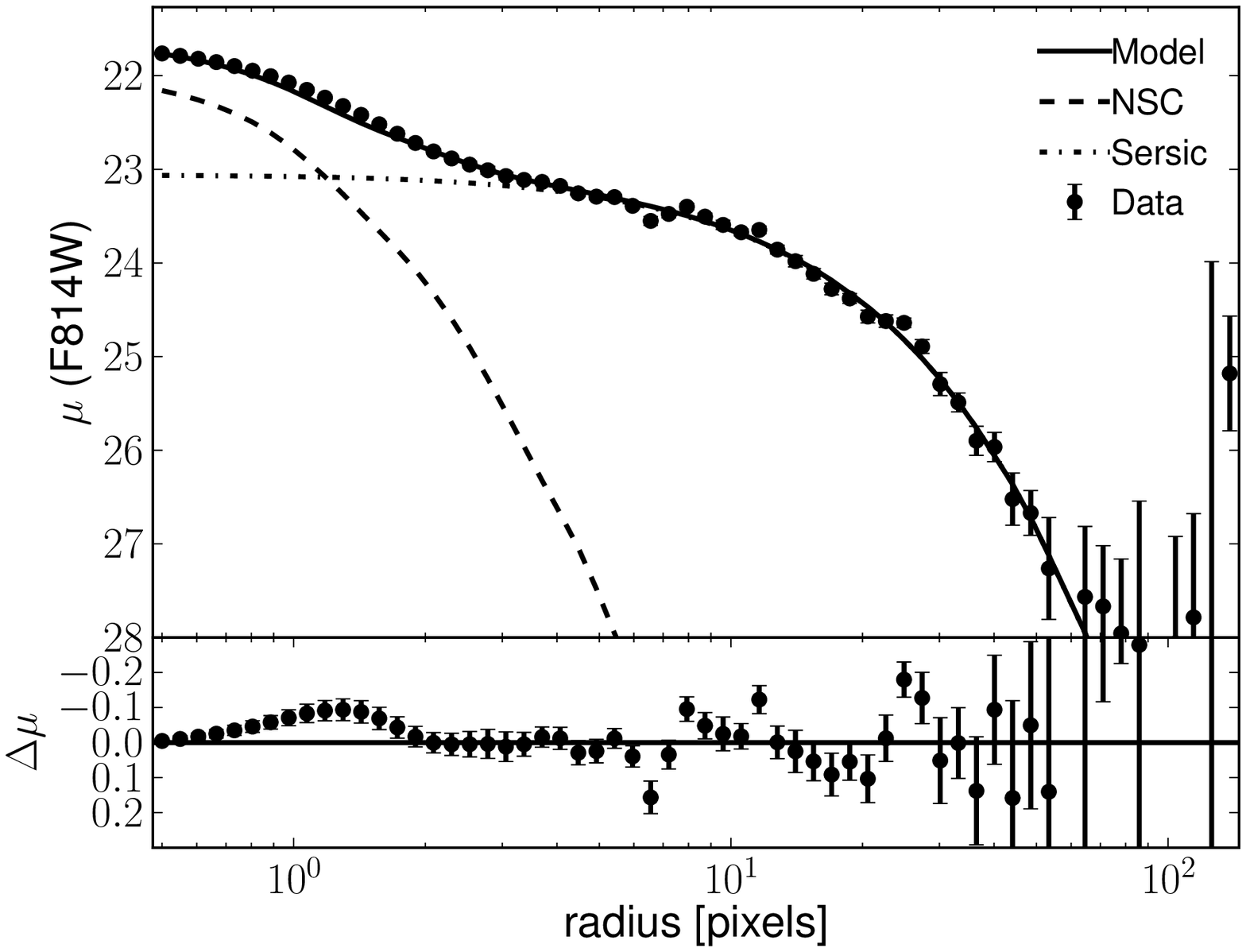}
\includegraphics[width=0.32\textwidth]{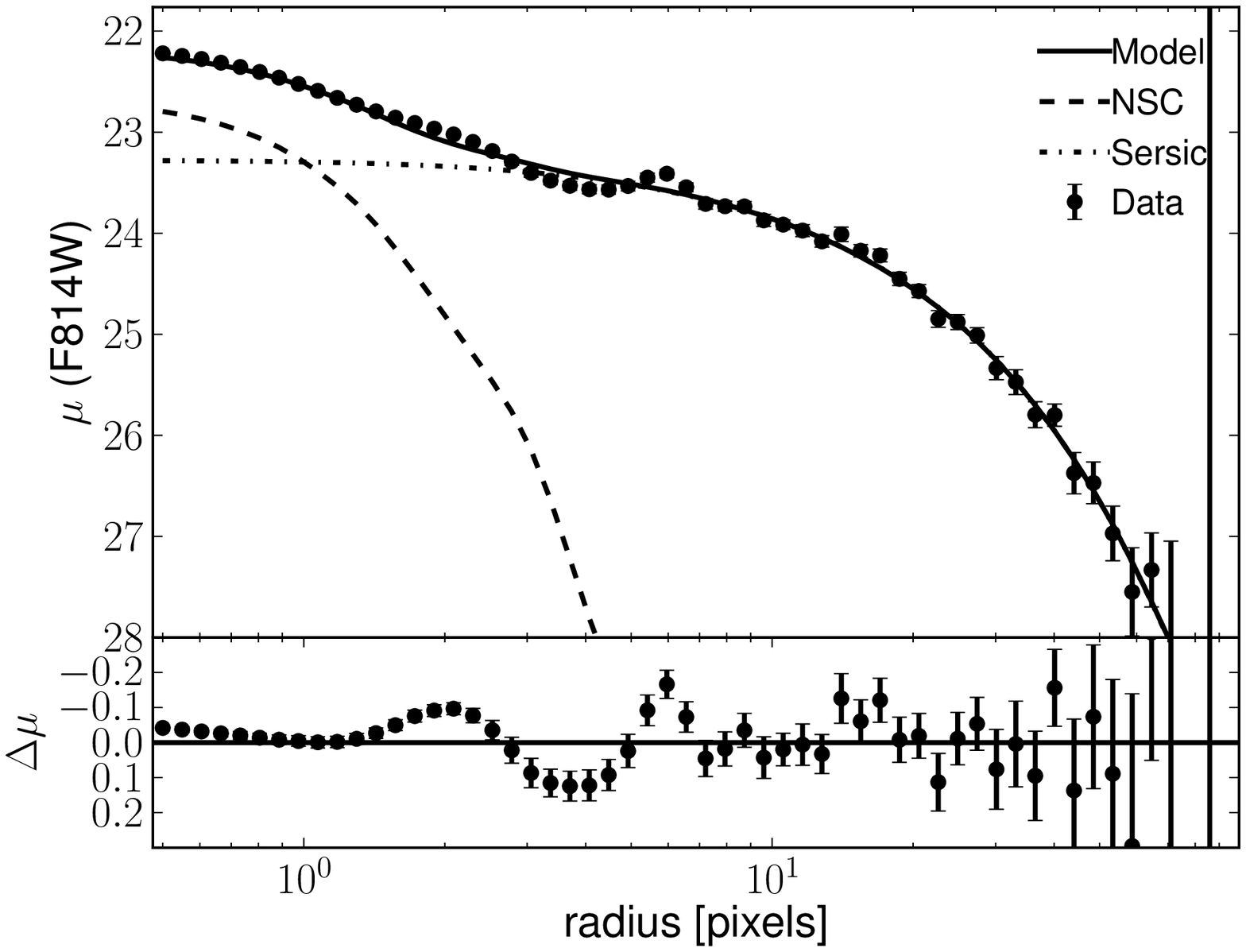}
\caption{Profiles of the brightest and faintest galaxies in the sample. From left to right, top to bottom: SDSSJ130018.54\_280549.7, LEDA126789, SDSSJ130041.19\_280242.4, SDSSJ130034.42\_275604.9, RB068, LEDA126815, COMAi125932.883p275800.05, COMAi125929.995p275348.12, COMAi125952.543p275824.21, COMAi13021.712p275650.16, COMAi125944.017p275615.29, COMAi125924.938p275320.35}
\end{figure}

\newpage
\section{Fitting code}\label{apx:bag}
We make use of our own custom fitting code, called \bagatelle, to measure the structural parameters of nuclei and host galaxies. Our code is a 2-dimensional fitting code, similar to \galfit\ \citep{PenHoImp02}, but based on a Bayesian framework for exploring the posterior distribution of the fitted parameters. It therefore allows comparison between simple and complex models (e.g. with and without nuclear cluster) and accurate estimates of covariances between parameters and more accurate error bars. It is thus similar to the \textsc{galphat} code of \citet{YooWeiKat11}.

\subsection{Model generation}
The basis for each model is a 1-dimensional surface brightness profile, for which we adopted variations of the S\'ersic profile (see Eq. \ref{eq:sersic}). We are forced to choose priors for each free parameter in the fit. These are not hard-coded, but we list them here. The priors on the host galaxy magnitude and effective radius are based on the output of \sex\ (we choose a 4.0 mag interval around the \sex\ magnitude and a factor 15 interval around the \sex\ radius). The S\'ersic index is allowed to vary between 0.3 and 10 for all galaxies. The nuclear cluster magnitude is allowed to vary between the lower-limit of the host galaxy magnitude and the detection limit of our data. For gaussian sources, we allowed the FWHM to vary between 0.4 and 15 pixels. We note that for good-enough data, the output parameters are basically insensitive to the choice of prior.

 Each 2-dimensional coordinate ($x, y$) corresponds to a surface brightness, where we assume that all isophotes are elliptical - i.e. we take into account the ellipticity and position angle of the model, but do not make use of generalized ellipses. The model centre falls exactly on a pixel centre (sub-pixel shifts of the model are treated during the convolution), and each pixel is subsampled by a factor $20\times20$, which is increased in the centre of the galaxy to a factor over 100, depending on the steepness of the profile.

\subsection{PSF convolution}
The model is then convolved by the PSF. For this, we first convolve the PSF with a two-dimensional modulated sinc function (to account for sub-pixel shifts), and then convolve the model with it. Convolution is done in the Fourier domain. We use the fast fourier transform modules of the FFTW3 library. We note that both the model and surrounding zero-pads are sufficiently big. After convolution, an optional sky background is added to the model.

\subsection{Comparison with data}
For each pixel, we calculate the probability that it was generated by the model. For this, we assume gaussian errors. The total likelihood is the product of the likelihoods of the individual pixels, except those that are flagged as bad.

\subsection{Likelihood exploration}
The likelihood is explored using the nested sampling algorithm \citep{Ski04}. The benefit of using nested sampling over ordinary MCMC codes is its ability to accurately calculate the Bayesian evidence for a model. The algorithm carries out a one-dimensional integral of likelihood samples over prior mass. New likelihood samples are generated using the ellipsoidal sampling algorithm of \citep{MukParLid06}. We do modify the standard way of calculating the mean and variance of the live sample for angle variables (in our case the PA of the galaxy). Instead, we calculate the mean angle of a collection of angles $\theta_i$ by the arctangent of the mean cosine and sine of the angles:
\begin{eqnarray}
\left<\theta\right> = \tan^{-1}\left(\frac{\left<\sin(\theta_i)\right>}{\left<\cos(\theta_i)\right>}\right),
\end{eqnarray}
where $\left<\ldots\right>$ denotes an average.

After the nested sampling has finished, we calculate the mean and variance of each free parameter by taking the mean and second moment of the posterior weight, as advertised in \citet{Ski04}. The code automatically generates a model and residual image.

\section{Comparison with Weinzirl et al.}\label{apx:comp_tw}
\begin{figure*}
\begin{minipage}{172mm}
\center
\scalebox{0.42}[0.42]{\includegraphics{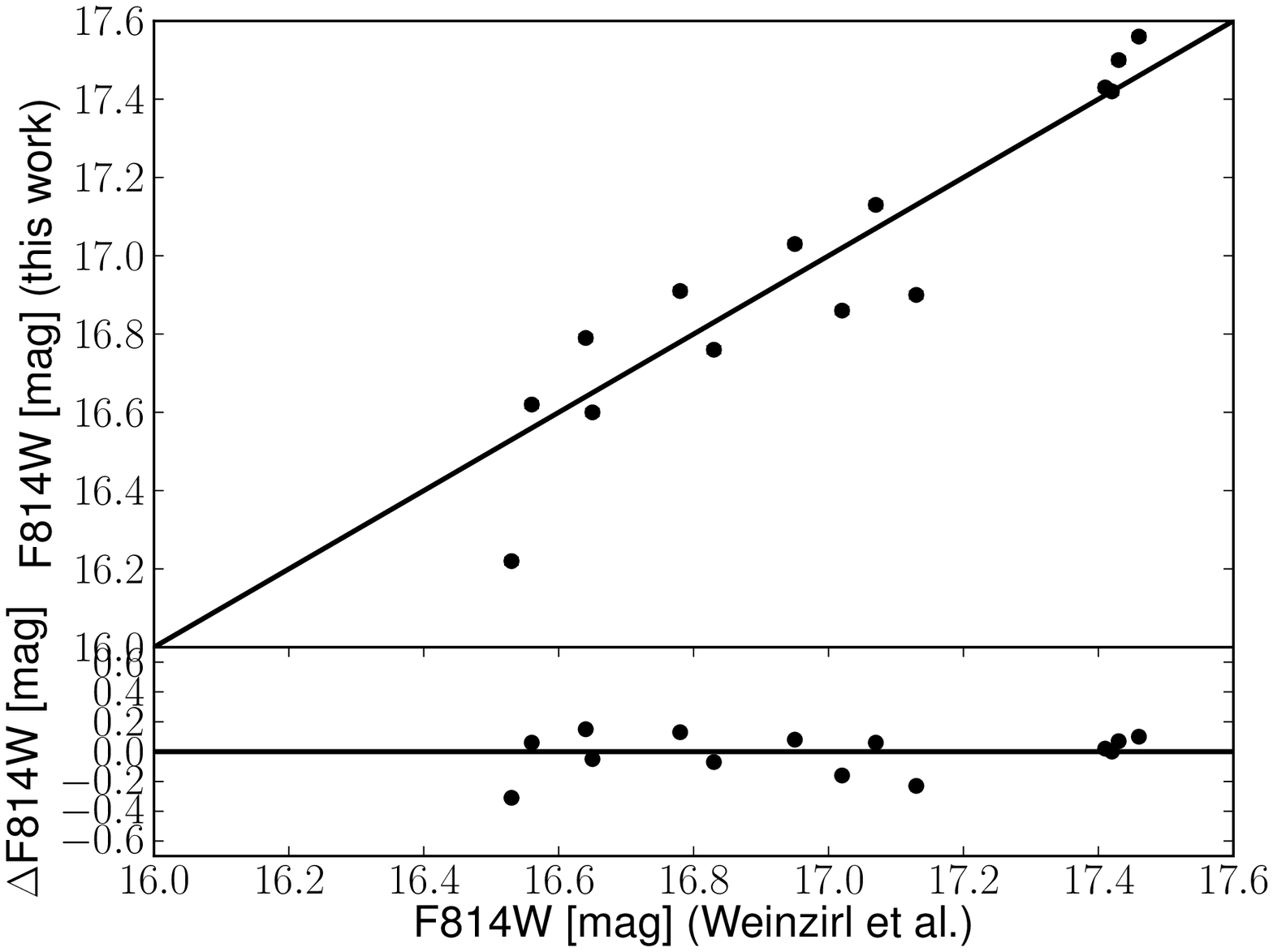}}
\scalebox{0.42}[0.42]{\includegraphics{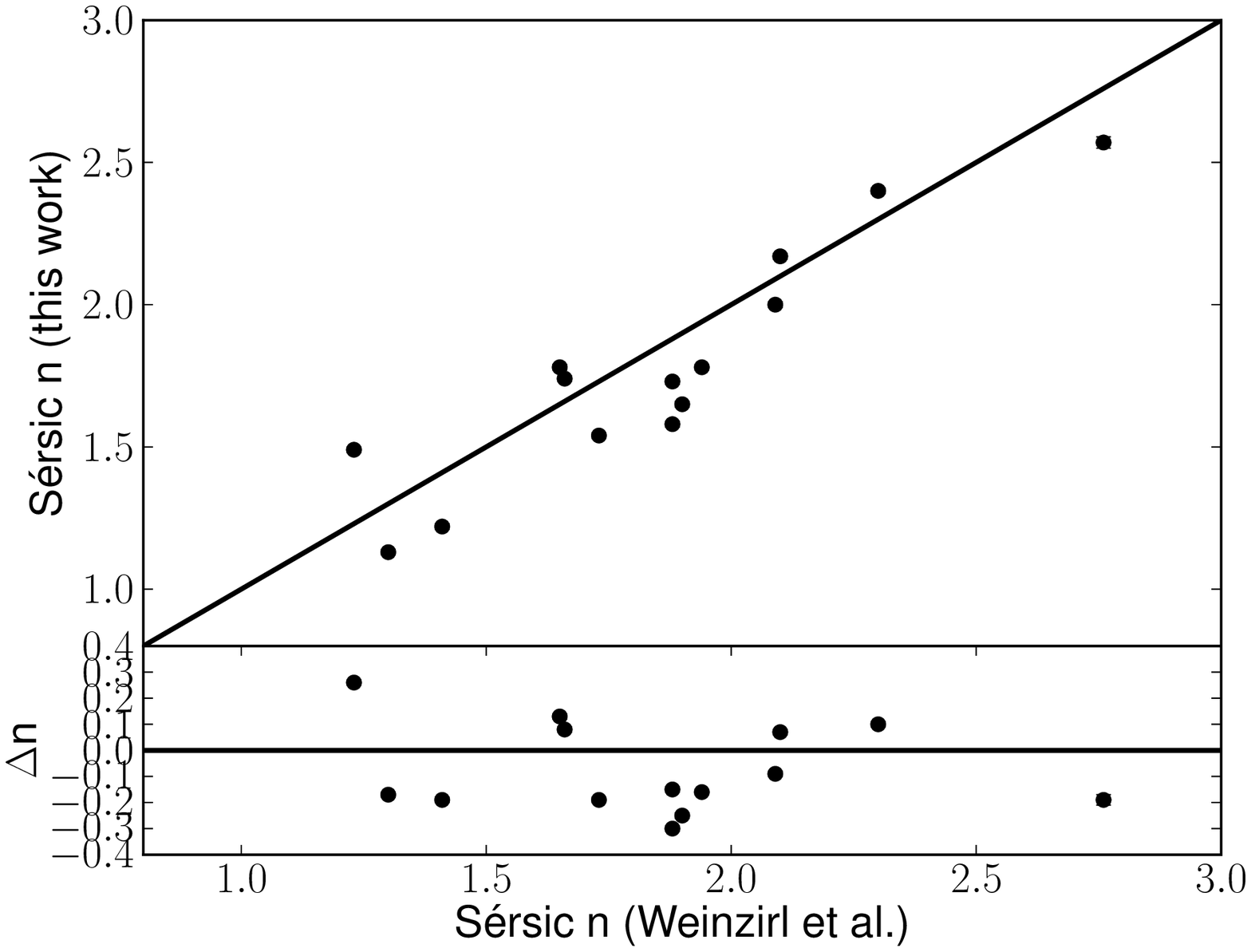}}\\
\scalebox{0.42}[0.42]{\includegraphics{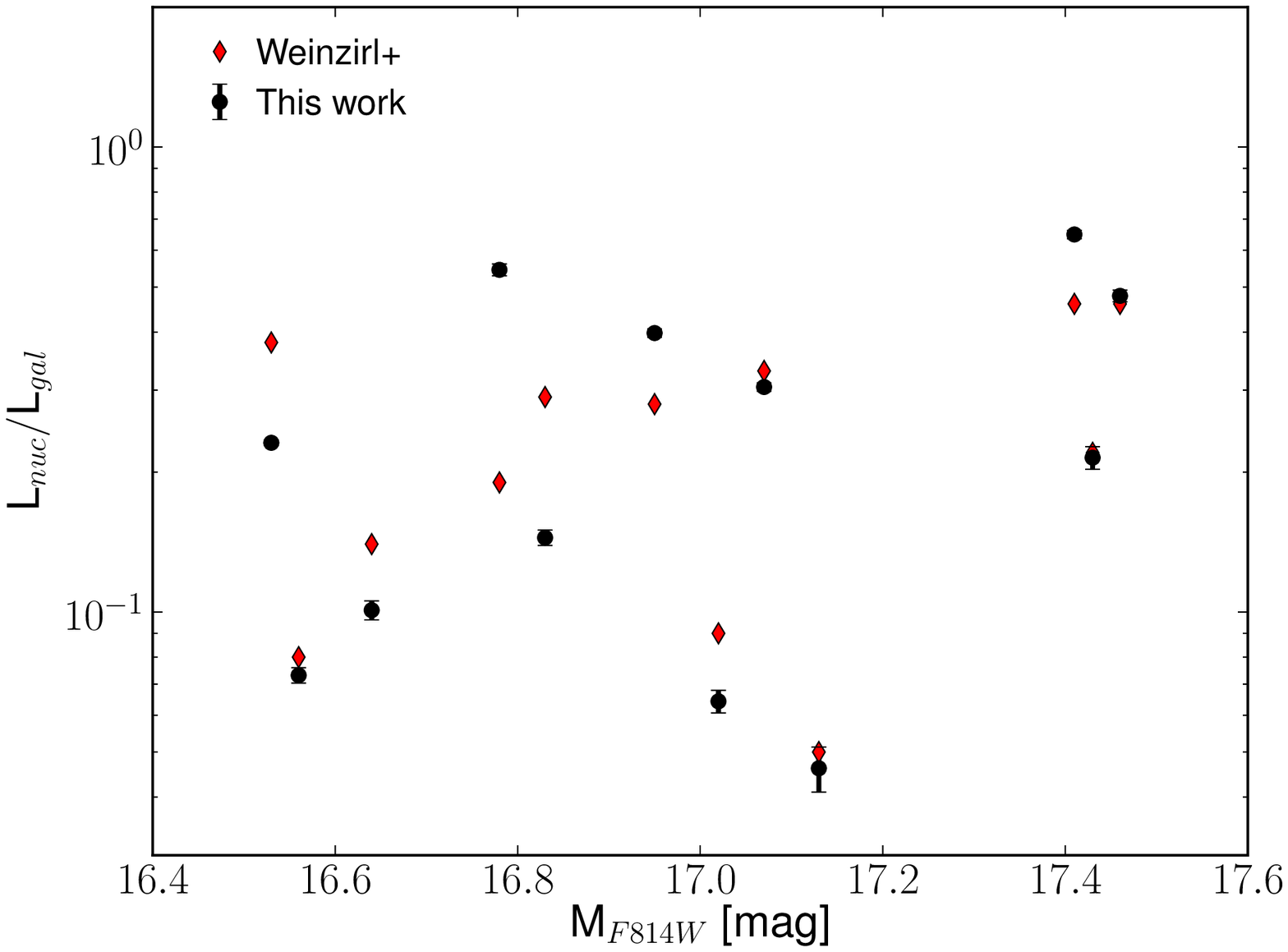}}
\caption{Comparison of the structural parameters for galaxies in common with Weinzirl et al. 2014. {\bf Left upper panel:} F814W host galaxy magnitude comparison. {\bf Right upper panel:} S\'ersic index from this work plotted against the S\'ersic index found by Weinzirl et al. {\bf Lower panel:} Fraction of the total luminosity emitted by the nucleus as a function of host galaxy magnitude.}\label{fig:comp_tw}
\end{minipage}
\end{figure*}
In Fig. \ref{fig:comp_tw} we compare the structural parameters fitted by \citet{ WeiJogNei14}with \galfit\ for the 15 galaxies in common. We note that we fit the nuclear star cluster in some cases by a gaussian and use a different PSF, which leads to a different luminosity than the point source (used in all cases) by Weinzirl et al. For the brightest source in common, we find a difference in the host galaxy magnitude of 0.3 mag. The average difference in magnitude is 0.01 $\pm 0.03$ mag, with a standard deviation of 0.13. Exluding the 0.3 mag offset source lowers the standard deviation to 0.1 mag. The difference in the logarithm of the S\'ersic index $\ln(n)$ is $0.04 \pm 0.03$, with a standard deviation of $0.10$. Despite the use of different codes to fit the galaxies, different masking schemes, different weighting and different PSFs, the comparison is excellent.
\twocolumn
\end{appendix}

\end{document}